\documentclass[prx,twocolumn,floatfix,superscriptaddress,longbibliography,nofootinbib]{revtex4-2}
\usepackage[utf8]{inputenc}
\usepackage{graphicx}
\usepackage{hyperref}
\usepackage{amssymb}
\usepackage{amsmath}
\usepackage{microtype}
\usepackage[table,dvipsnames]{xcolor}
\usepackage{braket}
\usepackage{dsfont}
\usepackage[export]{adjustbox}
\usepackage{booktabs}
\usepackage{multirow}
\usepackage{tabularx}
\usepackage{soul}
\usepackage{setspace}
\usepackage{newtxtext}
\usepackage[smallerops,varg,cmintegrals]{newtxmath}
\usepackage[inline]{enumitem}
\usepackage{mathtools}
\usepackage[caption=false,subrefformat=parens,labelformat=parens]{subfig}
\usepackage{xpatch}

\makeatletter
\patchcmd{\@ssect@ltx}
    {\addcontentsline{toc}{#1}{\protect\numberline{}#8}}
    {}
    {}
    {}
\makeatother

\hypersetup{
    colorlinks = true,
    linkcolor  = blue,
    citecolor  = blue,
    urlcolor   = blue,
    linktocpage= true
}
\numberwithin{equation}{section}

\newcommand{\DR}{\check{D}^\text{R}}
\newcommand{\DA}{\check{D}^\text{A}}
\newcommand{\DK}{\check{D}^\text{K}}

\renewcommand{\vec}[1]{\mathbf{#1}}

\newcommand{\ChiOne}{\chi^{(1)}_{x \mkern-1mu x }}
\newcommand{\ChiTwo}{\chi^{(2)}_{x \mkern-2mu x \mkern-2mu x }}
\newcommand{\ChiThree}{\chi^{(3)}_{x \mkern-2mu x \mkern-2mu x \mkern-2mu x}}
\newcommand{\ChiN}{\chi^{(n)}_{x \mkern-2mu x \cdots x}}

\newcommand{\CoNbO}{CoNb$_2$O$_6$}

\renewcommand\Im{\operatorname{Im}}

\newcommand{\fermionInt}{\int \mathcal{D}[\bar{\psi}, \psi]\, }

\DeclareMathOperator{\Pf}{Pf}       
\DeclareMathOperator{\Tr}{Tr}       

\DeclarePairedDelimiter\abs{\lvert}{\rvert}%
\DeclarePairedDelimiter\norm{\lVert}{\rVert}%

\renewcommand{\thesection}{\arabic{section}}
\renewcommand{\thesubsection}{\thesection.\arabic{subsection}}
\renewcommand{\thesubsubsection}{\thesubsection.\arabic{subsubsection}}
\makeatletter
\renewcommand{\p@subsection}{}
\renewcommand{\p@subsubsection}{}
\makeatother

\makeatletter
\g@addto@macro\bfseries{\boldmath}
\makeatother

\makeatletter
\let\oldabs\abs
\def\abs{\@ifstar{\oldabs}{\oldabs*}}
\let\oldnorm\norm
\def\norm{\@ifstar{\oldnorm}{\oldnorm*}}
\makeatother

\begin{document}

\title{Extracting spinon self-energies from two-dimensional coherent spectroscopy}
\author{Oliver Hart\,\href{https://orcid.org/0000-0002-5391-7483}{\includegraphics[width=6.5pt]{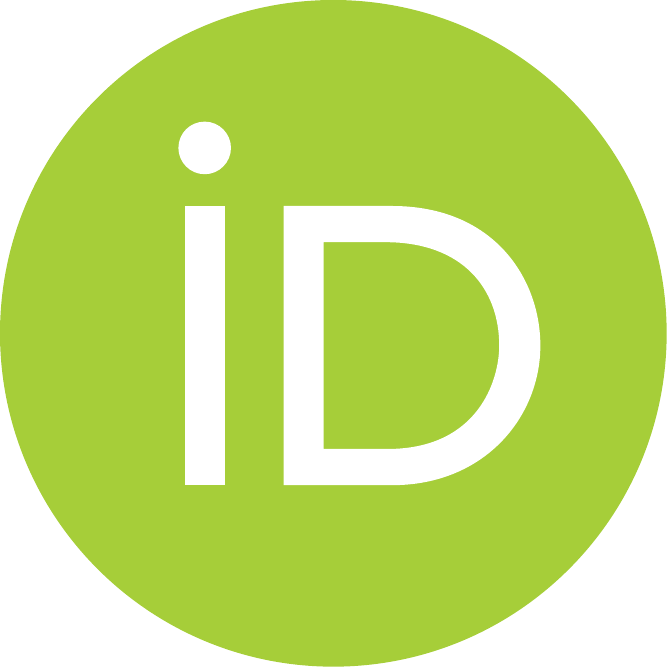}}}
\email{oliver.hart-1@colorado.edu}
\affiliation{Department of Physics and Center for Theory of Quantum Matter, University of Colorado, Boulder, Colorado 80309 USA}
\author{Rahul Nandkishore}
\affiliation{Department of Physics and Center for Theory of Quantum Matter, University of Colorado, Boulder, Colorado 80309 USA}
\affiliation{Department of Physics, Stanford University, Stanford, California 94305, USA}
\date{August 26, 2022}

\begin{abstract}
    \setstretch{1.1}
    Two-dimensional coherent spectroscopy (2DCS) is a nonlinear spectroscopy technique capable of identifying whether apparent continua in linear response are made out of multiplets of sharp deconfined quasiparticles. This makes it a potent tool for experimental identification of fractionalized phases. Previous discussions have focused on limits where the quasiparticles in question are infinitely long lived. In this manuscript we discuss 2DCS in the regime where the fractionalized quasiparticles can themselves decay. We introduce a powerful path integral based approach, whereby the computation of nonlinear susceptibilities reduces to an efficient exercise in diagrammatic perturbation theory. We apply this method to compute the 2DCS response of the one-dimensional transverse field Ising model, in the presence of integrability-breaking perturbations. We discuss aspects of the self-energy of the fractionalized quasiparticles that may be extracted via 2DCS, such as the momentum-dependent decay rate.
\end{abstract}

\maketitle


\tableofcontents

\section{Introduction}

Probing the linear response of equilibrium solid-state systems is the bread and butter of experimental condensed matter physics.
However, the relative simplicity of such experiments often comes at the cost of complete information.
Throughout the last century, theorists have endeavored to predict and characterize ever more exotic and intricate phases of matter, which may not possess unambiguous signatures at the level of linear response.
One particularly pertinent example relates to topologically-ordered phases of magnetic systems~\cite{balents2010frustrated,SavaryBalents2016,QSLStates2017,KnolleFieldGuide},
which feature a topological ground state degeneracy~\cite{WenDegeneracy}, long-range entanglement~\cite{HammaBipartite,KitaevPreskillEE,LevinWenEE}, and fractionalized quasiparticles, which cannot be created in isolation.
Only the latter property can be expected to give rise to meaningful experimental fingerprints, but the presence of a multi-particle continuum produces mundane, diffuse features in the dynamical spin structure factor~\cite{KnolleFieldGuide}, which can be difficult to differentiate from the more commonplace effects of, e.g., disorder and a finite intrinsic quasiparticle lifetime.
The dearth of sharp signatures has made the discovery of new candidate material realizations particularly challenging.

Thankfully, recent theoretical progress has been accompanied by a concomitant rise in experimental capability.
For instance, pump-probe spectroscopy, in which a system is brought strongly out of equilibrium and subsequently `probed' with a weaker pulse, sheds light on the nonequilibrium relaxation dynamics of solid-state systems on ultrafast time scales~\cite{BasovRevModPhys,GiannettiUltrafast}.
In this paper, we consider a related technique: two-dimensional coherent spectroscopy (2DCS)~\cite{Kuehn20112DCS,Woerner20132DCS}, in which multiple \emph{weak} pulses probe a system's nonlinear response about equilibrium via multi-time correlation functions.
While the system is not pumped into a far-from-equilibrium state, a system's nonlinear response properties nevertheless provide information beyond that revealed by linear response alone.

2DCS is a well-established technique in the optical and radio frequency ranges in the context of, e.g., biophysics and physical chemistry~\cite{Aue1976Two,Mukamel1999Principles,HammZanni2011,CundiffMukamel}, but it has only recently been applied to solid-state systems~\cite{Kuehn20112DCS,Woerner20132DCS,Lu2016_2DCS,Lu2017SpinWaves,Mahmood2021Observation} due to advancements in high-intensity terahertz (THz) sources.
On the theory side, the application of 2DCS to condensed matter systems remains in its infancy, but it has already been used to characterize and discover signatures of the Kitaev honeycomb model~\cite{Choi2DCSKitaev}, random quantum magnets~\cite{Parameswaran2020RandomTFIM}, gapped spin liquids and fractonic systems~\cite{Nandkishore2021Spectroscopic}, Luttinger liquids~\cite{Li2021Lensing}, and other systems~\cite{Fava2021Hydrodynamic,GerkenTopoPhases,PhucManyBody}. 
Most relevant to this manuscript is the ability of 2DCS to distinguish between ``inhomogeneous'' and ``homogeneous'' broadening~\cite{Mukamel1999Principles,HammZanni2011,CundiffMukamel}, which refer to broadening due to a continuum of sharp modes and broadening due to a finite lifetime, respectively.
As shown in Ref.~\cite{WanArmitage2019}, this allows 2DCS to partially address the problem of identifying fractionalization, which it does by disentangling the relative contributions of disorder, decay, and the multi-particle continuum.

The question that we set out to answer is: Precisely what information can be extracted from 2DCS in a many-body setting when a quasiparticle's finite lifetime is provided by realistic interactions. 
The main contributions of this paper in this direction are twofold. First, we develop a framework based on a real-time path integral approach capable of describing nonlinear response functions, which provides a convenient language to treat \emph{interacting} quantum many-body systems via diagrammatic many-body perturbation theory (MBPT).
This is significant, since many studies to date have focused either on noninteracting systems, or on models comprised of effective two-level systems (see, however, Refs.~\cite{Nandkishore2021Spectroscopic,Fava2021Hydrodynamic,PhucManyBody}).
Second, inspired by Ref.~\cite{WanArmitage2019}, we apply this technique to the one-dimensional (1D) transverse field Ising model (TFIM).
Specifically, we extend the results of Ref.~\cite{WanArmitage2019} by introducing experimentally-relevant perturbations that make the model interacting, endowing the quasiparticles with a nonzero self-energy, which we treat within MBPT.
 When the single-domain-wall dispersion relation overlaps with the three-domain-wall continuum, spontaneous decay of domain walls is possible via the emission of a pair of domain walls. Otherwise, nonzero temperatures, leading to a thermally-excited background of domain walls, are needed to give rise to decay due to scattering.
We consider different parameter regimes so as to produce both types of decay.
In both cases, the resulting frequency-dependent self-energy gives rise to additional spectral features, whose effect on the 2DCS spectrum we describe quantitatively.

The paper is structured as follows.
Due to the long and technical nature of the calculations involved in evaluating the nonlinear response in the presence of interactions,
we opt to provide a brief summary of our key physical results prior to expanding upon the details of the calculations used to obtain them.
This summary is given in Sec.~\ref{sec:summary-of-results}, which also includes a short introduction to the 2DCS protocol.
We then describe in detail the framework used to evaluate the nonlinear response in Sec.~\ref{sec:weak-interactions}. This starts with an introduction to the TFIM and the perturbations that we consider, followed by discussion of its noninteracting Green's functions on the closed, real-time (Keldysh) contour. Next, we describe how various nonlinear susceptibilities can be obtained via differentiation of a generating functional, which sets the stage for introducing interactions. The self-energy on the Keldysh contour is then evaluated to second order in the interactions. 
Finally, we examine how the mixed absorptive and dispersive character of the rephasing signal can be ameliorated by ``phase untwisting''.
We close with a discussion of our results in Sec.~\ref{sec:conclusions}.


\section{Summary of main results}
\label{sec:summary-of-results}

We work throughout the manuscript with the one-dimensional (1D) transverse field Ising model (TFIM) at nonzero temperatures in the presence of additional, experimentally-relevant perturbations.
The unperturbed Hamiltonian is given by
\begin{equation}
    \frac{\hat{H}_0}{J} = - \sum_{i=1}^{L} \hat{\sigma}_{i}^z \hat{\sigma}_{i+1}^z - g \sum_{i=1}^L \hat{\sigma}_i^x 
    \, ,
    \label{eqn:TFIM}
\end{equation}
where $g$ is the ratio of the static, `transverse' magnetic field strength to the Ising interactions and $\hat{\sigma}_i$ are the Pauli matrices for spin-$1/2$ degrees of freedom.
We consider exclusively the ferromagnetic regime of the model, $g<1$, in which Ising interactions between spins are predominant with respect to the transverse field.
In this regime, the quasiparticles can be viewed as domain walls dressed by quantum fluctuations, which can only be (locally) created in pairs.
Motivated in part by the prototypical material realization of the 1D TFIM, \CoNbO~\cite{coldea2010quantum,Morris2014,Cabrera2014Excitations,Kinross2014,Robinson2014breakdown,liang2015heat,Morris2021duality,Fava}, we consider
the addition of residual `XY'-like interactions between the spins~\cite{Robinson2014breakdown}: In addition to Ising interactions $\propto \hat{\sigma}^z_i \hat{\sigma}^z_j$ between neighboring spins $i$ and $j$, aligned along $z$, we look at the effect of weak interactions in the $xy$ plane of the form $\propto \hat{\sigma}^x_i \hat{\sigma}^x_j + \hat{\sigma}^y_i \hat{\sigma}^y_j$, breaking the perfect Ising anisotropy of Eq.~\eqref{eqn:TFIM}. Absent any perturbations, the 1D TFIM~\eqref{eqn:TFIM} is exactly solvable via a Jordan-Wigner transformation from spins to noninteracting fermions~\cite{Lieb1961Soluble,Pfeuty1970}. Under the same transformation, in addition to a renormalization of the single-particle band structure, the residual XY interaction maps to a density-density interaction between the Jordan-Wigner fermions, which provides the fermions with a nonvanishing self-energy whose effects we study quantitatively.
While additional perturbations are likely necessary for an accurate description of \CoNbO~\cite{Robinson2014breakdown,Morris2021duality}, XY-like interactions give rise to the salient phenomenology without additional complexities. Further, the framework that we present is perfectly capable of handling additional perturbations, which merely give rise to a modification of the single-particle dispersion and the interaction vertices\footnote{In order for this description to be effective, the perturbations should map to \emph{local} interactions in the fermionic language. A notable exception is a `longitudinal' field, parallel to the Ising easy axis, which inherits its nonlocality from the Jordan-Wigner string in Eq.~\eqref{eqn:jordan-wigner-transform}.}
(see Sec.~\ref{sec:interaction-MBPT}).

\begin{figure}
    \centering
    \subfloat[\label{fig:Ek-no-overlap}]{%
        \includegraphics[width=0.36\linewidth]{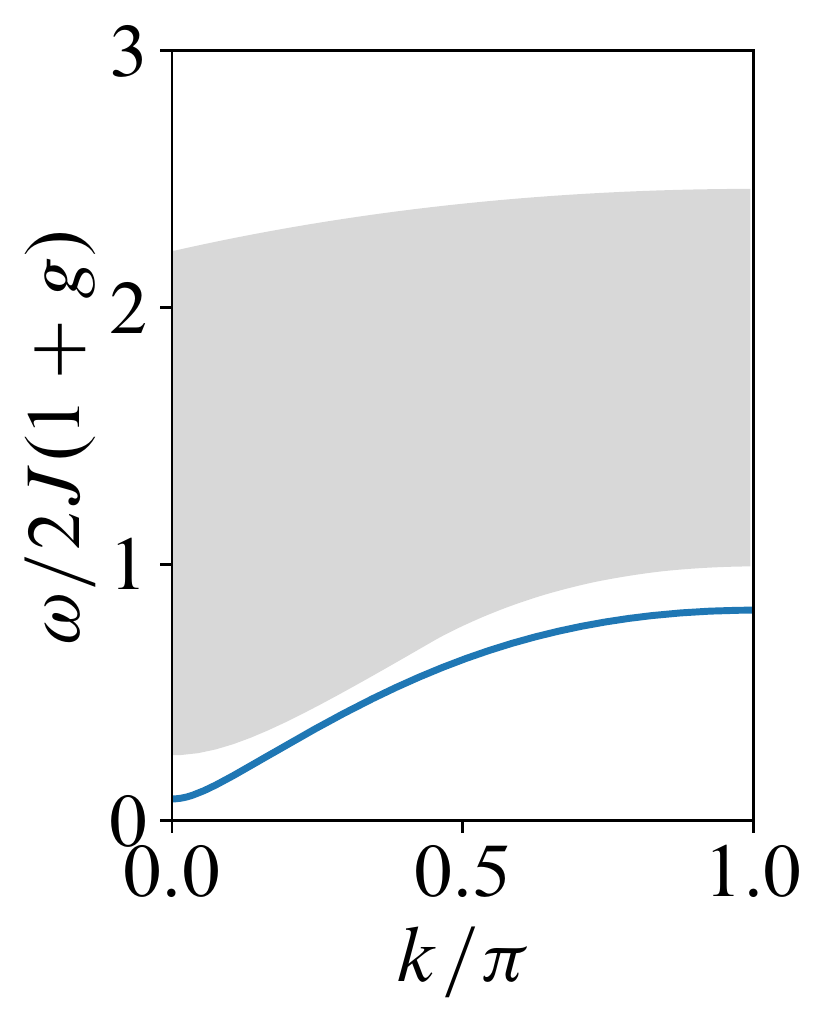}}%
    \hspace{1cm}%
    \subfloat[\label{fig:Ek-overlap}]{%
        \includegraphics[width=0.36\linewidth]{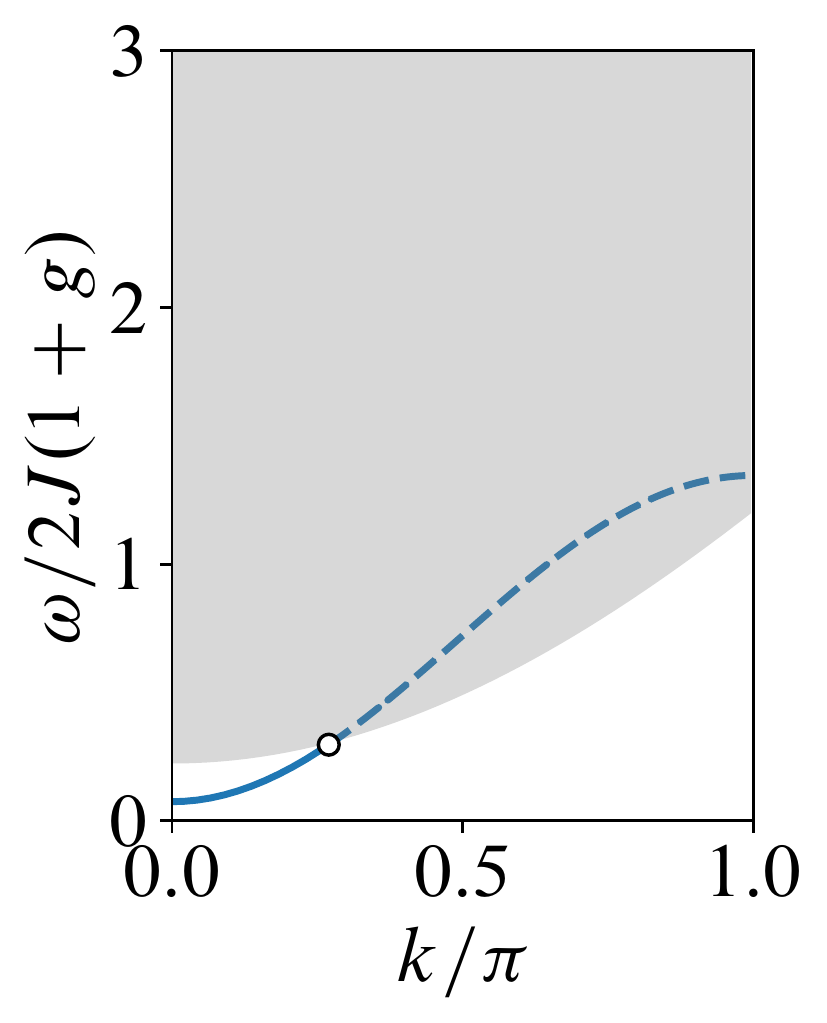}}%
    \caption{(a) Single-particle dispersion $E_k$ of the nearest-neighbor Ising model at first order in the interactions within the ferromagnetic regime (blue line) and the corresponding three-domain-wall continuum, $E_{p_1} + E_{p_2} + E_{k-p_1-p_2}$ for all pairs of momenta $p_1, p_2$ (gray shaded region). Absent next-nearest-neighbor contributions, the two do not overlap, forbidding spontaneous decay of isolated domain walls. (b) The analogous dispersion relation and three-domain-wall continuum in the presence of next-nearest-neighbor fermion processes that permit the single-particle mode to enter the continuum.}
    \label{fig:spontaneous-decay}
\end{figure}

Deep within the ferromagnetic regime, the spectrum of domain walls obtained from Eq.~\eqref{eqn:TFIM} is gapped and remains well separated from the three-domain-wall (3DW) continuum. As the critical point of the noninteracting model, $g = 1$, is approached from below, the single-particle dispersion gets closer to, but never touches, the 3DW continuum, due to the relativistic character of the dispersion at long wave lengths. This picture remains true when the XY-like interactions described above are introduced perturbatively. Consequently, it is not possible for the domain walls to decay spontaneously into three domain walls at zero temperature, necessitating nonzero temperatures to obtain a finite lifetime due to scattering from thermally-excited domain walls.
To study the impact of \emph{spontaneous} decay on the nonlinear response, we may modify the single-particle dispersion such that overlap with the 3DW continuum occurs. Practically, this can be accomplished by adding next-nearest-neighbor terms to the fermionic Hamiltonian. In the original spin degrees of freedom, such a modification is effected by three-body terms of the form
$\hat{\sigma}^z_i \hat{\sigma}_{i+1}^x \hat{\sigma}_{i+2}^z$.
Examples of dispersion relations that avoid and permit spectral overlap are depicted in Fig.~\ref{fig:spontaneous-decay}.
While such three-body interactions may be subleading in a typical experiment, they illustrate the physics at play in a more generic setting exhibiting spectral overlap.

\begin{figure*}[t]
    \centering
    \subfloat[\label{fig:free-2DCS-spectrum}]{%
        \includegraphics[width=0.27\linewidth]{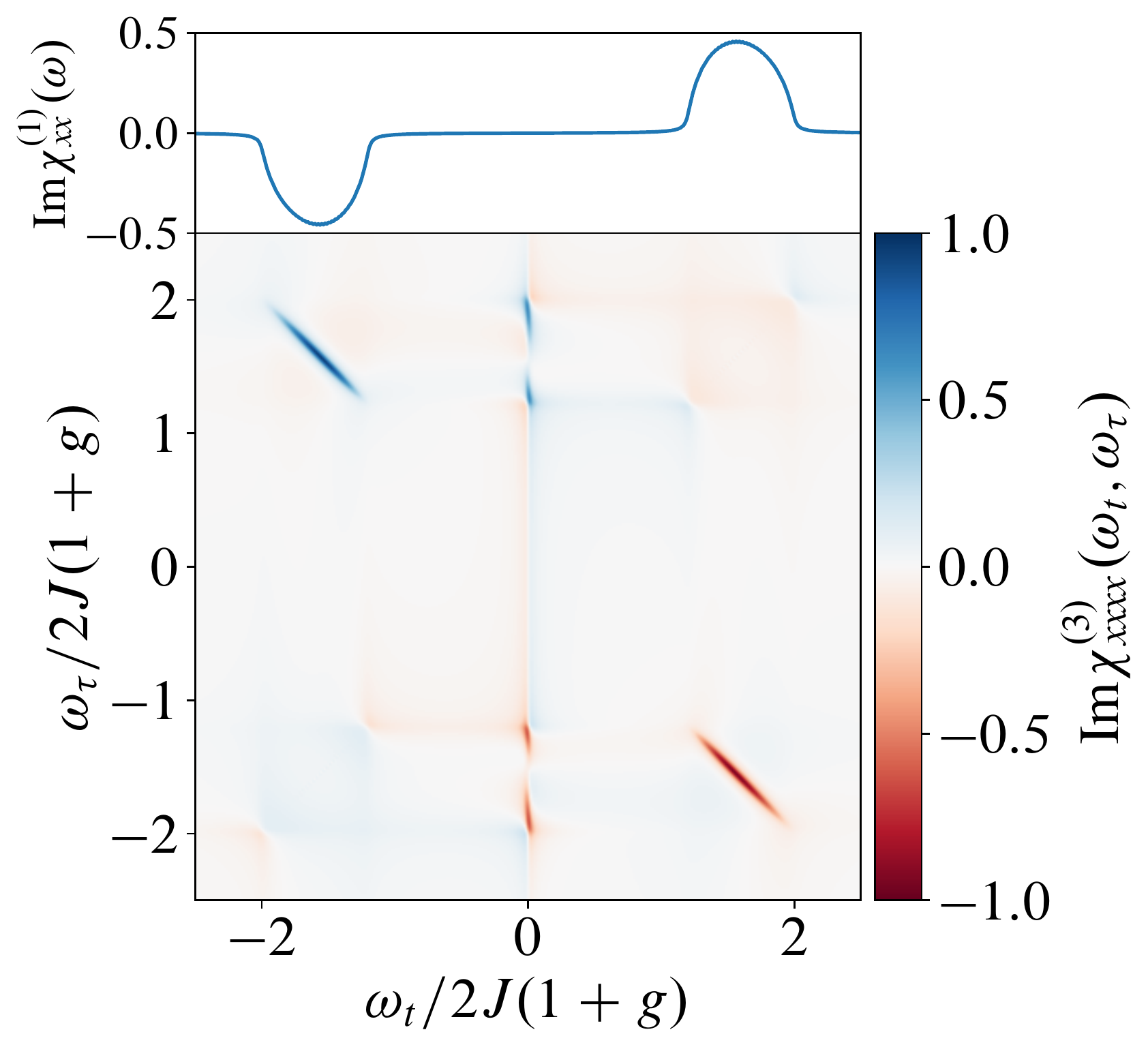}%
    }\hspace{2.5cm}%
    \subfloat[\label{fig:interacting-2DCS-spectrum}]{%
        \includegraphics[width=0.27\linewidth]{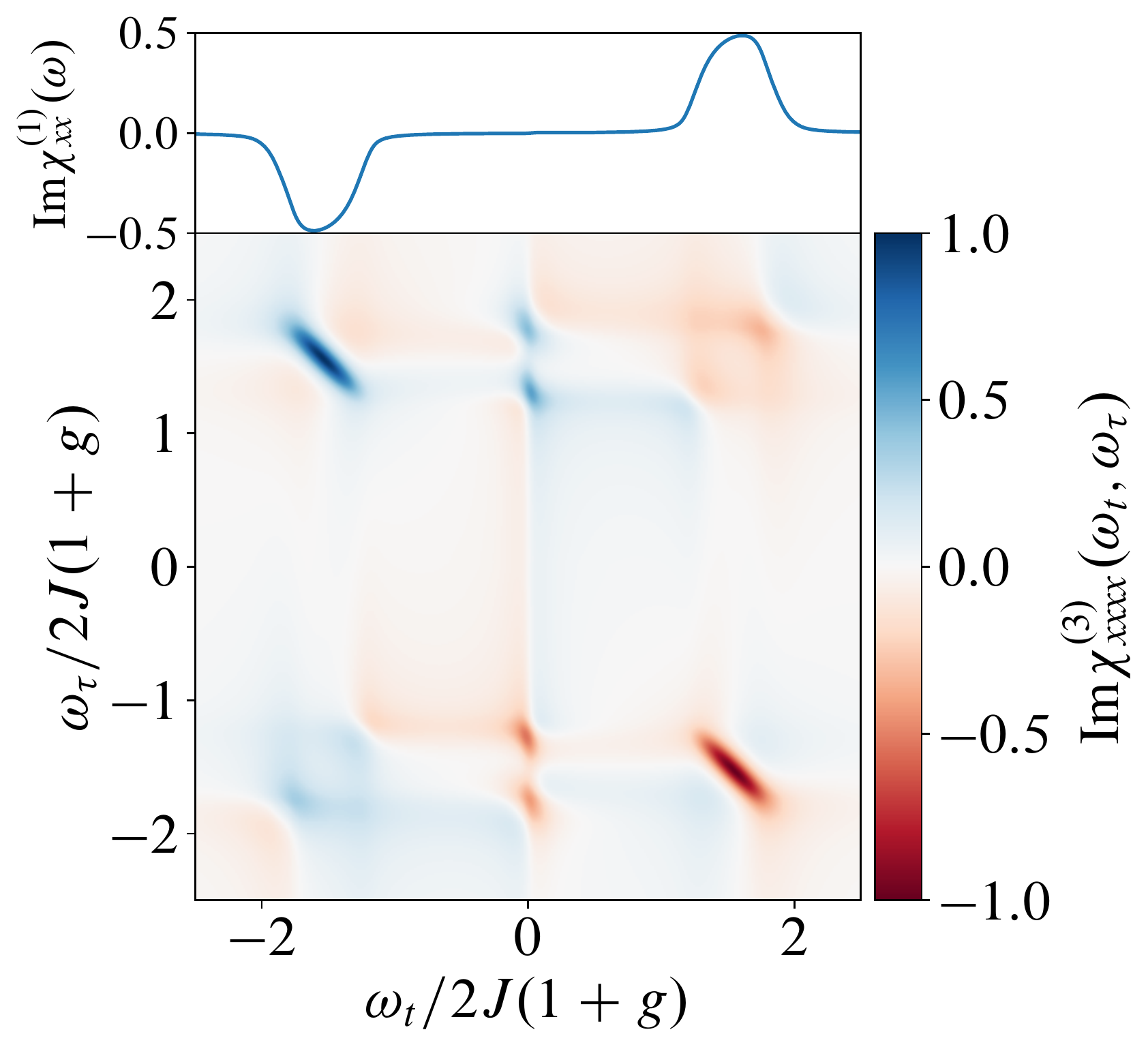}%
    }\\
    \subfloat[\label{fig:free-2DCS-overlap}]{%
        \includegraphics[width=0.27\linewidth]{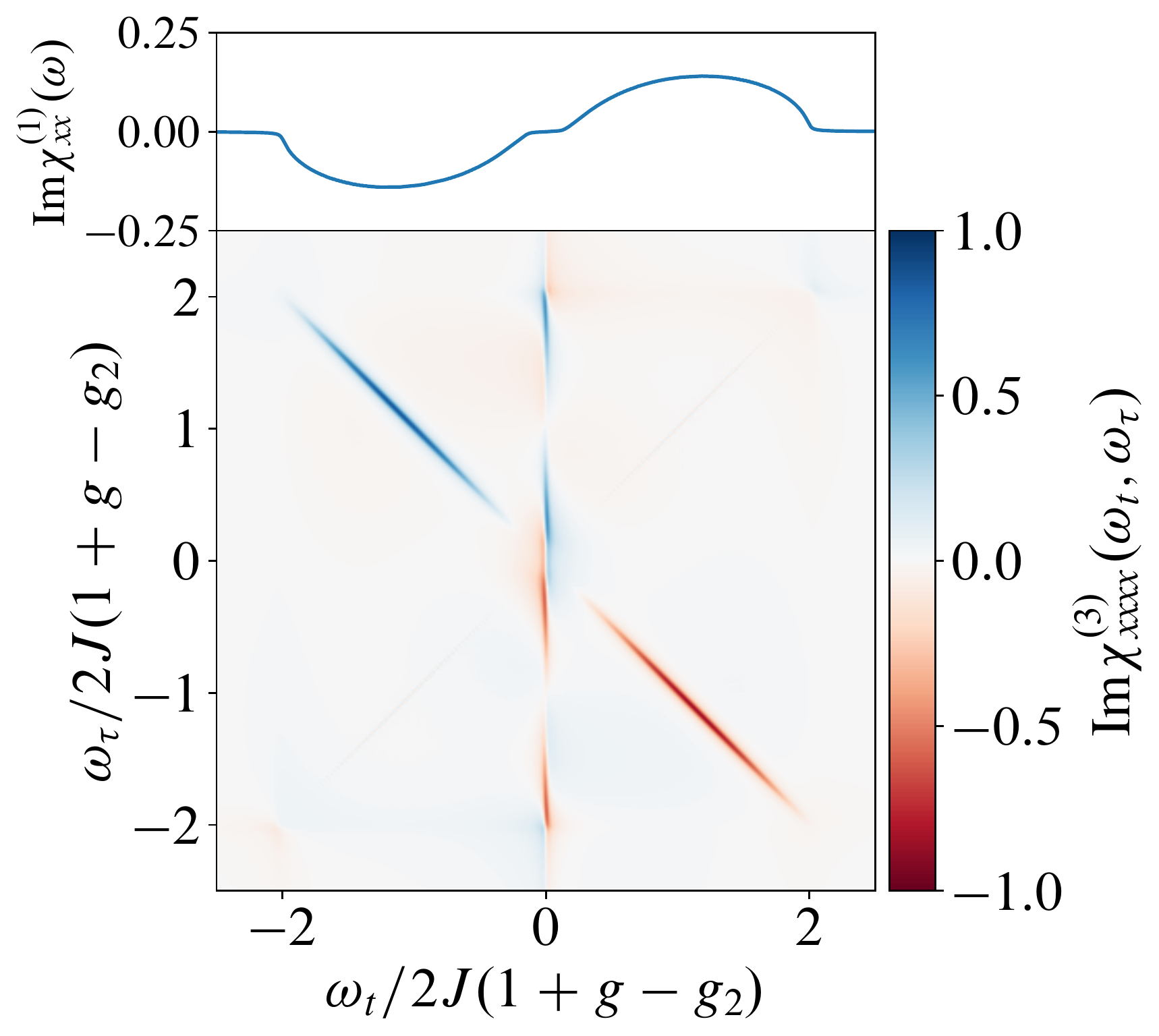}%
    }\hspace{2.5cm}%
    \subfloat[\label{fig:interacting-2DCS-overlap}]{%
        \includegraphics[width=0.27\linewidth]{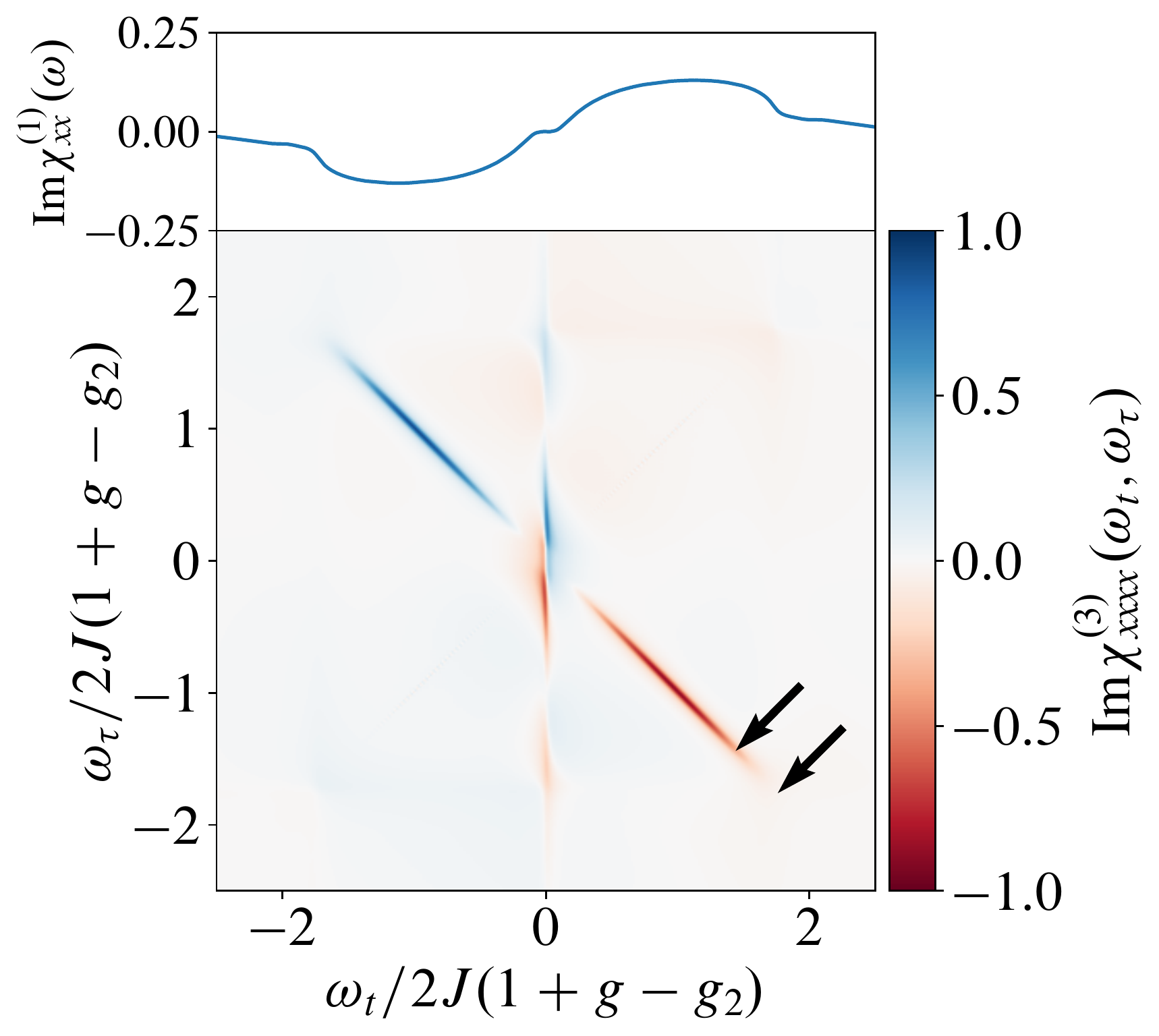}%
    }
    \caption{Imaginary part of the Fourier-transformed linear susceptibility per site, $\ChiOne(t)$, (top panel) and Fourier-transformed third-order susceptibility, $\ChiThree(t, t, t+\tau)$, (bottom panel) for both noninteracting [(a), (c)] and interacting [(b), (d)] systems. Two distinct cases are considered. (a), (b): The single-particle dispersion does not overlap with the three-domain-wall continuum. (c), (d): Overlap is induced by including next-nearest-neighbor interactions between spins [see Fig.~\ref{fig:spontaneous-decay}]. The noninteracting systems [(a), (c)] exhibit sharp streaks along the antidiagonal, $\omega_t = -\omega_\tau$, due to the rephasing process described in the main text. In the presence of interactions, but absent spectral overlap [(b)], spontaneous decay is forbidden. However, interactions between particles and nonzero temperature conspire to furnish the domain walls with a finite lifetime, leading to broadening of the antidiagonal streaks. At zero temperature when spectral overlap is present [(d)], the broadening is confined to frequencies where domain walls can decay spontaneously, leaving the complementary frequencies sharp. The locations highlighted by the arrows in (d) and Fig.~\protect\subref*{fig:DR-interacting-overlap} map onto one another. In (a), (b), the system size is $L=100$, and the system is deep within the ferromagnetic phase, $g=1/4$, at temperature $T=J/2$. In (c), (d) $L=200$, $g=2/5$, and $g_2 = -0.48$. Interaction strengths $\lambda=0.15J$ and $\lambda = 0.3J$ are used in (b) and (d), respectively. In all plots the color bar is normalized by the maximum absolute value of $\Im \ChiThree(\omega_t, \omega_\tau)$.}
    \label{fig:noninteracting-vs-interacting}
\end{figure*}

Since the ferromagnetic phase of Eq.~\eqref{eqn:TFIM} exhibits fractionalized excitations, linear response probes the creation and subsequent dynamics of a \emph{pair} of quasiparticles. `Broadening' of the absorption spectrum therefore has two distinct contributions, between which the linear response cannot distinguish: (i) `inhomogeneous' broadening due to the two-domain-wall continuum and (ii) `homogeneous' broadening due to the finite lifetime of the quasiparticles. This effect is illustrated by the qualitative similarity between the top panels of Figs.~\subref*{fig:free-2DCS-spectrum} and \subref*{fig:interacting-2DCS-spectrum}, which depict the linear response of noninteracting and interacting systems, respectively. To isolate these two in principle distinct contributions, one can probe the system's nonlinear response via 2DCS~\cite{WanArmitage2019}.

\subsection{2DCS protocol}

Motivated by Refs.~\cite{WanArmitage2019,Nandkishore2021Spectroscopic}, we consider the nonlinear response of the model to magnetic field pulses $h(t)$ in the `transverse' direction (i.e., parallel to the transverse field, which is oriented along $x$ in our conventions), which couple to the $x$ component of the magnetization, $\hat{M}^x$.
Specifically, the system's nonlinear response is probed by the two-dimensional coherent spectroscopy (2DCS) protocol~\cite{Kuehn20112DCS,Woerner20132DCS,Lu2016_2DCS,Lu2017SpinWaves,Mahmood2021Observation}, which consists of three separate pulse sequences, which we refer to as `A', `B', and `AB'.
First, a pulse A is applied to the system at time $s=0$, $h_\text{A}(s) = A_0 \delta(s)$.
Second, a separate pulse B at $s=\tau$, with $\tau > 0$,\footnote{One can also permit negative $\tau$. See, for instance, the Methods of Ref.~\cite{Mahmood2021Observation}. A protocol that utilizes negative $\tau$ is also explored in Sec.~\ref{sec:phase-untwisting} in the context of ``untwisting'' the rephasing signal.} is applied to the system in isolation, $h_\text{B}(s) = A_\tau \delta(s-\tau)$.
Finally, in a third experiment, both pulses are applied to the system in tandem, $h_\text{AB}(s) = A_0 \delta(s) +  A_\tau \delta(s-\tau)$.
In each of the three experiments, the system is measured at a later time $s=t+\tau$ (i.e., a time $t>0$ after the second pulse).
To find the nonlinear response of some observable $\hat{M}$, the responses to the two individual pulses, A and B, are subtracted from the response to the sum, AB, which removes the linear response component (as well as contributions to the nonlinear response that do not involve cross terms between the pulses A and B):
\begin{equation}
    M_\text{2DCS}(t+\tau) = M_\text{AB}(t+\tau) - M_\text{A}(t+\tau) - M_\text{B}(t+\tau)
    \, .
\end{equation}
A particularly natural and experimentally accessible choice for the observable $\hat{M}$ is the total magnetization of the system.
In this case, for the unperturbed TFIM, only the transverse component of the total magnetization, $\hat{M}^x = \tfrac12 \sum_i \hat{\sigma}_i^x$, exhibits a nonzero response. 
One finds that the measured magnetization response has contributions from both the second- ($\ChiTwo$) and third-order ($\ChiThree$) susceptibilities:
\begin{align}
    M_\text{2DCS}^x(t+\tau) = \,
       &A_0 A_\tau \ChiTwo (t, t+\tau) \notag\\
    +  &A_0^2 A_\tau \ChiThree (t, t+\tau, t+\tau) \notag\\
    +  &A_0A_\tau^2 \ChiThree ( t, t,t+\tau)
    \, ,
    \label{eqn:Mx-2DCS}
\end{align}
because the system does not posses any symmetries that enforce the second-order susceptibility, $\ChiTwo$, to vanish.
The time argument convention used in Eq.~\eqref{eqn:Mx-2DCS} is defined such that the $n$th-order response of $M^x(t) = \sum_n M_n^x(t)$ is recovered by convolution, i.e.,
\begin{equation}
    M^x_n(t) = \int \chi^{(n)}_{x\cdots x}(t-t_1, \ldots, t-t_n) h(t_1) \cdots h(t_n)
    \, ,
\end{equation}
where the integration is over times $t_1, \ldots, t_n$ and the $n$th-order susceptibility $\chi^{(n)}_{x\cdots x}$ is causal in the sense that time arguments must remain ordered: $\chi^{(n)}_{x\cdots x} \propto \theta(t-t_1) \theta(t_1-t_2)\cdots\theta(t_{n-1}-t_n)$.
Since the two contributions from $\ChiThree$ scale differently with the strengths of the two pulses, $A_0$ and $A_\tau$, they can -- at least in principle -- be isolated from one another~\cite{WanArmitage2019,Nandkishore2021Spectroscopic}.
As a result, we will examine the behavior of the two time argument sequences, $\ChiThree(t, t+\tau, t+\tau)$ and $\ChiThree(t, t, t+\tau)$, separately.
It should be noted, however, that any individual experiment with $A_0$ and $A_\tau$ fixed will include a contribution to the third-order response from both time arguments appearing in Eq.~\eqref{eqn:Mx-2DCS}.


\subsection{Third-order response}

While all contributions to Eq.~\eqref{eqn:Mx-2DCS} in principle contain new information beyond the linear response of the magnetization, $\hat{M}^x$, of particular importance is the contribution from the third line, proportional to $\ChiThree(t, t, t+\tau)$.
As shown in Ref.~\cite{WanArmitage2019}, for the unperturbed TFIM, this particular sequence of time arguments gives rise to a rephasing process analogous to the spin echo in the context of nuclear magnetic resonance (NMR): The phase accumulated during time evolution for a time $\tau$ can -- in the absence of any dephasing mechanism -- be entirely countered by time evolution during the time delay $t$ when the two delays are comparable, $\tau \simeq t$.
If the Fourier transform of this response function is evaluated from times $t, \tau$ to conjugate frequencies $\omega_t, \omega_\tau$, respectively, the rephasing process gives rise to a sharp streak along the antidiagonal in the $\omega_t$-$\omega_\tau$ plane, i.e., along $\omega_t = - \omega_\tau$.
In the absence of dissipation, and thence perfect rephasing, this streak is infinitely narrow in the transverse direction, parallel to $\omega_t = \omega_\tau$. In Ref.~\cite{WanArmitage2019}, it was shown that adding dissipation by hand leads to a nonzero broadening of the streak in the $\omega_t = \omega_\tau$ direction that directly probes the `decoherence time' (also known as $T_2$) of the effective two-level systems that comprise the noninteracting spectrum. The nonlinear response can therefore qualitatively distinguish between the presence and the absence of lifetime broadening in systems exhibiting a two-particle continuum.
We scrutinize the fate of, and elucidate what quantitative information can be garnered from, the rephasing signal when the finite lifetime of excitations is provided by realistic interactions between the quasiparticles.

\begin{figure*}[t]
    \centering
    \subfloat[\label{fig:DR-free}]{%
        \includegraphics[height=0.20\linewidth]{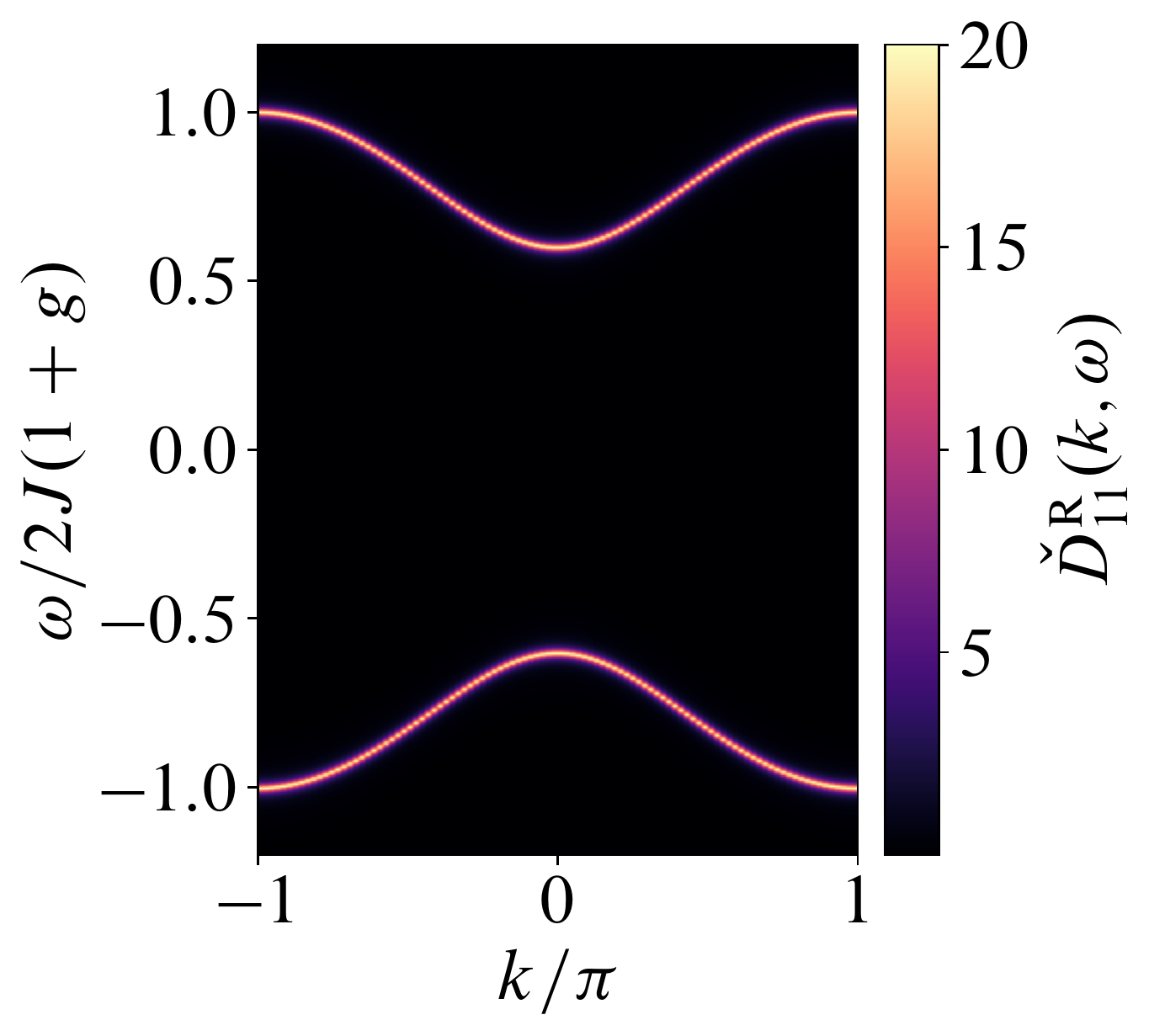}%
    }
    \hspace{0.2cm}%
    \subfloat[\label{fig:DR-interacting}]{%
        \includegraphics[height=0.20\linewidth]{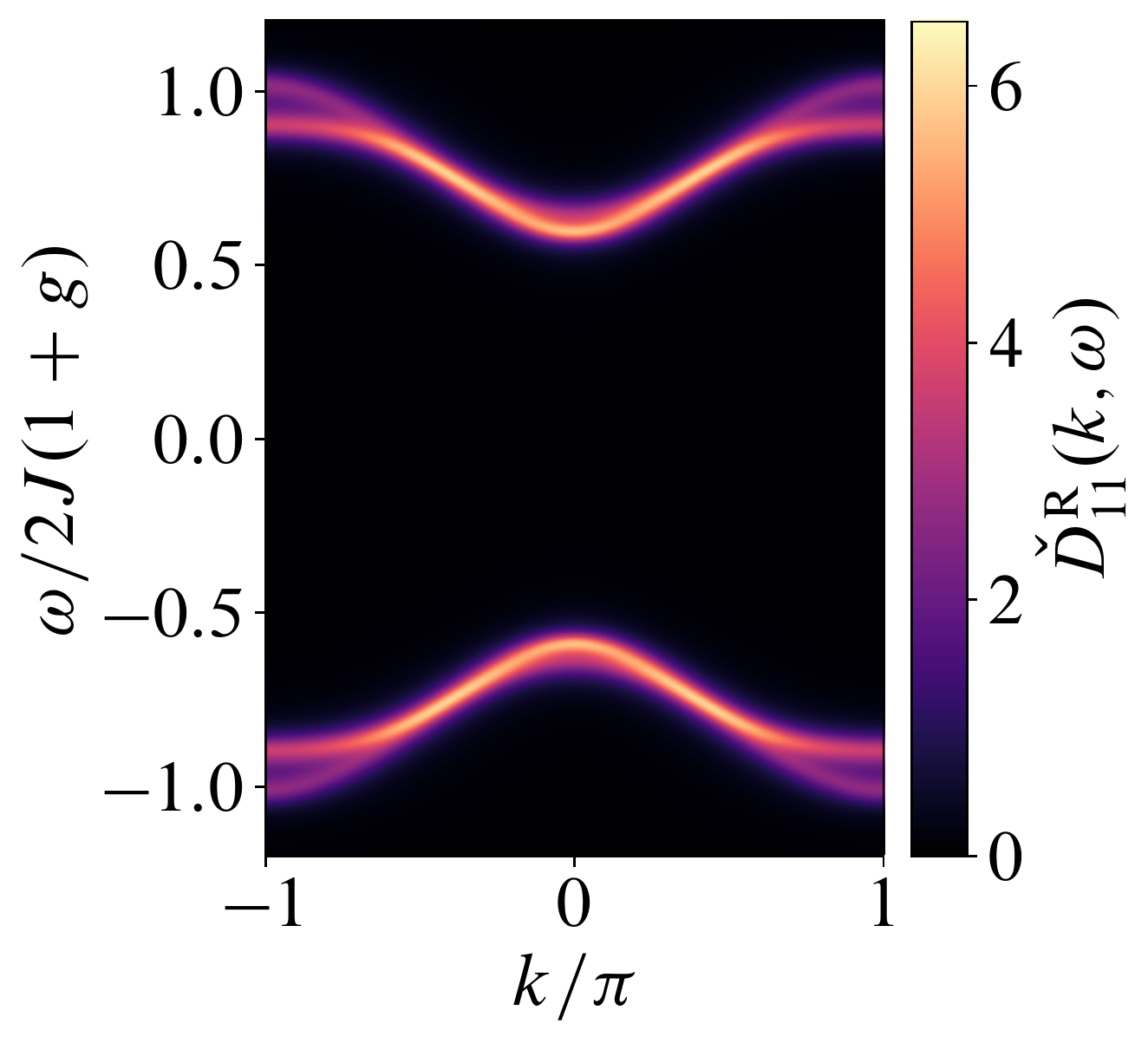}
    }%
    \hspace{1.4cm}%
    \subfloat[\label{fig:DR-free-overlap}]{%
        \includegraphics[height=0.20\linewidth]{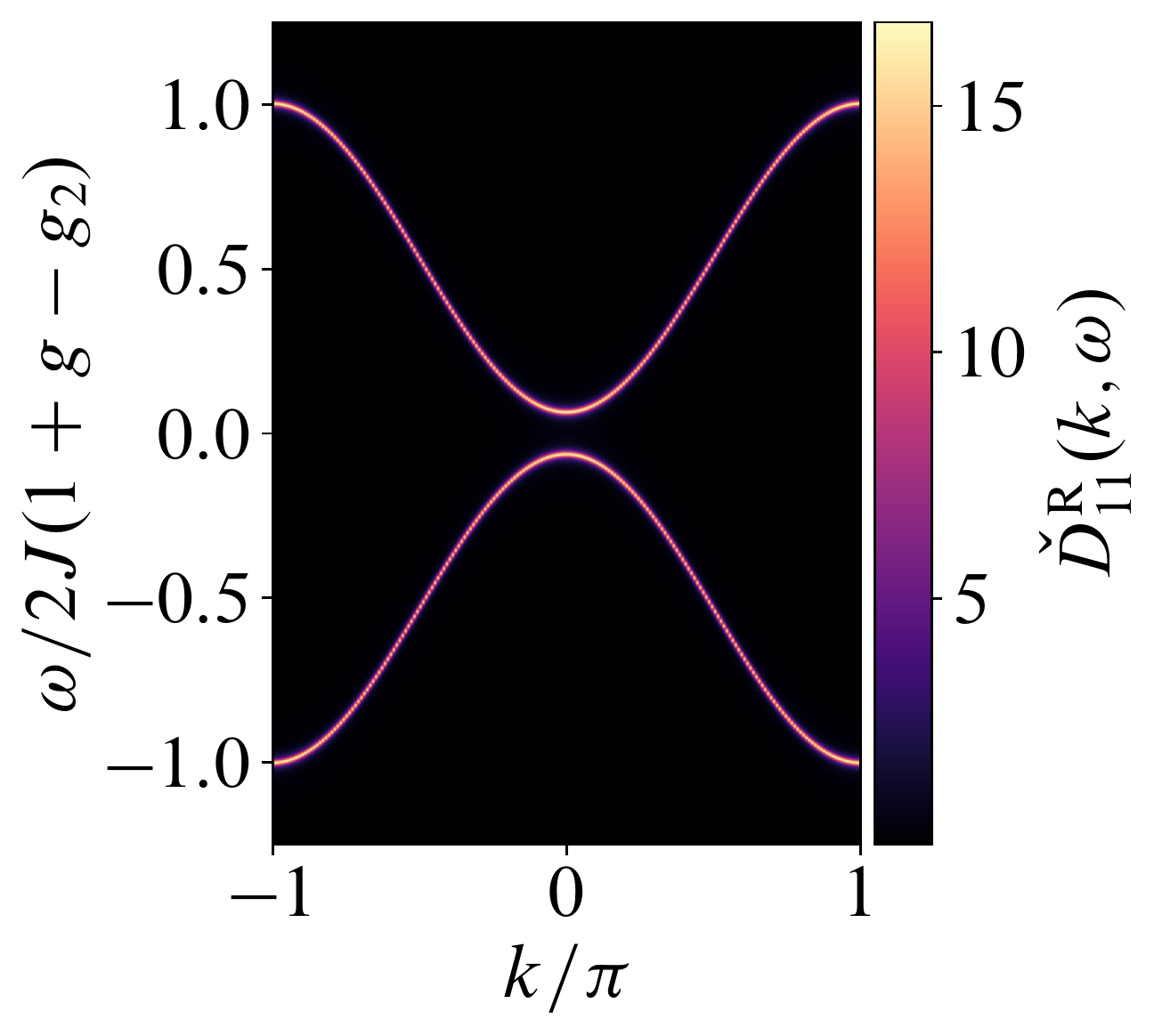}%
    }
    \hspace{0.2cm}%
    \subfloat[\label{fig:DR-interacting-overlap}]{%
        \includegraphics[height=0.20\linewidth]{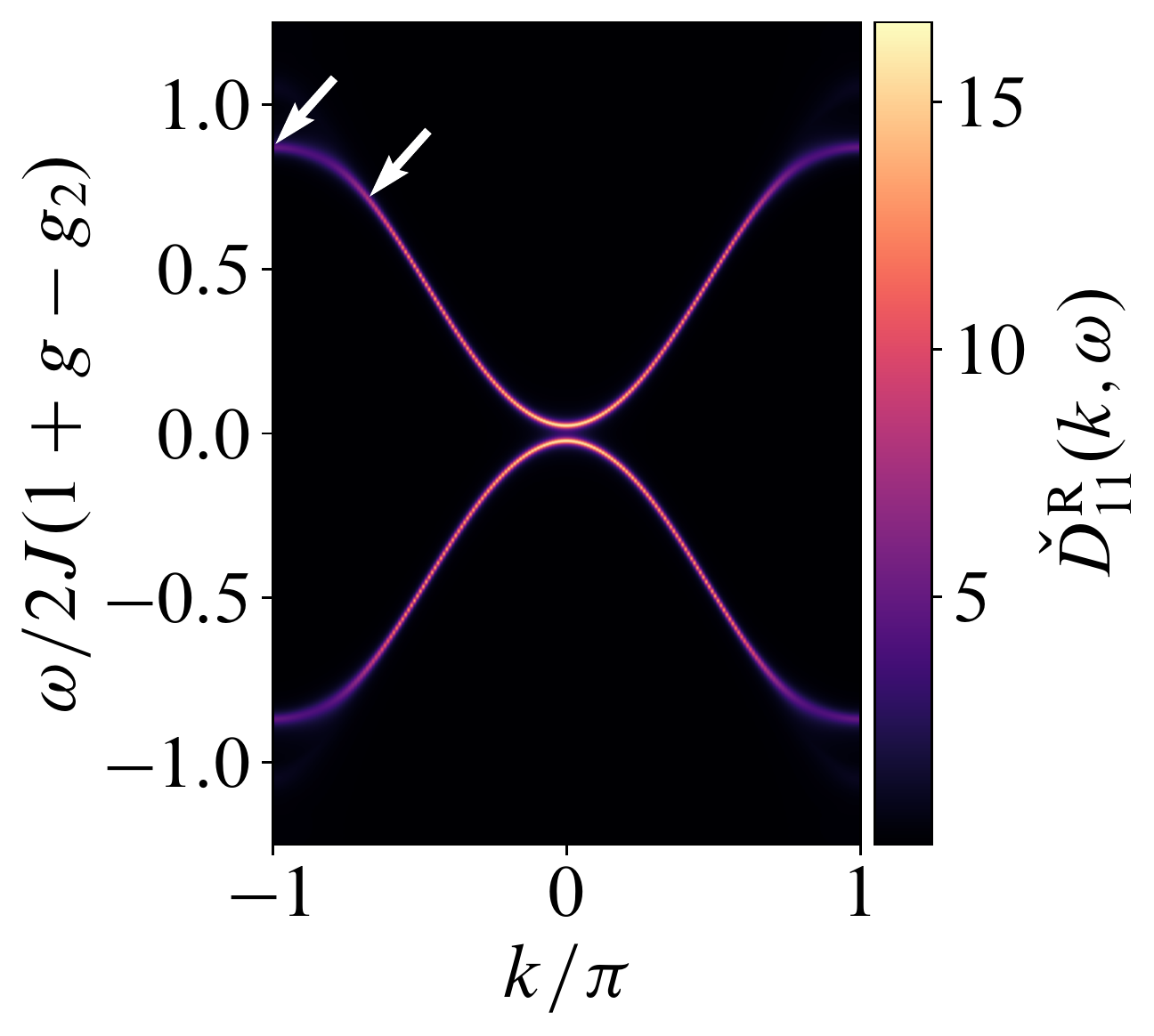}%
    }
    \caption{(a), (c): The diagonal entry of the retarded single-particle Green's function $\check{D}^\text{R}(k, \omega)$ [see Eq.~\eqref{eqn:Majorana-DR}] of the unperturbed transverse field Ising chain, without [(a)] and with [(c)] next-nearest-neighbor interactions between spins. The panels (a), (b), (c), (d) correspond to the same systems as Fig.~\ref{fig:noninteracting-vs-interacting}. Absent perturbations, the model maps to free fermions and $\check{D}^\text{R}(k, \omega)$ exhibits poles at energies corresponding to the quasiparticle dispersion, $\omega= \pm \epsilon_k$ [given in Eq.~\eqref{eqn:JW-energies}]. (b), (d): The same component of the Green's function at second order in the fermionic interactions. The quasiparticles are endowed with a finite lifetime and the single-particle Green's function develops nontrivial spectral features. Absent spectral overlap but at non-zero temperatures [(b)], broadening occurs throughout the spectrum, but is most pronounced at the zone boundaries, where the broadening becomes anomalous. When spectral overlap occurs but at zero temperature [(d)], broadening is confined to the frequencies where spontaneous decay is permitted. The locations highlighted by the arrows in (d) and Fig.~\protect\subref*{fig:interacting-2DCS-overlap} map onto one another. All plots use the same parameters as their counterparts in Fig.~\ref{fig:noninteracting-vs-interacting}.}
    \label{fig:greens-functions}
\end{figure*}


\subsection{Adding interactions}

To tackle the problem of introducing interactions, we introduce a powerful real-time path integral approach.
Specifically, one of the principal contributions of this work is to reduce the problem of evaluating the nonlinear response to performing a derivative expansion of a generating functional. Schematically, the $n$th order response functions are obtained via
\begin{equation}
    \ChiN (t_1; t_2, \ldots, t_n) = \frac{1}{2i} \left.  \frac{\delta^4 Z[h^\text{cl}, h^\text{q}]}{\delta h^\text{cl}_n \cdots \delta h^\text{cl}_2\delta h^\text{q}_1} \right\rvert_{h^\text{q} = h^\text{cl} =0}
    \, ,
    \label{eqn:chi-n-schematic}
\end{equation}
where the notation is explained in detail in Sec.~\ref{sec:nonlinear-suscept}.
At the noninteracting level, the generating function $Z[h^{\text{cl}}, h^{\text{q}}]$ can be evaluated \emph{exactly}, and the response to arbitrary orders in the driving field can easily be obtained by taking as many derivatives as one requires.
This procedure circumvents the calculation of the $n$-fold nested commutators that appear in generalized Kubo formulae, which can be tedious to evaluate.
In the presence of interactions, the nonlinear response functions in Eq.~\eqref{eqn:chi-n-schematic} have a diagrammatic interpretation, and the standard methods of diagrammatic many-body perturbation theory can be applied immediately to incorporate certain interaction effects nonperturbatively.

The Fourier transform of the third-order susceptibility, $\ChiThree(t, t, t+\tau)$, of the \emph{noninteracting} TFIM without [with] next-nearest-neighbor interactions between spins is shown in Fig.~\subref*{fig:free-2DCS-spectrum} [Fig.~\subref*{fig:free-2DCS-overlap}].
In both cases, sharp streaks can be observed in the lower right and upper left quadrants at energies corresponding to twice the quasiparticle spectrum $\omega_t =\omega_\tau = \pm 2\epsilon_k$, since the uniform magnetic field pulses create two oppositely-propagating domain walls.
Absent next-nearest-neighbor interactions, the streaks span the frequencies $2(1-g) < \epsilon_k < 2(1+g)$, which coincide with the range of frequencies that the absorption spectrum $\Im \ChiOne(\omega)$ is nonvanishing.
For the purposes of plotting, we couple the systems to a zero-temperature bath with a frequency- and momentum-independent spectral function, $J_k(\omega) = \gamma$, providing them with a finite but large lifetime set by $\sim 1/\gamma$, which gives the width of the streaks observed in Figs.~\subref*{fig:free-2DCS-spectrum} and \subref*{fig:free-2DCS-overlap}.
The absence of momentum dependence implies that the width of the streaks in the $\omega_t = \omega_\tau$ direction does not vary with the choice of cut.
The amplitude of the streak, on the other hand, does vary with momentum (and, hence, with the choice of cut). The amplitude is set by the optical matrix element; at zero temperature, the amplitude of the streak at $\omega_t = \pm 2\epsilon_k$ is set by $\propto \nu(\epsilon_k) \sin^4(\vartheta_k)$, where $\nu(\epsilon)$ is the single-particle density of states and $\vartheta_k$ is the angle entering the Bogoliubov transformation that diagonalizes the single-particle Hamiltonian~\cite{WanArmitage2019} [see also Eq.~\eqref{eqn:chi-3-sequential}].

Upon introducing the residual XY interactions, there are certain aspects of these interactions that can be treated \emph{exactly}. Namely, taking inspiration from Ref.~\cite{Robinson2014breakdown}, 
in order to maintain normal ordering of the interactions \emph{and} simultaneously diagonalize the bilinear contribution to the Hamiltonian, the Bogoliubov transformation should be performed in a self-consistent manner. 
This exact effect and the first-order contribution to the self-energy conspire to give a renormalization of the single-particle spectrum $\epsilon_k$; at this order in perturbation theory, the energy of the single-particle excitations is modified, but they remain exact quasiparticles with an infinite lifetime.
Correspondingly, the locations of the streaks in the $\omega_t$-$\omega_\tau$ plane are modified due to the renormalization of the spectrum $\epsilon_k$, and the amplitude profile along the antidiagonal is also modified, a consequence of the renormalized Bogoliubov parameter $\vartheta_k$ and the single-particle density of states $\nu(\epsilon)$.
While these modifications are important for a full \emph{quantitative} understanding of the 2DCS response, both of these effects are already seen at the level of linear response, and, consequently, this information can already be extracted from the first-order susceptibility, $\ChiOne(t)$.

At second order in the interactions, the system presents additional spectral features that the linear response is effectively blind to.
The retarded Green's function is shown in Fig.~\ref{fig:greens-functions} for the noninteracting systems [\subref{fig:DR-free}, \subref{fig:DR-free-overlap}], and when interactions are accounted for by including self-energy corrections to second order in the interactions [\subref{fig:DR-interacting}, \subref{fig:DR-interacting-overlap}]. In both cases, a frequency- and momentum-independent lifetime misses important aspects of the response function.
Namely, when spectral overlap is absent, but the system is subjected to nonzero temperatures [Fig.~\subref*{fig:DR-interacting}], while broadening is present throughout the spectrum, it becomes most pronounced near the edges of the Brillouin zone. Further, near the zone boundaries, this broadening becomes anomalous in the sense that the quasiparticle peak splits in two. As we show in Sec.~\ref{sec:self-energy}, this anomalous broadening is a consequence of the sharp frequency dependence of the self-energy in the vicinity of $k=\pi$.
When spectral overlap is present and there are few thermally excited domain walls [Fig.~\subref*{fig:DR-interacting-overlap}], quasiparticle decay proceeds via spontaneous emission of domain wall pairs.
In this case, the broadening exhibits strong momentum dependence since spontaneous decay can only occur at energies and momenta where it is kinematically allowed.
Again, this leads to broadening that is most pronounced near the zone boundaries, while the complementary momenta near the zone center remain sharp.
While we have access to these single-particle Green's functions analytically and numerically, they correspond to \emph{string} correlation functions in terms of the original, local spin degrees of freedom $\hat{\sigma}_i$ appearing in Eq.~\eqref{eqn:TFIM}, and are therefore inaccessible in a conventional experimental setting (see, however, Ref.~\cite{BlochStringOrder}).
Instead, as mentioned previously, linear response of the experimentally-accessible $\hat{M}^x$ probes the broad two-domain-wall continuum, and the lifetime broadening of the Green's function does not lead to a qualitative change in the already broad response (see top panels of Fig.~\ref{fig:noninteracting-vs-interacting}). Consequently, even identifying the presence or absence of lifetime broadening is challenging, and the intricate features discussed above relating to the frequency and momentum dependence of the self-energy will remain hidden to the linear response.

The corresponding 2DCS spectra for the systems exhibiting quasiparticle decay are shown in Figs.~\subref*{fig:interacting-2DCS-spectrum} and \subref{fig:interacting-2DCS-overlap}.
Focusing first on Fig.~\subref*{fig:interacting-2DCS-spectrum}, we observe that the broadening of the single-particle excitations, present throughout the spectrum in Fig.~\subref*{fig:DR-interacting}, gives the streaks a nonzero width in the $\omega_t = \omega_\tau$ direction.
The anomalous broadening near the zone boundaries, however, is seemingly absent.
As we explain in Appendix~\ref{sec:anomalous-broadening}, this is a consequence of insufficient `contrast' between the maxima, rather than a fundamental insensitivity of the 2DCS spectrum.
Indeed, in Appendix~\ref{sec:anomalous-broadening} we detail how the anomalous broadening would manifest in the nonlinear response if its effects were amplified.
In Fig.~\subref*{fig:interacting-2DCS-overlap} the strong momentum dependence of the Green's function due to spontaneous decay can be observed, with the location of the strongest broadening in the $\omega_t = \omega_\tau$ direction indicated between the two arrows [which are also shown in Fig.~\subref*{fig:DR-interacting-overlap}].
These results highlight an important point: While the 2DCS spectrum (within a self-energy approximation) can be written entirely in terms of single-particle Green's functions [Eq.~\eqref{eqn:nonlinear-functional-deriv}], the inverse problem -- extracting \emph{generic} single-particle properties from the 2DCS spectrum -- can be challenging.
In summary, the typical scale of broadening as a function of \emph{momentum} can be gleaned from the diagonal width of the rephasing `streak' in the 2DCS spectrum. More intricate spectral features related to the \emph{frequency} dependence of the self-energy, such as the anomalous broadening in Fig.~\subref*{fig:DR-interacting-overlap}, are more elusive, but may also be accessible if one is lucky with matrix elements, as we shall discuss.


\section{Quasiparticle scattering from interactions}
\label{sec:weak-interactions}

\subsection{Model}

Our idealized system consists of the following three terms
\begin{equation}
    \hat{H}(t) = \hat{H}_0 + \hat{H}_\text{int} + \hat{H}_\text{drive}(t)
    \, .
\end{equation}
The noninteracting Hamiltonian, $\hat{H}_0$, corresponds to the transverse field Ising model (TFIM) in one spatial dimension, and is given in Eq.~\eqref{eqn:TFIM}.
It is well known that Eq.~\eqref{eqn:TFIM} maps to free fermions via a standard Jordan-Wigner (JW) transformation~\cite{Lieb1961Soluble,Pfeuty1970}, and exhibits a ferromagnetic--paramagnetic quantum phase transition at $g=1$, characterizing the Ising universality class~\cite{SachdevQPT}.
We will focus solely on the ferromagnetic phase, $g<1$, in which the system possesses a nonzero gap and hosts fractionalized, domain-wall-like excitations.

We assume that interactions between fermions arise from imperfect Ising anisotropy between neighboring spins, giving rise to residual `XY'-like interactions in the plane perpendicular to the Ising easy axis:
\begin{equation}
    \hat{H}_\text{int} = \frac12 \sum_{i \neq j} U_{ij} \left[ \hat{\sigma}_i^x \hat{\sigma}_j^x + \hat{\sigma}_i^y \hat{\sigma}_j^y \right]
    \, .
    \label{eqn:H-int}
\end{equation}
For simplicity, we will take the interactions to be truncated at nearest-neighbor distance with strength $\lambda$, i.e., $U_{ij} = \lambda (\delta_{i,j+1}+\delta_{i,j-1})$.
Under this assumption, the XY interaction maps to a density-density coupling between the JW fermionic degrees of freedom, in addition to a bilinear fermionic term that renormalizes the single-particle band structure.
Finally, the system is perturbed about equilibrium with a time-varying, but spatially-homogeneous, magnetic field $\vec{h}(t)$
\begin{equation}
    \hat{H}_\text{drive} = - \frac12 \sum_{i=1}^L \vec{h}(t) \cdot \hat{\boldsymbol{\sigma}}_i 
    \, .
    \label{eqn:H-drive}
\end{equation}
In what follows we will take the driving field to be parallel to the static field in~\eqref{eqn:TFIM}, i.e.,
$\vec{h}(t) \cdot \hat{\boldsymbol{\sigma}}_i = h(t)\hat{\sigma}^x_i$.
In principle, we can allow $h(t)\to h(x_i, t)$ to vary in space as well as time, thereby probing the response of the system at nonzero momenta.
However, experiments do not -- at least at present -- have access to such momentum-resolved measurements, so we therefore focus on the translationally-invariant drive in Eq.~\eqref{eqn:H-drive}.
Since the static transverse field in Eq.~\eqref{eqn:TFIM} can be treated exactly via the JW transformation, there are no restrictions placed on $g$ other than the requirement that the system belongs to the ferromagnetic phase\footnote{Since we will be working at nonzero temperatures, the value of $g$, which determines the gap through Eq.\eqref{eqn:H-free-fermion}, will, in turn, determine the temperature regime over which our perturbative analysis is quantitatively valid.}.
However, both the characteristic scale of the interactions and the driving field are assumed weak, $gJ, J \gg \lambda, h$.
The former permits many-body perturbation theory in the interactions to be applied, while the latter is a prerequisite for the validity of (non)linear response theory.

The exact solution of Eq.~\eqref{eqn:TFIM} is obtained by introducing a standard JW transformation from spin-$1/2$ degrees of freedom to fermions
\begin{subequations}
\begin{align}
    \hat{\sigma}_i^x &= 1 - 2\hat{c}_i^\dagger \hat{c}_i^{\phantom{\dagger}} \\
    \hat{\sigma}_i^z &= - \prod_{j < i} (1 - 2\hat{n}_j) ( \hat{c}_i^\dagger + \hat{c}_i^{\phantom{\dagger}})
    \, .
\end{align}
\label{eqn:jordan-wigner-transform}%
\end{subequations}
This transformation is then followed by a Fourier transform, $\hat{c}_k = \sum_i e^{-ikx_i} \hat{c}_i/\sqrt{L}$, and a Bogoliubov rotation to new fermions $\hat{\gamma}_k$ through $\hat{c}_k = u_k \hat{\gamma}_k + iv_k \hat{\gamma}_{-k}^\dagger$, where
$u_k = \cos\tfrac12\vartheta_k$, $v_k = \sin\tfrac12\vartheta_k$, and $\tan\vartheta_k = \sin k/(g-\cos k)$~\cite{SachdevQPT}.
Up to a constant energy shift, this sequence of transformations brings the free Hamiltonian into the canonical free-fermion form
\begin{equation}
    \hat{H}_0 = \sum_k \epsilon_k (\hat{\gamma}_k^\dagger \hat{\gamma}^{\phantom{\dagger}}_k - \tfrac12 )
    \, ,
    \label{eqn:H-free-fermion}
\end{equation}
where the quantization of the quasimomentum $k$ is dictated by the boundary conditions%
\footnote{For convenience, we will impose antiperiodic boundary conditions on the JW fermions, such that the quasimomentum $k$ is quantized as $k_n = \frac{2\pi}{L}(n+\tfrac12)-\pi$, for $L$ even~\cite{Pfeuty1970}. If periodic boundary conditions are imposed on the physical spins, $\hat{\sigma}_i$, then periodic (antiperiodic) boundary conditions must be imposed on the odd (even) fermion parity sectors. The ground state is known to belong to the \emph{even} parity sector.}
on the real-space fermions, $\hat{c}_i$.
The spectrum of the free Bogoliubov fermions, $\hat{\gamma}^{\phantom{\dagger}}_k$, is given by
\begin{equation}
    \epsilon_k = 2J \sqrt{1 + g^2 - 2g\cos k}
    \, ,
    \label{eqn:JW-energies}
\end{equation}
which is gapped for all $g \neq 1$, with a gap that scales as $2J|1-g|$ in the thermodynamic limit.
In the strict limit $g\to 0^+$, the ground states of the model are the `cat' states $\propto \ket{\Uparrow} \pm \ket{\Downarrow}$, where, e.g., $\ket{\Uparrow}$ is the fully magnetized state with all spins pointing along $z$.
Applying $\hat{\sigma}_i^x$ to such a state flips the $i$th spin and thus creates a \emph{pair} of domain wall excitations on the two adjacent bonds.
Each member of the pair is subsequently able to propagate freely throughout the system.

 If the noninteracting Hamiltonian $\hat{H}_0$ is modified to include translationally-invariant next-nearest-neighbor fermion terms, it remains exactly solvable via the sequence of transformations described above. Specifically, as long as the Hamiltonian can still be brought into the standard form
\begin{equation}
    \hat{H}_0 = \frac{1}{2} J \sum_k
    \begin{pmatrix}
        \hat{c}_k^{\dagger} & \hat{c}_{-k}
    \end{pmatrix}
    \begin{pmatrix}
        \alpha_k & -i\beta_k \\
        i\beta_k & -\alpha_k
    \end{pmatrix}
    \begin{pmatrix}
        \hat{c}_k \\
        \hat{c}^\dagger_{-k}
    \end{pmatrix}
    \label{eqn:free-fermion-generic}
\end{equation}
in momentum space,
we obtain a dispersion relation of the form $\epsilon_k^2 = \alpha_k ^2 +  \beta_k^2 $. In the specific case that the noninteracting Hamiltonian is changed to $\hat{H}_0 \to \hat{H}_0 - g_2 J \sum_i  \hat{\sigma}^z_i \hat{\sigma}_{i+1}^x \hat{\sigma}_{i+2}^z  $, the coefficients entering Eq.~\eqref{eqn:free-fermion-generic} are modified according to $\alpha_k \to \alpha_k - 2g_2 \cos(2k)$, $\beta_k \to \beta_k + 2g_2 \sin(2k)$. This modification is used in Fig.~\subref*{fig:Ek-overlap} to engineer a single-particle dispersion that enters the 3DW continuum.


\subsection{Path integral representation}
\label{sec:path-integral-approach}


\subsubsection{Free Green's functions}

\begin{figure}
    \centering
    \includegraphics{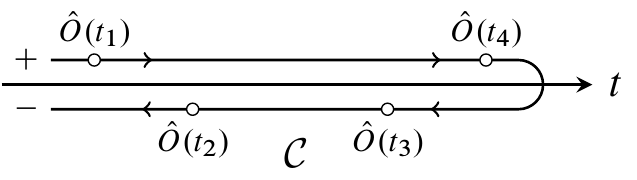}
    \caption{Schematic depiction of the Keldysh contour $\mathcal{C}$. Both the upper ($+$) and lower ($-$) branches span the entire real time axis. Also shown is a particular time ordering of operators $\hat{O}(t_i)$ that appears in the generalized Kubo formula for the third-order response, which can naturally be ordered on the Keldysh contour. The classical and quantum components of the field $h^{\text{cl/q}}(t) = \frac12 [h^+(t) \pm h^-(t)]$ are related to the average and difference between the two branches, respectively.}
    \label{fig:keldysh-contour}
\end{figure}

To compute the various linear and nonlinear susceptibilities, we find it convenient to use a real-time path integral approach, using the Keldysh contour $\mathcal{C}$~\cite{Kamenev2011}.
The generating functional $Z[h^\text{cl}, h^\text{q}]$, which contains complete information about the system's response properties to all orders in the drive, is written in terms of the classical (cl) and quantum (q) components of the drive, defined by $h^{\text{cl/q}}(t) = \frac12 [h^+(t) \pm h^-(t)]$.
The fields $h^\pm(t)$ are the generating fields on the upper and lower branches of $\mathcal{C}$, respectively:
\begin{equation}
    Z[h^\text{cl}, h^\text{q}] = \fermionInt e^{iS_0[\bar{\psi}, \psi] + iS_\text{int}[\bar{\psi}, \psi] + iS_\text{drive}[\bar{\psi}, \psi] }
    \, .
    \label{eqn:generating-functional-def}
\end{equation}
The upper and lower branches of the Keldysh contour $\mathcal{C}$ are depicted schematically in Fig.~\ref{fig:keldysh-contour}. The figure also highlights why a closed time contour approach is particularly convenient: In the generalized Kubo formulae for nonlinear susceptibilities, one encounters expectation values such as $\langle \hat{O}(t_1) \hat{O}(t_4) \hat{O}(t_3) \hat{O}(t_2) \rangle$, where $t_n > t_{n-1}$.
While this expression is not time ordered, it can naturally be path ordered on the Keldysh contour $\mathcal{C}$.
In the Keldysh formalism, the functional~\eqref{eqn:generating-functional-def} satisfies the identity $Z[h^\text{cl}, 0]=1$ since, in the absence of any quantum component, $h^{\text{q}}=0$, there is an exact cancellation between the forward and backward time evolutions: $\hat{U}(t_1, t_2) \hat{U}(t_2, t_1) = \hat{\mathds{1}}$.
The noninteracting, Grassmann-valued action $S_0[\bar{\psi}, \psi]$ is in general described by the free Green's function
\begin{equation}
    S_0[\bar{\psi}, \psi]
    =
    \int \mathrm{d}x \, \mathrm{d}x' \,
    \bar{\psi}_\alpha (x) [{G}_0^{-1}]_{\alpha\beta}(x, x') \psi_\beta(x)
    \, ,
    \label{eqn:noninteracting-action}
\end{equation}
where the multi-indices $\alpha, \beta$ (which are implicitly summed over) represent all internal degrees of freedom, including the contour index,
and the variables $x$, $x'$ denote space-time indices.
The fermionic Green's function has the following structure
\begin{equation}
    \langle \psi_\alpha (x) \bar{\psi}_\beta (x') \rangle = i{G}_0^{\alpha\beta}(x, x') =i
    \begin{pmatrix}
        G^\text{R} & G^\text{K} \\
        0 & G^\text{A}
    \end{pmatrix}_\text{K}
    \, ,
    \label{eqn:free-fermion-Greens}
\end{equation}
where the subscript `$\text{K}$' denotes that the depicted matrix structure is with respect to the `Keldysh' indices only; space-time and internal indices are left implicit.
In the momentum/frequency representation, the Keldysh components of $G_0$, corresponding to the Hamiltonian~\eqref{eqn:H-free-fermion}, are given explicitly by
\begin{subequations}
\begin{align}
    G^\text{R/A}(k, \omega) &= (\omega-\epsilon_k \pm i0^+)^{-1} \\
    G^\text{K}(k, \omega) &= -2\pi i F(\omega) \delta(\omega - \epsilon_k) \label{eqn:fermionic-GK}
    \, ,
\end{align}
\label{eqn:fermionic-greens}%
\end{subequations}
where $F(\omega) = 1- 2n_\text{F}(\omega) = \tanh{}(\frac12\beta\omega)$ denotes the equilibrium fermionic distribution function at temperature $T=\beta^{-1}$.
The retarded/advanced (R/A) Green's functions, $G^{\text{R/A}}(k, \omega)$, are analytic in the upper/lower half plane, and therefore correspond to upper/lower triangular matrices in the time domain.
In the absence of interactions, the energies $\epsilon_k$ are given by Eq.~\eqref{eqn:JW-energies}. The inverse Green's function that appears in Eq.~\eqref{eqn:noninteracting-action} is found by inverting the matrix in Eq.~\eqref{eqn:free-fermion-Greens} with components~\eqref{eqn:fermionic-greens}.


\subsubsection{Majorana fermions}

Since the interaction term~\eqref{eqn:H-int} conserves only fermion parity, $\hat{P} = (-1)^{\hat{N}_f}$, not fermion number, $\hat{N}_f = \sum_i \hat{c}^\dagger_i \hat{c}^{\phantom{\dagger}}_i$,
we find that it is convenient to work in an operator basis that corresponds to Majorana fermions. At the operator level, we introduce two Hermitian operators $\hat{\phi}_{ap}$ for each momentum $p$
\begin{subequations}
\begin{align}
    \hat{\phi}_{1p} &= \frac{1}{\sqrt{2}} \left(  \hat{c}_{-p}^\dagger +  \hat{c}_p^{\phantom{\dagger}} \right) = \frac{1}{\sqrt{2}} e^{i\vartheta_p/2} \left(  \hat{\gamma}_{-p}^\dagger +  \hat{\gamma}_p^{\phantom{\dagger}} \right) \\
    \hat{\phi}_{2p} &= \frac{i}{\sqrt{2}} \left(  \hat{c}_{-p}^\dagger - \hat{c}_p^{\phantom{\dagger}} \right) = \frac{i}{\sqrt{2}} e^{-i\vartheta_p/2} \left(  \hat{\gamma}_{-p}^\dagger - \hat{\gamma}_p^{\phantom{\dagger}} \right) 
    \, .
\end{align}
\label{eqn:k-space-Majorana}%
\end{subequations}
which correspond to the Fourier transform of the real-space Majorana operators, e.g., $\hat{c}_i =  (\hat{\phi}_{1i} + i\hat{\phi}_{2i})/\sqrt{2}$, with $\hat{\phi}_{ai} = \hat{\phi}_{ai}^\dagger$ and normalization $\hat{\phi}_{ai}^2 = 1/2$.
While Majorana fermions have no Grassmann number analogue (since anticommuting Grassmann numbers must square to zero), 
it is nevertheless possible to perform the transformation in Eq.~\eqref{eqn:k-space-Majorana} at the level of Grassmann numbers%
\footnote{The transformation is given by, e.g., $\phi_{1p} = \frac{1}{\sqrt{2}}e^{i\vartheta_p/2}(\bar{\psi}_{-p} + \psi_p)$. The `barred' fields, $\bar{\phi}_{ap}$ are not independent from the $\phi_{ap}$. If they are introduced, they are identically equal to $\bar{\phi}_{ap} \equiv \phi_{a,-p}$.}
to express the Green's function~\eqref{eqn:fermionic-greens} in the `Majorana basis'~\cite{ShankarQFT}.
Grassman integration rules for these variables are outlined in Appendix~\ref{sec:integration-rules}.
While this choice is essentially immaterial at the level of the noninteracting action, when interactions are accounted for in Sec.~\ref{sec:self-energy}, the Majorana basis will circumvent the need to introduce multiple interaction vertices, thereby simplifying the obtained expressions.
Using this representation, the noninteracting action is expressed in terms of the free Majorana Green's function as
\begin{equation}
    iS_0[\phi^\alpha] 
    = -\frac12 \sum_k
    \int \mathrm{d}t \, \mathrm{d}t' \,
    \phi^\alpha_{-k}(t) [\breve{D}_0^{-1}]^{\alpha\beta}_{k}(t, t') \phi^\beta_{k}(t')
    \, ,
\end{equation}
The multi-indices $\alpha$, $\beta$ now run over both Keldysh (cl, q) and Majorana $(1, 2)$ indices.
The breve ``$\breve{\phantom{o}}$'' is used to denote matrices in $\text{Keldysh}\otimes\text{Majorana}$ space, while the check ``$\check{\phantom{o}}$'' is reserved for matrices with respect to Majorana indices only.
In contrast to~\eqref{eqn:free-fermion-Greens}, the causality structure of the matrix $\breve{D}_0$ is now bosonic in character:
\begin{equation}
    \langle \phi_\alpha (t) \phi_\beta(t') \rangle = \breve{D}_{\alpha\beta}(t,t') =
    \begin{pmatrix}
        \DK & \DR \\
        \DA & \check{0}
    \end{pmatrix}_\text{K}
    \, .
\end{equation}
This structure derives from using the Keldysh rotation for bosons $\phi^{\text{cl/q}}(t) = [\phi^+(t) \pm \phi^-(t)] / \sqrt{2}$.
The structure of the Green's functions with respect to Majorana indices can be derived directly from the fermionic Green's functions in Eq.~\eqref{eqn:fermionic-greens} using the transformation~\eqref{eqn:k-space-Majorana} treated as a linear transformation between Grassmann numbers.
Expressed in terms of real time, we find
\begin{equation}
    \DR(k,t) = \theta(t)
    \begin{pmatrix}
        \cos\epsilon_k t & e^{i\vartheta_k} \sin \epsilon_k t \\
        -e^{-i\vartheta_k} \sin \epsilon_k t & \cos\epsilon_k t
    \end{pmatrix}
    \, ,
    \label{eqn:Majorana-DR}
\end{equation}
where the momentum dependence of $\breve{D}_{\alpha\beta}(k, t)$ is defined by $\langle \phi^\alpha_k(t) \phi^\beta_{-k}(0) \rangle$.
The advanced component $\DA(k, t)$ then follows directly from the symmetry properties $\DA(k, t) = -[\DR(-k, -t)]^\mathsf{T}$, with $\mathsf{T}$ denoting transposition over Majorana indices.
The Heaviside step function ensures that the retarded and advanced Green's functions are appropriately causal.
Meanwhile, for the Keldysh component, we find
\begin{equation}
    \DK (k, t) = -iF_k 
    \begin{pmatrix}
        \sin\epsilon_k t & - e^{i\vartheta_k} \cos\epsilon_k t \\
        e^{-i\vartheta_k} \cos\epsilon_k t & \sin\epsilon_k t
    \end{pmatrix}
    \, ,
    \label{eqn:Majorana-DK}
\end{equation}
where $F_k$ is shorthand for $F(\epsilon_k)$, which derives from the fermionic distribution function in Eq.~\eqref{eqn:fermionic-GK}, since the two are connected to one another via the linear transformation in Eq.~\eqref{eqn:k-space-Majorana}.
The Keldysh component satisfies $\DK(k, t) = - [\DK(-k, -t)]^\mathsf{T}$, where we have used antisymmetry of the Bogoliubov parameter $\vartheta_{-k} = -\vartheta_{k}$.
The phase factors $e^{i\vartheta_k}$ can be removed from Eqs.~\eqref{eqn:Majorana-DR} and~\eqref{eqn:Majorana-DK}, if desired, by rotating the fields into the `diagonal' basis
$\varphi_{1p} = e^{-i\vartheta_p/2} \phi_{1p}$ and $\varphi_{2p} = e^{i\vartheta_p/2} \phi_{2p}$.
This transformation is equivalent to the Bogoliubov transformation from $k$-space fermions $\hat{c}_k$ to the operators $\hat{\gamma}_k$ that diagonalize $\hat{H}_0$.
If this transformation is performed, the phases factors $e^{i\vartheta_k}$ are transferred from the Green's functions in Eqs.~\eqref{eqn:Majorana-DR}
and~\eqref{eqn:Majorana-DK} to the interaction and drive vertices. We opt to leave the phases in the Green's functions~\eqref{eqn:Majorana-DR}
and~\eqref{eqn:Majorana-DK}.


\subsubsection{Magnetization}

We now move to expressing the time-dependent drive~\eqref{eqn:H-drive} in the Majorana basis~\eqref{eqn:k-space-Majorana}.
At the operator level, the total magnetization evaluates to $\hat{M}^x = \frac{1}{2} \sum_i \hat{\sigma}_i^x = -\sum_p i \hat{\phi}_{1,-p}\hat{\phi}_{2p}$.
On the Keldysh contour $\mathcal{C}$, the corresponding contribution to the action becomes
\begin{equation}
    S_\text{drive} = \int_{-\infty}^\infty \mathrm{d}t\,  \sum_{p>0}
    \left[
        h^\text{q} \phi^\mathsf{T}_{-p} \breve{M}^\text{q} \phi^{\phantom{^\mathsf{T}}}_{p} +
        h^\text{cl} \phi^\mathsf{T}_{-p} \breve{M}^\text{cl} \phi^{\phantom{^\mathsf{T}}}_{p}
    \right]  
    \, ,
    \label{eqn:magnetization-vertices}
\end{equation}
where $(M^\text{q})^{ab}_{cd} = \tau^2_{ab}\sigma^0_{cd}$, and $(M^\text{cl})^{ab}_{cd} = \tau^2_{ab}\sigma^1_{cd}$, with $\boldsymbol{\tau}$ and $\boldsymbol{\sigma}$ the Pauli matrices with respect to Majorana and Keldysh indices, respectively.
Note that the choice of operators $\hat{\phi}_{ap}$ in Eq.~\eqref{eqn:k-space-Majorana} leads to magnetization vertices $\breve{M}^{\text{cl/q}}$ that are independent of momentum, and that the symmetry properties of the integrand have allowed us to sum over positive momenta only, removing an overall factor of $1/2$.
Finally, we mention for completeness that, since the fermionic fields are evaluated at equal times in the continuum notation of Eq.~\eqref{eqn:magnetization-vertices}, the constant term generated by normal ordering $\hat{M}^x$ has been removed in order to generate the correct expectation values. This subtlety is described in further detail in Ref.~\cite{Kamenev2011}.


\subsection{Nonlinear susceptibilities}
\label{sec:nonlinear-suscept}


\subsubsection{The generating functional \texorpdfstring{$Z[h^\text{cl}, h^\text{q}]$}{}}

We are interested in the response of the system's magnetization $\hat{M}^x$ to the driving field given in Eq.~\eqref{eqn:H-drive}.
The expectation value of the magnetization $\hat{M}^x = \frac12 \sum_{i} \hat{\sigma}_i^x$ is, by construction, given conveniently by functional differentiation of the generating functional $Z[h^\text{cl}, h^\text{q}]$.
The classical component of $M^x(t)$ couples to the \emph{quantum} component of the generating field, such that the expectation value is recovered by evaluating
\begin{equation}
    \langle \hat{M}^x(t) \rangle = \left. \frac{1}{2i} \frac{\delta Z[h^\text{cl}, h^\text{q}]}{\delta h^\text{q}(t)} \right\rvert_{h^\text{q}=0}
    \, .
    \label{eqn:exact-magnetisation}
\end{equation}
Although formal, this expression is \emph{exact} to all orders in $h^\text{cl}(t)$, i.e., Eq.~\eqref{eqn:exact-magnetisation} contains complete information about the system's linear and nonlinear response properties. The calculation of the system's response properties is therefore reduced to evaluating the generating functional $Z[h^{\text{cl}}, h^{\text{q}}]$. 
To obtain the linear and the various higher-order (i.e., nonlinear) susceptibilities, Eq.~\eqref{eqn:exact-magnetisation} is written as functional series expansion in the driving field $h^\text{cl}(t)$. For instance, at the linear level
\begin{equation}
    \langle \delta \hat{M}^x(t_1) \rangle =
    \int_{-\infty}^{\infty} \mathrm{d}t_2 \, \ChiOne (t_1, t_2)
    h^\text{cl}(t_2) 
    + \ldots
    \, ,
\end{equation}
where the linear susceptibility $\ChiOne (t_1, t_2)$ is identified from the functional derivative expansion of Eq.~\eqref{eqn:exact-magnetisation} as
\begin{equation}
    \ChiOne(t_1, t_2) = \frac{1}{2i} \left.  \frac{\delta^2 Z[h^\text{cl}, h^\text{q}]}{\delta h^\text{cl}(t_2)\delta h^\text{q}(t_1)} \right\rvert_{h^\text{q} = h^\text{cl} =0}
    \, .
    \label{eqn:chi-1}
\end{equation}
The notation $\delta \hat{M}^x(t_1)$ denotes the removal of the equilibrium magnetization, $\delta \hat{M}^x \equiv \hat{M}^x - \langle \hat{M} \rangle_0$ (i.e., the expectation value $\langle \cdots \rangle_0$ is evaluated in the absence of the driving field).
In thermal equilibrium, the derivative appearing in Eq.~\eqref{eqn:chi-1} contains only time differences, and $\ChiOne(t_1, t_2)$ can be redefined through $\ChiOne(t_1, t_2) \to \ChiOne(t_1-t_2)$.
The generalization of this result to higher orders in $h^\text{cl}(t)$ is straightforward.
Of particular importance to the problem at hand is the third-order susceptibility, which is given by
\begin{equation}
    \ChiThree (t_1; t_2, t_3, t_4) = \frac{1}{2i} \left.  \frac{\delta^4 Z[h^\text{cl}, h^\text{q}]}{\delta h^\text{cl}_4\delta h^\text{cl}_3\delta h^\text{cl}_2\delta h^\text{q}_1} \right\rvert_{h^\text{q} = h^\text{cl} =0}
    \, ,
    \label{eqn:chi-3-from-functional}
\end{equation}
where $h_n^{a}$ is shorthand for $h^a(t_n)$, with $a=\text{cl},\text{q}$.
These expressions are equivalent to the generalized Kubo formulae~\cite{ChouUnified} used in, e.g., Refs.~\cite{WanArmitage2019,Choi2DCSKitaev,Nandkishore2021Spectroscopic}.
In the above symmetric representation, the third-order contribution to the magnetization response is simply
\begin{equation}
    \langle \hat{M}^x(t_1) \rangle^{(3)}
    =
    \frac{1}{3!}
    \int
    \prod_{n=2}^4 \mathrm{d}t_n \,
    \ChiThree (t_1; t_2, t_3, t_4)
    \prod_{n=2}^4 h(t_n)
    \, ,
\end{equation}
where each time integration spans the entire real axis.
An alternative -- and often more convenient -- representation takes advantage of (i) symmetry of the integrand under permutations of $\{ t_n \}_{n>1}$, and (ii) causality, to restrict the domain of integration to $t_1 > t_2 > t_3 > t_4$
\begin{equation}
    \langle \hat{M}^x(t_1) \rangle^{(3)}
    = \!
    \int\limits_{t_n > t_{n+1}} \!\!\!
    \, \prod_{n=2}^4 \mathrm{d}t_n \,
    \ChiThree (t_1; t_2, t_3, t_4)
    \prod_{n=2}^4 h(t_n)
    \, .
    \label{eqn:M3-restricted-domain}
\end{equation}
Finally, since the response function is a function of time \emph{differences} only -- at least in thermal equilibrium -- we can once again redefine the third-order susceptibility according to
\begin{equation}
    \ChiThree (t_1; t_2, t_3, t_4) \to \ChiThree (t_1-t_2, t_1-t_3, t_1-t_4)
    \, .
    \label{eqn:time-difference-convention}
\end{equation}
Since $t_n > t_{n+1}$ in our convention, the time differences on the right hand of Eq.~\eqref{eqn:time-difference-convention} side increase when the time arguments are read from left to right.

\begin{figure}[t]
    \centering
    \includegraphics[width=0.65\linewidth,valign=c]{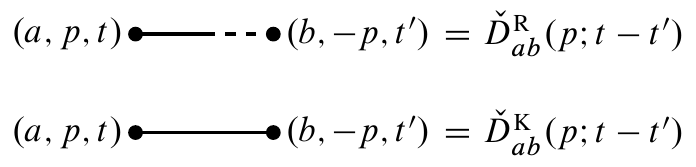}
    \hfill
    \includegraphics[width=0.325\linewidth,valign=c]{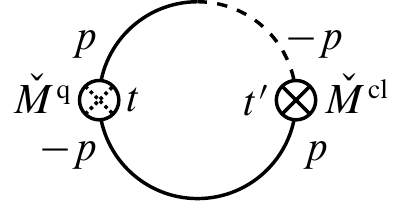}
    \caption{\textbf{Left}: Notation for the Green's functions. Dashed (solid) lines correspond to the Majorana field $\phi^\text{q}_a(t)$ ($\phi^\text{cl}_a(t)$). \textbf{Right}: The only diagram that contributes to the linear response of the magnetization.}
    \label{fig:linear-feynman}
\end{figure}


\subsection{Discussion}

At this point, we pause to reflect on what has been accomplished in the prior Sections \ref{sec:path-integral-approach} and \ref{sec:nonlinear-suscept}.
Rather than working with the nested commutators that appear in generalized Kubo formulae, we have framed the problem in the language of a real-time path integral on the Keldysh contour.
For noninteracting theories, the generating functional $Z[h^\text{cl}, h^\text{q}]$ can be evaluated \emph{exactly} (in the noninteracting TFIM, it can be expressed in terms of a Pfaffian, as shown in Appendix~\ref{sec:integration-rules}).
Once $Z[h^\text{cl}, h^\text{q}]$ has been obtained, all linear and nonlinear response proprieties follow immediately via functional differentiation, providing a simple recipe for deducing the nonlinear response properties to arbitrary order in the driving field.
In addition to this practical benefit, the formalism additionally offers conceptual benefits, since the nonlinear susceptibilities can be represented using a familiar diagrammatic language (see, e.g., Figs.~\ref{fig:linear-feynman} and \ref{fig:nonlinear-feynman}).
These practical and conceptual benefits become even more pronounced once interactions are introduced to the model.
As we will show in the following sections, the present formalism allows us to write the nonlinear response directly in terms of the free Majorana Green's functions, which can subsequently be augmented via introducing a self-energy, immediately accounting for some of the most salient features furnished by interactions. Other interaction effects can also be conveniently represented in the diagrammatic language.
Of course, not all problems will lend themselves to such a perturbative treatment, and one may have to resort to, e.g., the nested commutator approach.


\subsection{Noninteracting results}

We begin by evaluating the linear and nonlinear response properties of the noninteracting model (i.e., $\hat{H}_\text{int}=0$), which constitute the zeroth-order term in the perturbative expansion in the interaction strength $\lambda$.
As studied in Ref.~\cite{WanArmitage2019}, the results for $\ChiThree$ are still highly nontrivial, even at the noninteracting level.
We begin by rederiving the results of Ref.~\cite{WanArmitage2019} explicitly as a demonstration of the techniques developed in prior sections.


\subsubsection{Linear response}

\begin{figure}
    \centering
    \includegraphics[width=0.35\linewidth]{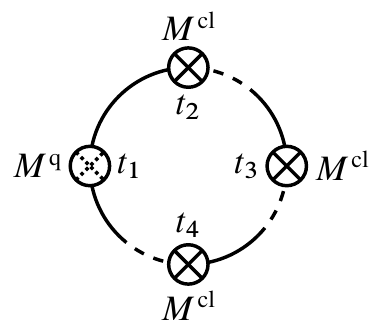}%
    \hspace{20pt}%
    \includegraphics[width=0.35\linewidth]{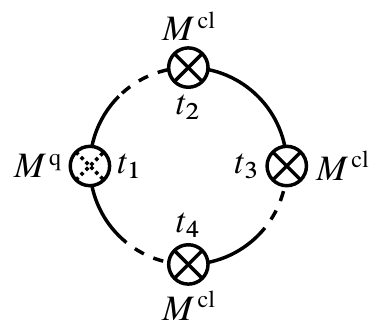}%
    \\
    \includegraphics[width=0.35\linewidth]{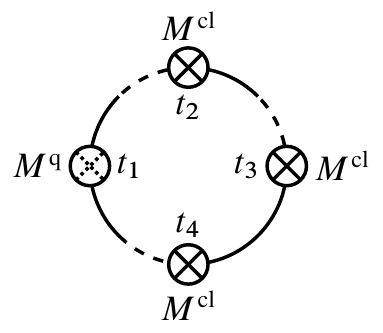}%
    \hspace{20pt}%
    \includegraphics[width=0.35\linewidth]{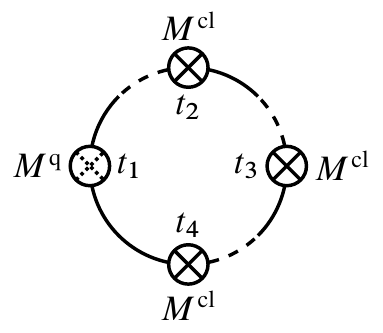}%
    \caption{The four connected Feynman diagrams that contribute to the third-order nonlinear susceptibility, $\ChiThree (t_1; t_2, t_3, t_4)$, the response at $t_1$ to perturbations at times $(t_2, t_3, t_4)$. The diagrams are shown for a particular permutation of the time arguments $(t_2, t_3, t_4)$; analogous diagrams can be written for the other permutations. The momentum labels have been omitted.}
    \label{fig:nonlinear-feynman}
\end{figure}

To first order in the drive, we must evaluate the functional derivative in Eq.~\eqref{eqn:chi-1}.
Making use of Wick's theorem for Majorana fermions, we find that
\begin{multline}
    \left. \frac{\delta^2 Z[h^\text{cl}, h^\text{q}]}{\delta h^\text{cl}(t') \delta h^\text{q}(t)}\right\rvert_{h^\text{q}=h^\text{cl}=0}
    =\\
    \sum_{p>0} \breve{M}^\text{q}_{\mu\nu} \breve{D}_{\nu\lambda}(p; t, t') \breve{M}^\text{cl}_{\lambda\sigma} \breve{D}_{\sigma\mu}(p; t', t) 
    \, ,
    \label{eqn:linear-response-unsimplified}
\end{multline}
where, as usual, the Greek indices run over both Keldysh and Majorana indices.
Any contributions from disconnected diagrams can be shown to vanish identically by virtue of the identity $Z[h^\text{cl}, 0]=1$; the nonequilibrium counterpart of the linked cluster theorem~\cite{Kamenev2011}.
Explicitly, such diagrams either vanish by virtue of having vanishing temporal support, $\propto \theta(t)\theta(-t)$, or as a consequence of the result $D^\text{R}(t, t) + D^{\text{A}}(t, t)=0$ when the Green's functions are evaluated at equal times.
An alternative derivation of the result~\eqref{eqn:linear-response-unsimplified} is presented in Appendix~\ref{sec:integration-rules}, where we take derivatives of the exact generating functional $Z[h^\text{cl}, h^\text{q}]$.
When expressed in terms of the Keldysh components, making the causality structure explicit, the above evaluates to
\begin{equation}
    \ChiOne(t,t') =
    -i\sum_{p>0}
    \Tr[\check{\tau}^2 \DR(p; t, t')\check{\tau}^2 \DK (p; t', t)]
    \, ,
\end{equation}
where the trace is over Majorana indices only.
The corresponding Feynman diagram is shown in Fig.~\ref{fig:linear-feynman}.
Substituting the expressions for the components of $\breve{D}_0$ in Eqs.~\eqref{eqn:Majorana-DR} and~\eqref{eqn:Majorana-DK} under the trace,
we arrive at the final expression
\begin{equation}
    \ChiOne(t-t') = 2 \theta(t-t') \sum_{p>0} F_p \sin^2 \vartheta_p \sin[2\epsilon_p (t-t')] 
    \, .
    \label{eqn:TFIM-linear-response-xx}
\end{equation}
At zero temperature, $F_p=1$, since the spectrum $\epsilon_k$ is strictly positive%
\footnote{Since we impose antiperiodic boundary conditions and even $L$, the $k=0$ mode is not compatible with the boundary conditions and all energies in a system of finite size are strictly positive, $\epsilon_k>0, \,\forall k$, even at $g=1$, where the gap vanishes only in the thermodynamic limit.}, and we recover the results of Ref.~\cite{WanArmitage2019}.
In the ferromagnetic regime, $g<1$, the Fourier transform of Eq.~\eqref{eqn:TFIM-linear-response-xx}, $\ChiOne(\omega)$, exhibits a broad continuum as a consequence of the two-domain-wall continuum. The spatially homogeneous nature of the magnetization pulses enforces that domain wall pairs must have net zero momentum, $k_1+k_2=0$, but the magnitude of the individual momenta $\abs{k_1} = \abs{k_2} = k$ remains unconstrained, giving rise to a broad response at $\omega = \pm 2\epsilon_{k}$.
The extension of the result to nonzero temperatures is trivial: For $T>0$, the weight of the contribution from quasimomentum $p$ is reduced from unity to $\tanh(\beta\epsilon_p/2)$, with the states near the minimum of the dispersion being most strongly modified.


\subsubsection{Third-order response}

The response properties at third order are more interesting, since they contain the ``spinon echo''~\cite{WanArmitage2019} (or rephasing) signal.
We leave a discussion of the second-order response to Appendix~\ref{sec:second-order};
while the second-order response does contribute to the magnetization response under the 2DCS protocol [Eq,.~\eqref{eqn:Mx-2DCS}], it does not contain significant new information beyond linear response.
The Feynman diagrams that result from differentiating $Z[h^\text{cl}, h^\text{q}]$ four times are numerous.
Four of them are shown in Fig.~\ref{fig:nonlinear-feynman}, while the rest can be obtained via permutations of the indices $(t_2, t_3, t_4)$ [the resulting $\ChiThree (t_1; t_2, t_3, t_4)$ is therefore symmetric in the indices $t_2, t_3, t_4$].
Again, the contributions from disconnected diagrams can be shown to vanish identically as a direct consequence of the identity $Z[h^\text{cl}, 0]=1$.
The diagrams in Fig.~\ref{fig:nonlinear-feynman} evaluate to
\begin{align}
    &\left. \frac{\delta^4 Z[h^\text{cl}, h^\text{q}]}{\delta h^\text{cl}(t_4) \delta h^\text{cl}(t_3) \delta h^\text{cl}(t_2) \delta h^\text{q}(t_1)}\right\rvert_{h^\text{q}=h^\text{cl}=0} 
    =  -\sum_{p>0}  \notag \\
    \big\{ &\Tr[\check{\tau}^2  \check{D}_{-p}^{\text{K}}(t_{12}) \check{\tau}^2 \check{D}_{-p}^\text{A}(t_{23}) \check{\tau}^2 \check{D}_{-p}^\text{A}(t_{34}) \check{\tau}^2  \check{D}_{-p}^\text{A}(t_{41})] \notag\\
    +&\Tr[\check{\tau}^2  \check{D}_{-p}^{\text{R}}(t_{12}) \check{\tau}^2 \check{D}_{-p}^{\text{K}}(t_{23}) \check{\tau}^2 \check{D}_{-p}^\text{A}(t_{34}) \check{\tau}^2  \check{D}_{-p}^\text{A}(t_{41})] \notag\\
    +&\Tr[\check{\tau}^2  \check{D}_{-p}^{\text{R}}(t_{12}) \check{\tau}^2 \check{D}_{-p}^{\text{R}}(t_{23}) \check{\tau}^2 \check{D}_{-p}^{\text{K}}(t_{34}) \check{\tau}^2  \check{D}_{-p}^\text{A}(t_{41})] \notag\\
    +&\Tr[\check{\tau}^2  \check{D}_{-p}^{\text{R}}(t_{12}) \check{\tau}^2 \check{D}_{-p}^{\text{R}}(t_{23}) \check{\tau}^2 \check{D}_{-p}^{\text{R}}(t_{34}) \check{\tau}^2  \check{D}_{-p}^{\text{K}}(t_{41})] \big\} \notag \\
    + &\text{ permutations of $(t_2, t_3, t_4)$}
    \, ,
    \label{eqn:nonlinear-functional-deriv}
\end{align}
where we have employed the shorthand notation $t_{ij} \equiv t_i - t_j$ (as expected, the response is a function of time differences only).
We can immediately see that~\eqref{eqn:nonlinear-functional-deriv} is appropriately causal: In each of the four terms the $\theta$-functions demand that the time arguments $t_2$, $t_3$, and $t_4$ must precede the `measurement' time $t_1$.
This result can be expressed more compactly using a sequential parametrization of the times.
We define new sequential time arguments $s_n$ via
$t_n = \sum_{i\geq n} s_i$ (equivalently, $s_n = t_{n} - t_{n+1}$ for $n<4$).
One of the benefits of this representation is that the two time argument orderings that appear in the 2DCS response~\eqref{eqn:Mx-2DCS} are particularly simple in this language: $\ChiThree(t, t, t+\tau)$ is described by $(s_1, s_2, s_3) = (t, 0, \tau)$ and $\ChiThree(t, t+\tau, t+\tau)$ by $(s_1, s_2, s_3) = (t, \tau, 0)$. 
In terms of these new variables we are able to write Eq.~\eqref{eqn:nonlinear-functional-deriv} compactly as
\begin{equation}
\begin{aligned}
    \ChiThree (s_1, s_2, s_3) = -2 & \theta(s_1) \theta(s_2) \theta(s_3) \sum_{p>0} \\
    F_p \big\{&\sin^2 (2\vartheta_p) \sin[2\epsilon_p (s_1 + s_2 + s_3)] \\
              -&\sin^2 (2\vartheta_p) \sin[2\epsilon_p (s_2 + s_3)] \\
              +2&\sin^4 (\,\vartheta_p\,)\, \sin[2\epsilon_p (s_1 + s_3)]  \\
              +2&\sin^4 (\,\vartheta_p\,)\, \sin[2\epsilon_p (s_1 - s_3)]\big\}
    \, .
\end{aligned}
\label{eqn:chi-3-sequential}
\end{equation}
Again, at strictly zero temperature, $F_p=1$ and this expression is in agreement with the results of Ref.~\cite{WanArmitage2019},
with the final line of~\eqref{eqn:chi-3-sequential} corresponding to the rephasing signal for the state labeled by momentum $p$.
Note that Eq.~\eqref{eqn:chi-3-sequential} employs an alternative convention for the time arguments with respect to Eq.~\eqref{eqn:time-difference-convention}.
The extension to nonzero temperatures simply re-weights the contributions from the various momenta according to their corresponding energy.

When the final line of Eq.~\eqref{eqn:chi-3-sequential} is evaluated with the time arguments that correspond to the third-order susceptibility $\ChiThree(t, t, t+\tau)$, it gives $\sim \sin^4(\vartheta_p) \sin[2\epsilon_p (t-\tau)]$.
Taking the Fourier transform over time arguments $t$ and $\tau$ to conjugate variables $\omega_t$ and $\omega_\tau$, respectively, the imaginary part of the response, $\Im \ChiThree(\omega_t, \omega_\tau)$, exhibits sharp streaks along the antidiagonal $\omega_t = -\omega_t$ [see Figs.~\subref*{fig:free-2DCS-spectrum} and \subref{fig:free-2DCS-overlap}].
These streaks occur at the frequencies that were observed in the broad linear response spectrum~\eqref{eqn:TFIM-linear-response-xx}, i.e., $\omega_t = \pm 2\epsilon_p$.
In the noninteracting expression, Eq.~\eqref{eqn:chi-3-sequential}, these streaks are infinitely sharp in the direction transverse to the streaks, i.e., $\omega_t = \omega_\tau$.
This is a consequence of the perfect rephasing that occurs in the absence of any dissipation, since the Green's functions exhibit undamped oscillations. We will see in the next section that the signal broadens along the $\omega_t = \omega_\tau$ direction in the presence of interaction-induced quasiparticle decay, and that the width will provide quantitative information about the interaction-induced self-energy.

While we focus in this manuscript on the information that can be extracted from the rephasing signal, we also briefly consider the pump-probe response in Appendix~\ref{sec:pump-probe}. In contrast to the rephasing signal, the pump-probe response manifests in the complementary contribution to the 2DCS response, $\ChiThree(t, t+\tau, t+\tau)$, and appears along the $\omega_\tau=0$ axis. This contribution is also of experimental interest since, for two-level systems, the width of the pump-probe signal contains information about the $T_1$ time, which characterizes the rate of population decay~\cite{WanArmitage2019}.

\begin{figure}
    \centering
    \begin{minipage}{0.6\linewidth}
        \includegraphics[width=0.75\linewidth]{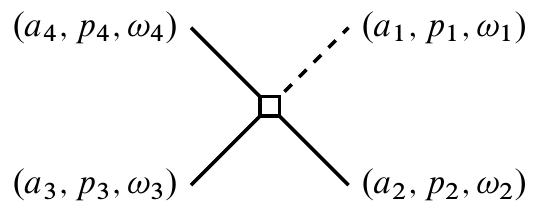}\\[5pt]
        \includegraphics[width=0.75\linewidth]{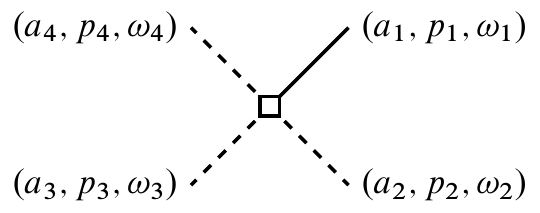}
    \end{minipage}%
    \begin{minipage}{0.35\linewidth}
        \includegraphics[width=0.45\linewidth]{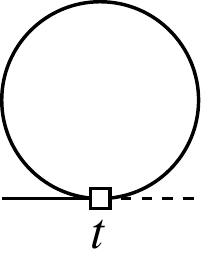}
    \end{minipage}
    \caption{\textbf{Left}: The two types of interaction vertex that appear in $S_\text{int}$ \eqref{eqn:S_int} in frequency/momentum space. \textbf{Right}: The `tadpole' diagram, which is the only diagram that contributes to the self-energy at first order in the interactions. The $\check{\Sigma}^\text{q,q}\equiv \check{\Sigma}^\text{K}$ component vanishes at this order. The diagram is local in time, i.e., independent of frequency, and leads to a renormalization of the single-particle spectrum.}
    \label{fig:first-order-selfenergy}
\end{figure}


\subsection{Self-energy corrections}
\label{sec:self-energy}

One of the benefits of formulating the nonlinear response in the language of path integrals is that the tools of many-body perturbation theory can be applied to the problem directly.
The first \emph{perturbative} correction to the noninteracting result that we will consider comes from one-particle irreducible (1PI) diagrams, i.e., we will perform a perturbative expansion of the self-energy $\breve{\Sigma}$.
However, there also are a number of contributions to the bilinear part of the action, arising from normal ordering the interaction terms, which can be accounted for \emph{exactly}.
These terms generated by normal ordering lead to a renormalization of the quasiparticle spectrum~\eqref{eqn:JW-energies}.

A nontrivial feature of the Keldysh formalism is that the self-energy (which, like the Green's function, is a matrix with Keldysh indices) possesses the same causality structure as the inverse Green's function.
Since we utilize the bosonic Keldysh rotation for the Majorana fields $\phi(t)$, the real-time self-energy assumes the form
\begin{equation}
    \breve{\Sigma}(k; t, t') =
    \begin{pmatrix}
        0 & \check{\Sigma}^\text{A} \\
        \check{\Sigma}^\text{R} & \check{\Sigma}^\text{K}
    \end{pmatrix}_\text{K}
    \, .
\end{equation}
The expression for the full Green's function is then $\breve{D} = (\breve{D}_0^{-1} - \breve{\Sigma})^{-1}$, which implies that, e.g.,
$\check{D}^\text{R} = ([\check{D}_0^\text{R}]^{-1} - \check{\Sigma}^\text{R})^{-1}$, while $\DK = \DR \hat{\Sigma}^\text{K} \DA$ if the Keldysh component of the \emph{inverse} Green's function $\check{D}_0^\text{K}$ is a pure regularization (i.e., if the noninteracting system has no intrinsic lifetime arising from, e.g., coupling to bath).


\subsubsection{Exact bilinear contributions}
\label{sec:nonperturbative}

In the presence of interactions, the way in which the fermionic operators are normal ordered prior to introducing the coherent state path integral allows different contributions to be taken into account exactly.
Normal ordering at the level of the real-space JW fermions merely gives rise to (at most) a sign: $\sum_i \hat{\sigma}_i^x \hat{\sigma}_{i+1}^x \supset - \sum_i \hat{c}_{i+1}^\dagger \hat{c}_i^\dagger \hat{c}_{i+1}^{\phantom{\dagger}} \hat{c}_i^{\phantom{\dagger}}$.
However, if the noninteracting Bogoliubov transformation is used [defined below Eq.~\eqref{eqn:jordan-wigner-transform}], then the interaction term does not remain normal ordered; if the transformed interaction in the $\hat{\gamma}_k$ fermions is subsequently (re-) normal ordered, new terms, bilinear in fermionic operators $\hat{\gamma}_k$, are generated.
This feature can be remedied by determining the Bogoliubov parameter $\vartheta_k$ self-consistently and \emph{nonperturbatively}.
In the manner of Ref.~\cite{Robinson2014breakdown}, $\vartheta_k$ is found (in the thermodynamic limit) by solving an integral equation that treats all bilinear terms generated via normal ordering exactly. This procedure is outlined below for completeness.

In momentum space, the Hamiltonian (absent the drive) can be written in standard form as
\begin{multline}
    \hat{H}_\lambda = \frac{1}{2} \sum_k
    \begin{pmatrix}
        \hat{c}_k^{\dagger} & \hat{c}_{-k}
    \end{pmatrix}
    \begin{pmatrix}
        \alpha_k & -i\beta_k \\
        i\beta_k & -\alpha_k
    \end{pmatrix}
    \begin{pmatrix}
        \hat{c}_k \\
        \hat{c}^\dagger_{-k}
    \end{pmatrix}
    \\
    -\frac{4\lambda}{L}\sum_{\{k_i\}}
    V_{k_1, k_2, k_3, k_4}
    \hat{c}^\dagger_{k_1} 
    \hat{c}^\dagger_{k_2}
    \hat{c}^{\phantom{\dagger}}_{-k_3}
    \hat{c}^{\phantom{\dagger}}_{-k_4}
    \, ,
    \label{eqn:static-H-kspace}
\end{multline}
where the bilinear term is expressed in terms of the functions 
\begin{subequations}
\begin{align}
    \alpha_k &= 2[g-(1-\lambda)\cos k] -4\lambda \\
    \beta_k  &= 2(1+\lambda) \sin k 
\end{align}
\end{subequations}
which include renormalization of the single particle band structure due to the term $\hat{\sigma}_i^y \hat{\sigma}_{i+1}^y$ and a diagonal chemical potential term arising from cross terms in $\hat{\sigma}_i^x \hat{\sigma}_{i+1}^x$.
 If next-nearest-neighbor fermion processes are additionally included, then $\alpha_k$ and $\beta_k$ are modified in the manner described below Eq.~\eqref{eqn:free-fermion-generic}, acquiring terms $\propto \cos(2k)$ and $\propto \sin(2k)$, respectively.
The interaction matrix elements are
\begin{multline}
    V_{k_1, k_2, k_3, k_4} = \delta_\text{P}\left( \sum_i k_i \right) \frac14 \big[ \cos(k_1 + k_3) - \cos(k_2 + k_3) \\
    +\cos(k_2+k_4) - \cos(k_1 + k_4) \big]
    \, ,
    \label{eqn:interaction-mel-fermion}
\end{multline}
where $\delta_\text{P}(x)$ is the $2\pi$-periodic Kronecker delta that enforces momentum conservation up to a reciprocal lattice vector.
As described above, the introduction of new fermions via the Bogoliubov transformation $\hat{c}_k = u_k \hat{\gamma}_k + iv_k \hat{\gamma}_{-k}^\dagger$, with
$u_k = \cos\tfrac12\vartheta_k$, $v_k = \sin\tfrac12\vartheta_k$, does not maintain normal ordering for a $\vartheta_k$ that diagonalizes the noninteracting part of Eq.~\eqref{eqn:static-H-kspace}. Instead, we choose $\vartheta_k$ such that it satisfies the nonlinear equation
\begin{equation}
    \tan\vartheta_k = \frac{\beta_k + B_k[\vartheta]}{\alpha_k + A_k[\vartheta]}
    \, .
    \label{eqn:self-consistent-Bogoliubov}
\end{equation}
The functionals $A_k[\vartheta]$ and $B_k[\vartheta]$ are defined by the terms generated by normal ordering the interaction in the new $\hat{\gamma}_k$ degrees of freedom. Explicitly,
\begin{subequations}
\begin{align}
    A_k[\vartheta] &= -16\lambda \int \frac{\mathrm{d}p}{2\pi} \, V_{k,p,-p,-k} \sin^2(\vartheta_p/2) \\
    B_k[\vartheta] &= -4 \lambda \int \frac{\mathrm{d}p}{2\pi} \, V_{p,-p,-k,k} \sin(\vartheta_p) \, .
\end{align}
\end{subequations}
A Bogoliubov parameter $\vartheta_k$ that satisfies the integral equation~\eqref{eqn:self-consistent-Bogoliubov} simultaneously diagonalizes the bilinear contribution \emph{and} ensures that the interaction term remains normal ordered. This allows the coherent state path integral to be constructed in the `diagonal' basis.
We resort to a numerical solution of the integral equation~\eqref{eqn:self-consistent-Bogoliubov} to obtain $\vartheta_k$ self-consistently using standard methods.
The quasiparticle spectrum $\epsilon_k$ is obtained using the resulting self-consistently determined $\vartheta_k$ through
\begin{equation}
    \epsilon_k = \sqrt{ (\alpha_k + A_k[\vartheta])^2 +  (\beta_k + B_k[\vartheta])^2 } 
    \, ,
    \label{eqn:self-consistent-energy}
\end{equation}
which generalizes the free dispersion relation in Eq.~\eqref{eqn:JW-energies}.

\begin{figure}[t]
    \centering
    \includegraphics[width=0.3\linewidth]{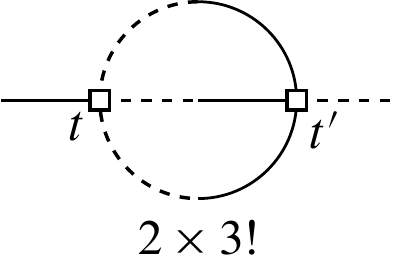}%
    \hspace{10pt}%
    \includegraphics[width=0.3\linewidth]{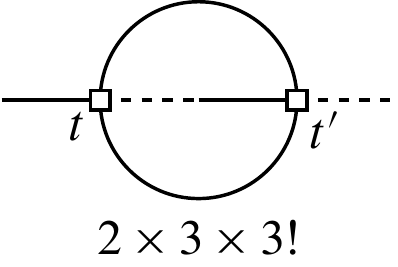}%
    \\
    \includegraphics[width=0.3\linewidth]{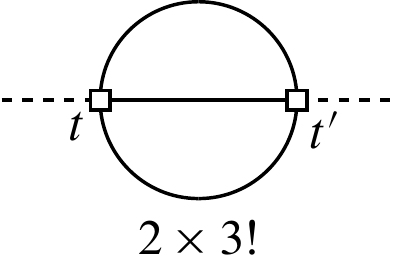}%
    \hspace{10pt}%
    \includegraphics[width=0.3\linewidth]{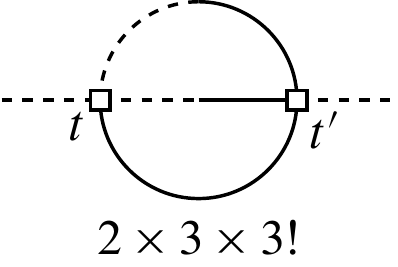}%
    \hspace{10pt}%
    \includegraphics[width=0.3\linewidth]{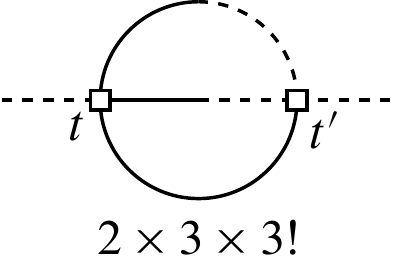}%
    \caption{Second-order contributions to the self-energy matrix $\breve{\Sigma}(k, \omega)$ on the Keldysh contour. \textbf{Top}: The two nonvanishing `sunset' diagrams that contribute to $\check{\Sigma}^{\text{cl},\text{q}} \equiv \check{\Sigma}^\text{A}$. \textbf{Bottom}: The three second-order `sunset' diagrams that contribute to $\check{\Sigma}^{\text{q},\text{q}} \equiv \check{\Sigma}^\text{K}$. In both rows the symmetry factors associated with each diagram are shown underneath.}
    \label{fig:second-order-selfenergy}
\end{figure}


\subsubsection{Perturbative corrections}
\label{sec:interaction-MBPT}

The preceding section dealt with treating certain components of the interactions exactly through the introduction of a self-consistently determined Bogoliubov parameter.
We now evaluate the contributions from the remaining four-fermion terms perturbatively.
To perform a perturbative expansion of the self-energy, we write the interaction term in the form
\begin{multline}
    S_\text{int}[\phi] = \int \frac{\mathrm{d}\omega_i}{2\pi} \, \sum_{\{p_i\}} V_{\{a_i\}}(\{p_i\}, \{\omega_i\} ) \times \\
        \left[  \phi^\text{cl}(1)\phi^\text{q}(2)\phi^\text{q}(3)\phi^\text{q}(4) + \phi^\text{q}(1)\phi^\text{cl}(2)\phi^\text{cl}(3)\phi^\text{cl}(4) \right]
    \, ,
    \label{eqn:S_int}
\end{multline}
with $\phi^\alpha(i)$ shorthand for the Majorana fields in the momentum/frequency representation, $\phi^\alpha_{a_i p_i}(\omega_i)$. 
The two interaction vertices are represented diagramatically in Fig.~\ref{fig:first-order-selfenergy}. 
By transforming Eq.~\eqref{eqn:interaction-mel-fermion} to Majorana fields, the (totally antisymmetric) interaction matrix elements can be shown to equal
\begin{multline}
    \frac{4\lambda}{3L}
    2\pi \delta\left(\sum_{i=1}^4 \omega_i\right)
    \delta_\text{P} \left( \sum_{i=1}^4 p_i \right) \cos\left(  \frac{1}{2} \sum_{i=1}^4 p_i  \right)\times \\
    \bigg\{
        \delta^{a_1 a_2} \delta^{a_3 a_4} \sin\left( \frac{p_1-p_2}{2} \right)\sin\left(\frac{p_3-p_4}{2} \right) \\
        + \text{cyc.~perm.~} [(a_2, p_2), (a_3, p_3), (a_4, p_4)]
    \bigg\}
    \, .
\end{multline}
The delta functions give rise to energy and momentum conservation.
Momentum conservation occurs modulo $2\pi$, i.e., $\sum_i k_i = 2\pi n$ with $n \in \mathbb{Z}$, which implies that the cosine term on the first line contributes only a sign $(-1)^n$.
The expression also vanishes identically if all $a_i$ are equal to one another.

The first contribution to the self-energy $\breve{\Sigma}(k, \omega)$ comes from a tadpole diagram at linear order in the interaction, which is shown in the right-hand side of Fig.~\ref{fig:first-order-selfenergy}.
Accounting for symmetry factors, the diagram evaluates to
\begin{equation}
    \check{\Sigma}^{\text{A}}_{ab}(p, \omega)
    =-\frac{16\lambda}{L}
    \sum_{k}  n_{k}
    \begin{pmatrix}
        0 & -e^{i\vartheta_k} \\
        e^{-i\vartheta_k} & 0
    \end{pmatrix}_{ab} \sin\left( \frac{p+k}{2} \right)^2
    ,
    \label{eqn:HF-correction}
\end{equation}
where $n_k \equiv n_\text{F}(\epsilon_k) = 1/(e^{\beta\epsilon_k}+1)$ is the Fermi-Dirac distribution and the Bogoliubov parameter $\vartheta_k$ is determined self-consistently via solving Eq.~\eqref{eqn:self-consistent-Bogoliubov}.
Note that the full self-energy matrix, which is (almost) diagonal in Fourier space, satisfies $\check{\Sigma}^\text{A}_{ab}(p_1, p_2; \omega_1, \omega_2) = 2\pi \delta(\omega_1 + \omega_2)\delta_\text{P}(p_1 + p_2) \check{\Sigma}_{ab}^\text{A}(p_2, \omega_2)$.
The Keldysh component, on the other hand, remains a pure regularization at this order, i.e., Eq.~\eqref{eqn:HF-correction} corresponds to a renormalization of the single-particle spectrum, but does not endow the quasiparticles with a finite lifetime at first order in the interactions.
The energy shift can be read off directly from Eq.~\eqref{eqn:HF-correction}:
\begin{equation}
    E_k = \abs{\epsilon_k + 16\lambda \int \frac{dp}{2\pi} \, n_p \sin^2\left(\frac{k+p}{2}\right) e^{i(\vartheta_p - \vartheta_k)} }
    \, ,
    \label{eqn:one-loop-dispersion}
\end{equation}
where the dispersion relation absent self-energy corrections (but including the nonperturbative corrections discussed in Sec.~\ref{sec:nonperturbative}), $\epsilon_k$, is given by Eq.~\eqref{eqn:self-consistent-energy}.
Note that normal ordering in the `diagonal' basis implies that the dispersion relation \eqref{eqn:self-consistent-energy} is, in fact, \emph{exact} to first order in the interactions at zero temperature. Only at $T>0$ is the dispersion relation modified due to the presence of a nonvanishing background of thermally excited quasiparticles, with the extrema of the dispersion being most strongly affected  due to the presence of the factor $n_\text{F}(\epsilon_p)$.
Note further that the expression \eqref{eqn:one-loop-dispersion} is substantially simpler than the equivalent expression that would have been obtained if we had worked in the complex fermion basis due to the reduced number of interaction vertices%
\footnote{Specifically, there are three interaction vertices (prior to introducing the contour index) since fermion number is not conserved; in addition to the number conserving term $\sim \hat{\gamma}^\dagger_{k_1}\hat{\gamma}^\dagger_{k_2} \hat{\gamma}^{\phantom{\dagger}}_{-k_3} \hat{\gamma}^{\phantom{\dagger}}_{-k_4}$, there are two additional terms $\sim\hat{\gamma}^\dagger_{k_1}\hat{\gamma}^\dagger_{k_2} \hat{\gamma}^{{\dagger}}_{k_3} \hat{\gamma}^{{\dagger}}_{k_4}$ and $ \sim \hat{\gamma}^\dagger_{k_1}\hat{\gamma}^\dagger_{k_2} \hat{\gamma}^{{\dagger}}_{k_3} \hat{\gamma}^{\phantom{\dagger}}_{-k_4}$, plus their Hermitian conjugates.}.
Equation~\eqref{eqn:HF-correction} satisfies $\check{\Sigma}^\text{A}(k, \omega) = -[\check{\Sigma}^\text{A}]^\mathsf{T}(-k, -\omega)$, so the retarded component is obtained via\footnote{If the time discretisation is taken care of more precisely, the retarded and advanced components of the self-energy in the real time representation should be appropriately causal, e.g., $\check{\Sigma}^\text{R}(t,t') \sim \delta^\text{R}(t-t')$, where $\delta^\text{R/A}(t) = \delta(t \mp 0^+)$. This structure is hidden in the continuum notation.} $\check{\Sigma}^{\text{R}}(p) = \check{\Sigma}^{\text{A}}(p)$.
For the calculation of the second-order correction, we will use the one-loop dispersion~\eqref{eqn:one-loop-dispersion}.
In Eq.~\eqref{eqn:one-loop-dispersion}, we have taken the thermodynamic limit $L\to\infty$, in which the summation can be replaced with an integral in the standard way: $\frac{1}{L}\sum_k \to \int_{-\pi}^\pi \frac{\mathrm{d}k}{2\pi}$.
For numerical calculations, however, we work with systems of finite size.

To calculate the second-order correction, we introduce the following notation for the Majorana Green's functions in momentum/frequency space.
The Fourier transform of~\eqref{eqn:Majorana-DR} and~\eqref{eqn:Majorana-DK} equal
\begin{subequations}
\begin{align}
    \check{D}^\text{R/A}(k,\omega) &=
    \frac{i}{2} \sum_{\sigma=\pm 1}
    \sigma \check{A}^\sigma(k) (\omega - \sigma E_k \pm i0)^{-1} \\
    \DK(k,\omega) &=
     \frac{1}{2}F_k  
     \sum_{\sigma=\pm 1} \check{A}^\sigma(k) 2\pi \delta(\omega-\sigma E_k)
     \, ,
\end{align}
\label{eqn:A-matrix-definition}%
\end{subequations}
respectively, where we have defined $\check{A}^\sigma = \left(\begin{smallmatrix} \sigma & i e^{i\vartheta_k} \\ -ie^{-i\vartheta_k} & \sigma \end{smallmatrix} \right)$.
This compact notation allows the various frequency integrations to be calculated exactly.
The contributions to the second-order self-energy are given diagrammatically in Fig.~\ref{fig:second-order-selfenergy}, which take the form of `sunset' diagrams (with additional Keldysh structure).
We find that the diagrams on the top line of Fig.~\ref{fig:second-order-selfenergy} evaluate to
\begin{widetext}
\begin{multline}
    \Sigma^\text{R}_{ab}(k, \omega) =
    -i\frac{2\cdot 3! }{8}
    \sum_{p_1, p_2, p_3}  V_{a a_1 a_2 a_3} (-k, p_1, p_2, p_3) V_{b b_1 b_2 b_3} (k, -p_1, -p_2, -p_3) \times \\
     \sum_{\substack{\sigma_1, \sigma_2, \sigma_3 \\ =\pm 1}} [\sigma_1\sigma_2\sigma_3 + \left( \sigma_1 F_{p_2}F_{p_3} + \text{cyc. perm.~} (1,2,3) \right)]  A_{a_1 b_1}^{\sigma_1}(p_1) A_{a_2 b_2}^{\sigma_2}(p_2) A_{a_3 b_3}^{\sigma_3}(p_3)
    (\omega - \sigma_1 E_{p_1} - \sigma_2 E_{p_2} - \sigma_3 E_{p_3} + i0^+ )^{-1}
    \, ,
    \label{eqn:Sigma-R-secondorder}
\end{multline}
where the weights $A_{ab}^{\sigma}(p)$ depend implicitly on the self-consistently-determined Bogoliubov parameter $\vartheta_p$.
The advanced component $\Sigma_{ab}^\text{A}(k, \omega)$ is obtained from the same expression via the replacement $\omega+i0^+ \to \omega-i0^+$, such that both $\Sigma^\text{R}$ and $\Sigma^\text{A}$ have the appropriate causality structure.
The Keldysh component follows similarly from the diagrams on the second line of Fig.~\ref{fig:second-order-selfenergy}
\begin{multline}
    \Sigma^\text{K}_{ab}(k, \omega) =
    -\frac{2\cdot 3!}{8}
    \sum_{p_1, p_2, p_3}  V_{a a_1 a_2 a_3} (-k, p_1, p_2, p_3) V_{b b_1 b_2 b_3} (k, -p_1, -p_2, -p_3) \times \\
     \sum_{\substack{\sigma_1, \sigma_2, \sigma_3 \\ =\pm 1}} [ F_{p_1} F_{p_2} F_{p_3} + \left( \sigma_1 \sigma_2 F_{p_3} + \text{cyc. perm.~} (1,2,3) \right)]  A_{a_1 b_1}^{\sigma_1}(p_1) A_{a_2 b_2}^{\sigma_2}(p_2) A_{a_3 b_3}^{\sigma_3}(p_3)
    2\pi \delta(\omega - \sigma_1 E_{p_1} - \sigma_2 E_{p_2} - \sigma_3 E_{p_3} )
    \, ,
    \label{eqn:Sigma-K-secondorder}
\end{multline}
\end{widetext}
where repeated matrix indices, $a_i$ and $b_i$, are implicitly summed over.
While the quantitative details of these expressions are not particularly illuminating,
we note that the absence of fermion
number conservation gives rise to contributions from fermion creation and annihilation processes, in addition 
to scattering processes. As a result, a particular term in the momentum summation is peaked at 
$\omega = \sigma_1 E_{p_1 } + \sigma_2 E_{p_2 } + \sigma_3 E_{p_3 } $ for all eight
choices of the sign variables $\sigma_i =\pm 1$.
Note also that momentum conservation requires that $k = p_1 + p_2 + p_3$
mod $2\pi$, which is enforced by the interaction vertices.
Finally, at zero temperature, $F_{p} = 1$, and therefore the factor $\prod_i \sigma_i + \sum_i \sigma_i$ that weights terms in~Eq.\eqref{eqn:Sigma-R-secondorder} vanishes
if not all $\sigma_i$ are equal to one another [the same is true for the analogous factor that appears in~\eqref{eqn:Sigma-K-secondorder}, which evaluates to $\left(\prod_i\sigma_i \right)(\prod_i \sigma_i + \sum_i \sigma_i)$].
At $T=0$, the combination of momentum and energy conservation (i.e., evaluating the self-energy on-shell)
then requires that
$E_k = E_{p_1} + E_{p_2} + E_{k-p_1-p_2}$,
 which corresponds to spontaneous creation of a domain wall pair with momenta $(p_1, p_2)$, while the original domain wall is scattered to $k-p_1-p_2$.
If the single-domain-wall dispersion does not enter the 3DW continuum, this equation has no solutions, and spontaneous decay cannot occur at $T=0$.
To obtain a finite lifetime, we therefore consider two cases.
First, if spontaneous decay is forbidden [Fig.~\subref*{fig:Ek-no-overlap}] then we consider
nonzero temperatures, which give rise to a nonvanishing density of excitations from which
the quasiparticles are able to scatter,
i.e., at $T > 0$, $F_p < 1$, and all combinations of the sign factors $\sigma_i$ can contribute, at least in principle.
Quasiparticle scattering, $E_{p_1} + E_{p_2} = E_{k} + E_{k-p_1-p_2}$, is, however, the dominant process.
Second, we consider a parameter regime in which the single-particle dispersion relation enters the continuum [Fig.~\subref{fig:Ek-overlap}], such that a nonvanishing background of excitations is no longer required, since the domain walls inside the continuum are able to lower their energy by spontaneously emitting additional domain walls. While the process is permitted kinematically by conservation of energy and momentum, there must also be nonvanishing matrix elements in Eq.~\eqref{eqn:Sigma-R-secondorder} for spontaneous decay to occur, with the rate of decay being proportional to both the interaction matrix elements and the multi-particle density of states.

The expressions~\eqref{eqn:Sigma-R-secondorder} and~\eqref{eqn:Sigma-K-secondorder} can be further
`simplified' by performing the summation over the Majorana indices $a_i$ and $b_i$ explicitly to obtain a closed-form expression for the momentum- and sign-dependent weights.
This procedure is described in further detail in Appendix~\ref{sec:self-energy-details} and is used to evaluate~\eqref{eqn:Sigma-R-secondorder} and~\eqref{eqn:Sigma-K-secondorder} efficiently in our numerical implementation.
Once conservation of momentum has been taken into account, the resulting two-dimensional integral over the unconstrained momenta can then be evaluated numerically to obtain the full $4 \times 4$ second-order self-energy matrix $\breve{\Sigma}(k, \omega)$.

\begin{figure}[t]
    \centering
    \includegraphics[width=0.5\linewidth]{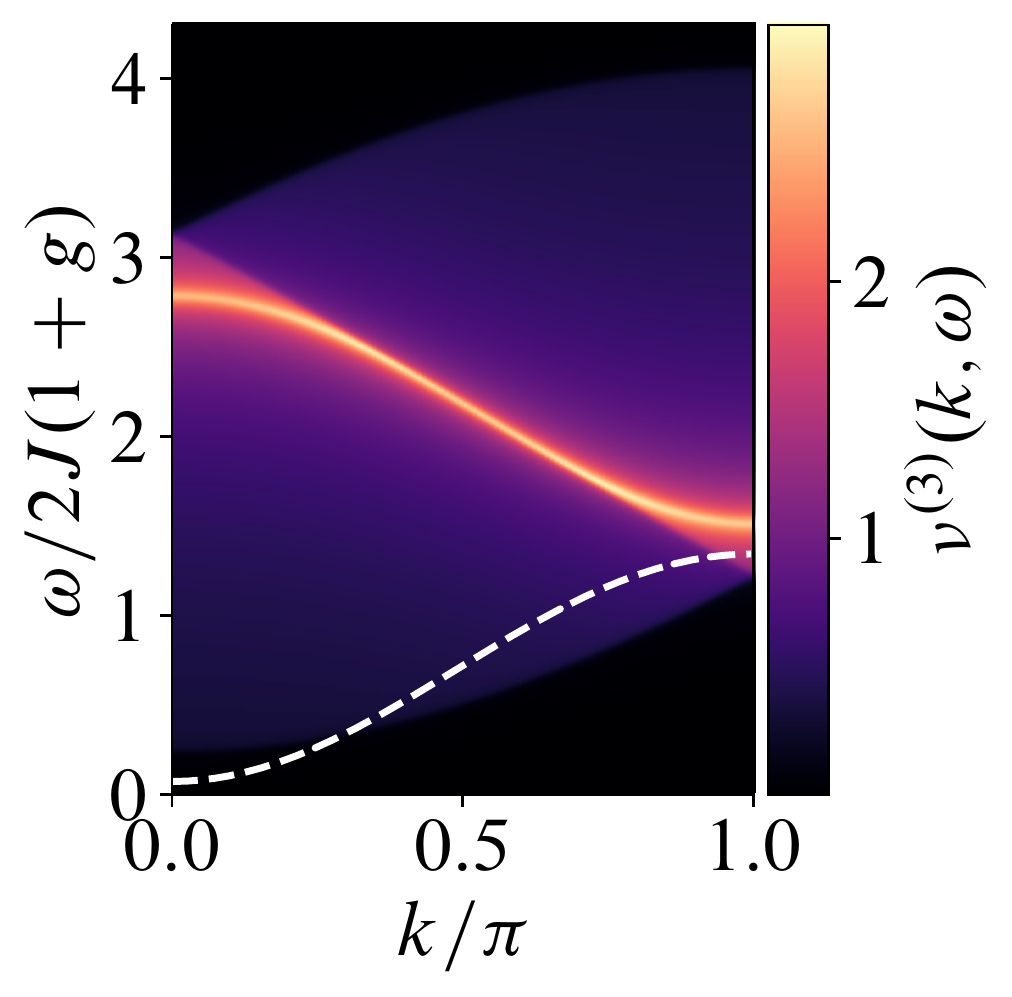}
    \caption{Momentum-resolved three-domain-wall density of states for the single-particle dispersion relation $E_k$, which is shown by the white dashed line. While the single-particle mode enters the continuum at $k \approx \pi/4$, the corresponding density of states is strongly suppressed. This behavior is inherited by the components of the second-order self-energy~\eqref{eqn:Sigma-R-secondorder} and \eqref{eqn:Sigma-K-secondorder}, leading to pronounced decay in the vicinity of the zone boundaries $k=\pi$.}
    \label{fig:3dw-density-of-states}
\end{figure}

The result of including the second-order self-energy on the Majorana Green's functions is shown in Fig.~\ref{fig:greens-functions}.
When the lifetime is provided by a thermally excited background of quasiparticles,
the frequency dependence of the second-order self-energy gives rise to nontrivial broadening of the single-particle poles [Fig.~\subref*{fig:DR-free} vs.~Fig.~\subref*{fig:DR-interacting}].
In addition to simple broadening throughout the spectrum, we observe ``anomalous'' broadening of the single-particle Green's function in the vicinity of $k=\pi$, where the quasiparticle peak splits in two.
The anomalous broadening is a consequence of the strong frequency dependence of $\check{\Sigma}^\text{R}(\pi, \omega)$, which varies sharply at the corresponding single-particle energy $\omega = E(k=\pi)$.
When the lifetime is instead provided by \emph{spontaneous} decay [Fig.~\subref*{fig:DR-free-overlap} vs.~Fig.~\subref*{fig:DR-interacting-overlap}], we once again see that the character of the broadening varies significantly with momentum, being most pronounced at the zone edges.
While the single-particle mode enters the 3DW continuum at around $k\approx \pi/4$, the 3DW density of states at these momenta is strongly suppressed, only becoming appreciable in the vicinity of the zone boundaries, as shown in Fig.~\ref{fig:3dw-density-of-states}.
The remainder of this section is devoted to characterizing which features of the single-particle Green's function(s) described above can be probed by third-order nonlinear spectroscopy.


\subsection{Nonlinear response of interacting system}

We now combine all ingredients derived thus far to evaluate the nonlinear response of the interacting TFIM chain.
Interaction effects are incorporated using the self-energy matrix derived in Sec.~\ref{sec:self-energy}, which includes both nontrivial real and imaginary parts that lead to decay when evaluated in the time domain.
The self-energy is included by augmenting the
diagrams in Fig.~\ref{fig:nonlinear-feynman}, i.e., all instances of the noninteracting Green's function, $\breve{D}_0$, are replaced by their dressed counterparts $\breve{D} = (\breve{D}_0^{-1} - \breve{\Sigma})^{-1}$.
As mentioned previously, we focus in particular on the contribution from $\ChiThree ( t, t,t+\tau)$ in Eq.~\eqref{eqn:Mx-2DCS}, since it contains the rephasing signal.
From the general expression presented in Eq.~\eqref{eqn:nonlinear-functional-deriv}, we find that
\begin{align}
    &\ChiThree ( t, t, t+\tau)
    = \frac{i}{2} \theta(t) \theta(\tau) \sum_{p}  \notag \\
    \big\{ & \Tr[\check{\tau}^2  \check{D}_{-p}^{\text{R}}(t) \check{\tau}^2 \check{D}_{-p}^{\text{R}}(0^+) \check{\tau}^2 \check{D}_{-p}^{\text{R}}(\tau) \check{\tau}^2  \check{D}_{-p}^{\text{K}}(-t-\tau)] \big\} \notag\\
    +&\Tr[\check{\tau}^2  \check{D}_{-p}^{\text{R}}(t) \check{\tau}^2 \check{D}_{-p}^{\text{R}}(0^+) \check{\tau}^2 \check{D}_{-p}^{\text{K}}(\tau) \check{\tau}^2  \check{D}_{-p}^\text{A}(-t-\tau)] \notag\\
    +&\Tr[\check{\tau}^2  \check{D}_{-p}^{\text{R}}(t) \check{\tau}^2 \check{D}_{-p}^{\text{R}}(\,\tau\,) \check{\tau}^2 \check{D}_{-p}^\text{K}(-\tau) \check{\tau}^2  \check{D}_{-p}^\text{A}(-t)] \notag\\
    +&\Tr[\check{\tau}^2  \check{D}_{-p}^{\text{R}}(t) \check{\tau}^2 \check{D}_{-p}^{\text{K}}(\,\tau\,) \check{\tau}^2 \check{D}_{-p}^\text{A}(-\tau) \check{\tau}^2  \check{D}_{-p}^\text{A}(-t)] \big\}
    \, ,
    \label{eqn:interacting-spinon-echo}
\end{align}
where $\check{D}^\text{R}_{-p}(0^+) = \mathds{I}$, the $2\times 2$ identity matrix, has been kept for clarity.
Note that the summation now runs over \emph{all} momenta, and that the permutations over the various time orderings have been accounted for.
It can be verified using the symmetry properties of $\breve{D}$ that~\eqref{eqn:interacting-spinon-echo} is pure real, as it must be.
The 2DCS spectrum is then found by transforming to variables $\omega_t$ and $\omega_\tau$, Fourier conjugate to $t$ and $\tau$, respectively.
That is, we are interested in
\begin{equation}
    \ChiThree(\omega_t, \omega_\tau) = \iint_{-\infty}^\infty \mathrm{d}t \mathrm{d}\tau \,
    e^{i(\omega_t t +\omega_\tau \tau)} \ChiThree ( t, t, t+\tau)
    \, .
    \label{eqn:2DCS-Fourier}
\end{equation}
In practice, we find that it is most convenient to evaluate~\eqref{eqn:2DCS-Fourier} by first performing a fast Fourier transform (FFT) of the frequency-space Green's functions to find the real-time Green's functions $\breve{D}(t, t')$, allowing us to evaluate~\eqref{eqn:interacting-spinon-echo} in the time domain. Equation~\eqref{eqn:2DCS-Fourier} can then be evaluated using an inverse FFT using the same frequency discretisation.
When evaluating the FFT, we compute the FFT of the \emph{difference} between the free and interacting Green's functions. This avoids introducing errors due to Gibbs' phenomenon associated with the discontinuities inherited by the step functions $\theta(t)$ enforcing causality in the retarded and advanced components of $\breve{D}(t, t')$.

The resulting 2DCS spectra for both flavors of decay, obtained from the imaginary part of Eq.~\eqref{eqn:2DCS-Fourier}, are shown in Figs.~\subref*{fig:interacting-2DCS-spectrum} and \subref{fig:interacting-2DCS-overlap}.
When the lifetime is provided by a thermally excited background of domain walls [Fig.~\subref*{fig:interacting-2DCS-spectrum}], we observe that the rephasing signal in the upper left and lower right quadrants is broadened throughout the spectrum.
The anomalous broadening seen in the single-particle Green's function [Fig.~\ref{fig:DR-interacting}], however, is seemingly absent.
There are a number of reasons for this absence, which we discuss in detail below.
Principally, the expression \eqref{eqn:interacting-spinon-echo} depends in a highly nonlinear way on the individual single-particle Green's functions.
Hence, if the temporal decay of the single-particle Green's functions is not well described by simple lifetime broadening, leading to exponential decay, the time arguments will combine in a highly nontrivial manner.
For the anomalous broadening observed in Fig.~\subref*{fig:DR-interacting}, the Green's functions can be approximated by a sum of two simple poles separated in frequency by $2\delta_k$. As we show in Appendix~\ref{sec:anomalous-broadening}, this simple modification of the single-particle Green's functions leads to \emph{two} additional peaks on either side of the rephasing streak (if the peaks are separated in frequency by $2\delta$ with broadening $\gamma$ in the Green's function, they are separated by $\sqrt{2}\delta$ with broadening $2\gamma$ in the rephasing streak). Since the separation of the peaks relative to their broadening is reduced in the rephasing streak, the anomalous broadening remains concealed in Fig.~\subref*{fig:interacting-2DCS-spectrum}.
Evidently, additional spectral features associated with nontrivial \emph{frequency} dependence of the self-energy can be more challenging to extract than simple Lorentzian broadening.

\begin{figure*}
    \centering
    \subfloat[\label{fig:interacting-2DCS-twisted}Raw 2DCS spectrum]{%
        \includegraphics[width=0.23\linewidth]{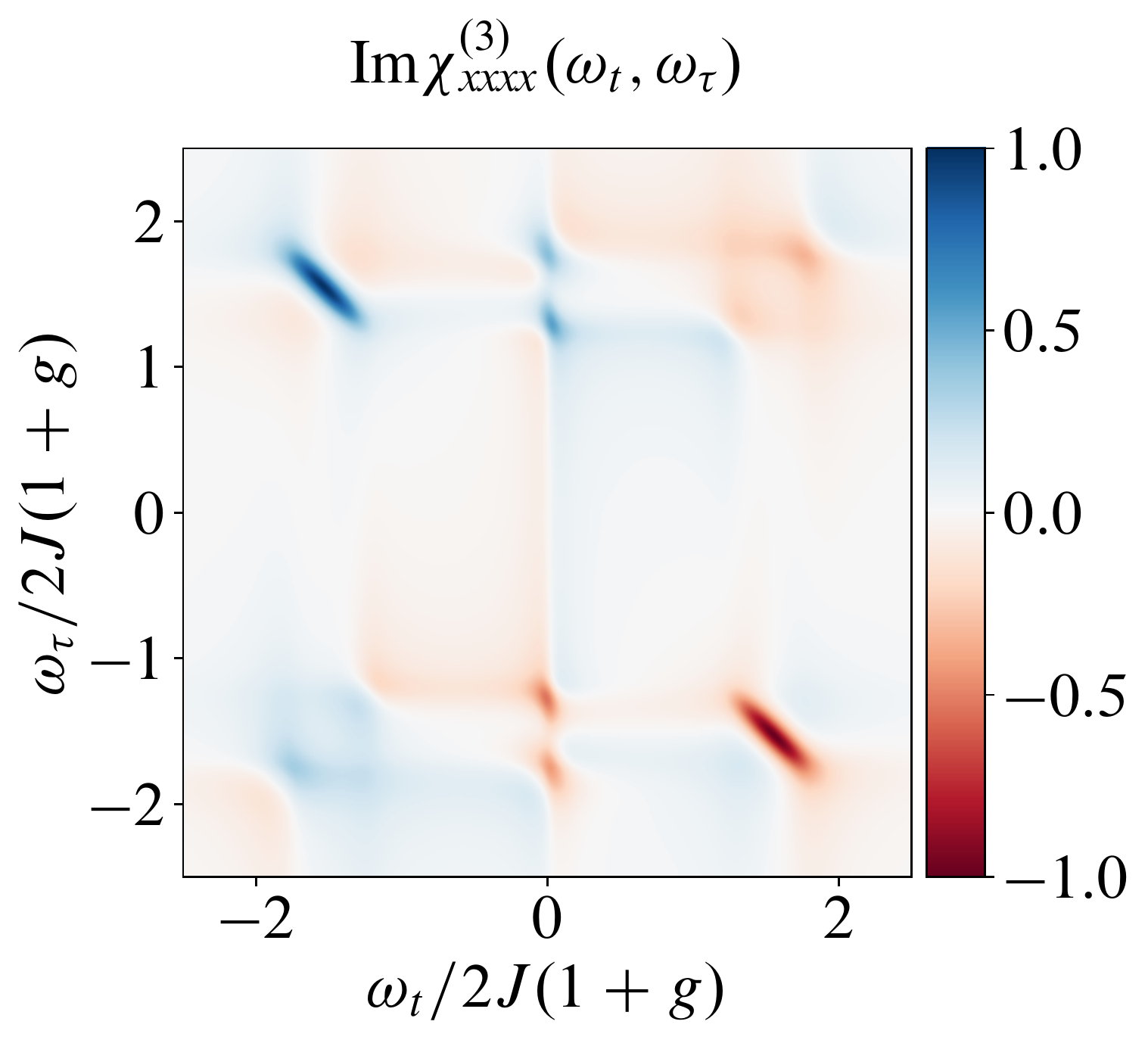}%
    }%
    \hspace{0.3cm}%
    \subfloat[\label{fig:interacting-2DCS-partial}Partially untwisted]{%
        \includegraphics[width=0.23\linewidth]{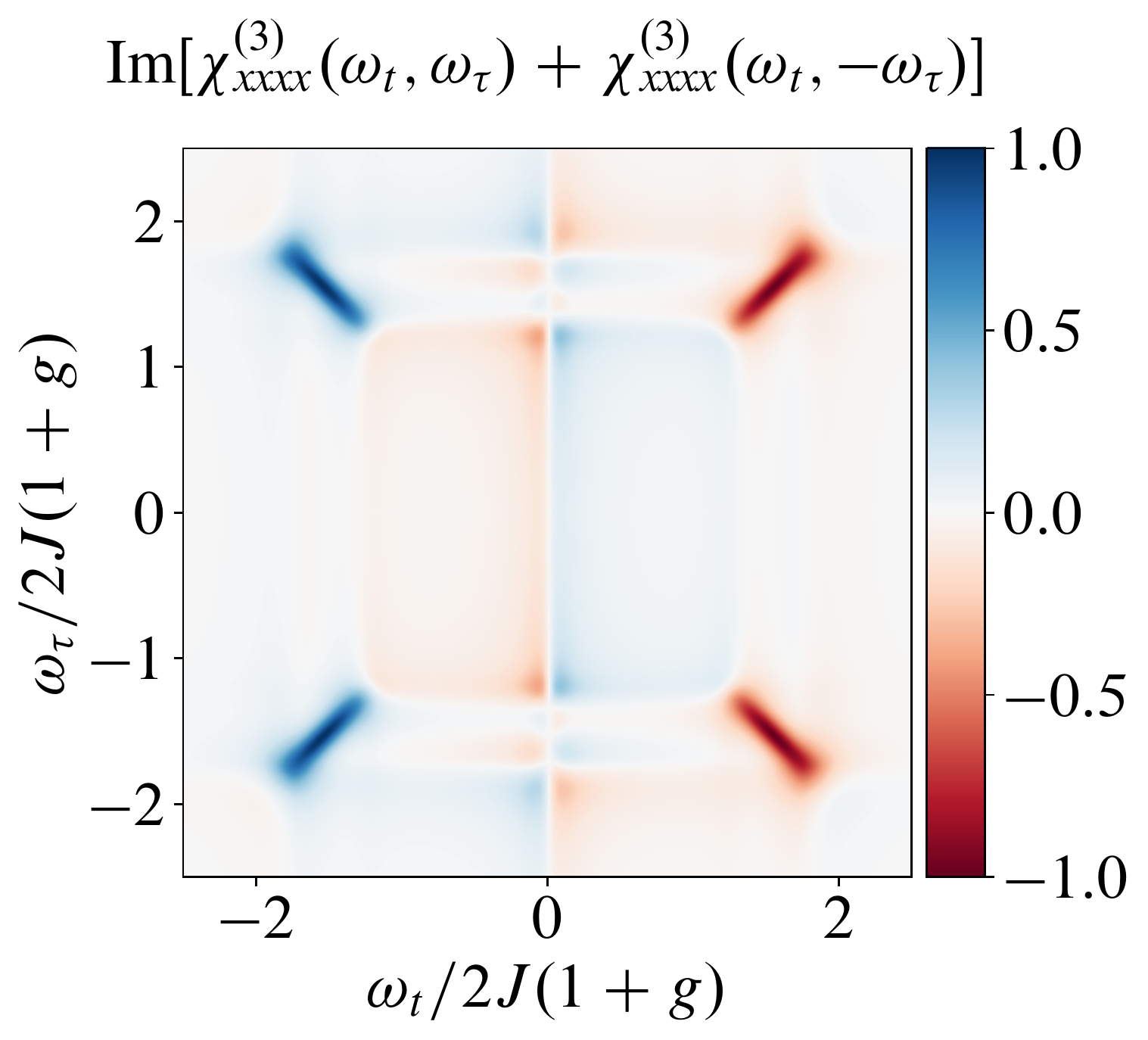}%
    }%
    \hspace{0.3cm}%
    \subfloat[\label{fig:interacting-2DCS-untwisted}Fully untwisted]{%
        \includegraphics[width=0.23\linewidth]{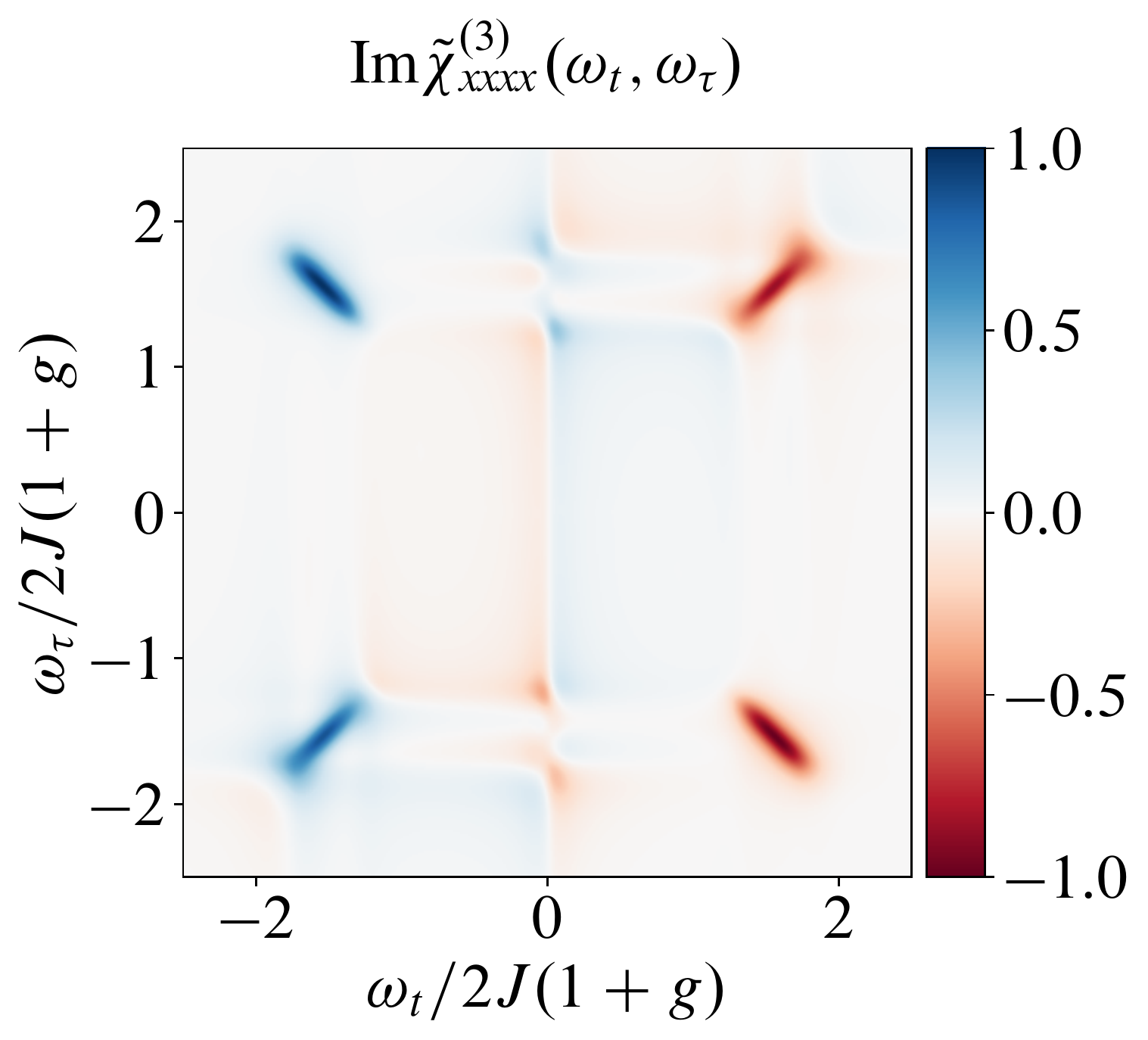}%
    }%
    \hspace{0.3cm}
    \subfloat[\label{fig:extracted-lifetime}Extracted lifetime]{%
        \includegraphics[width=0.188\linewidth]{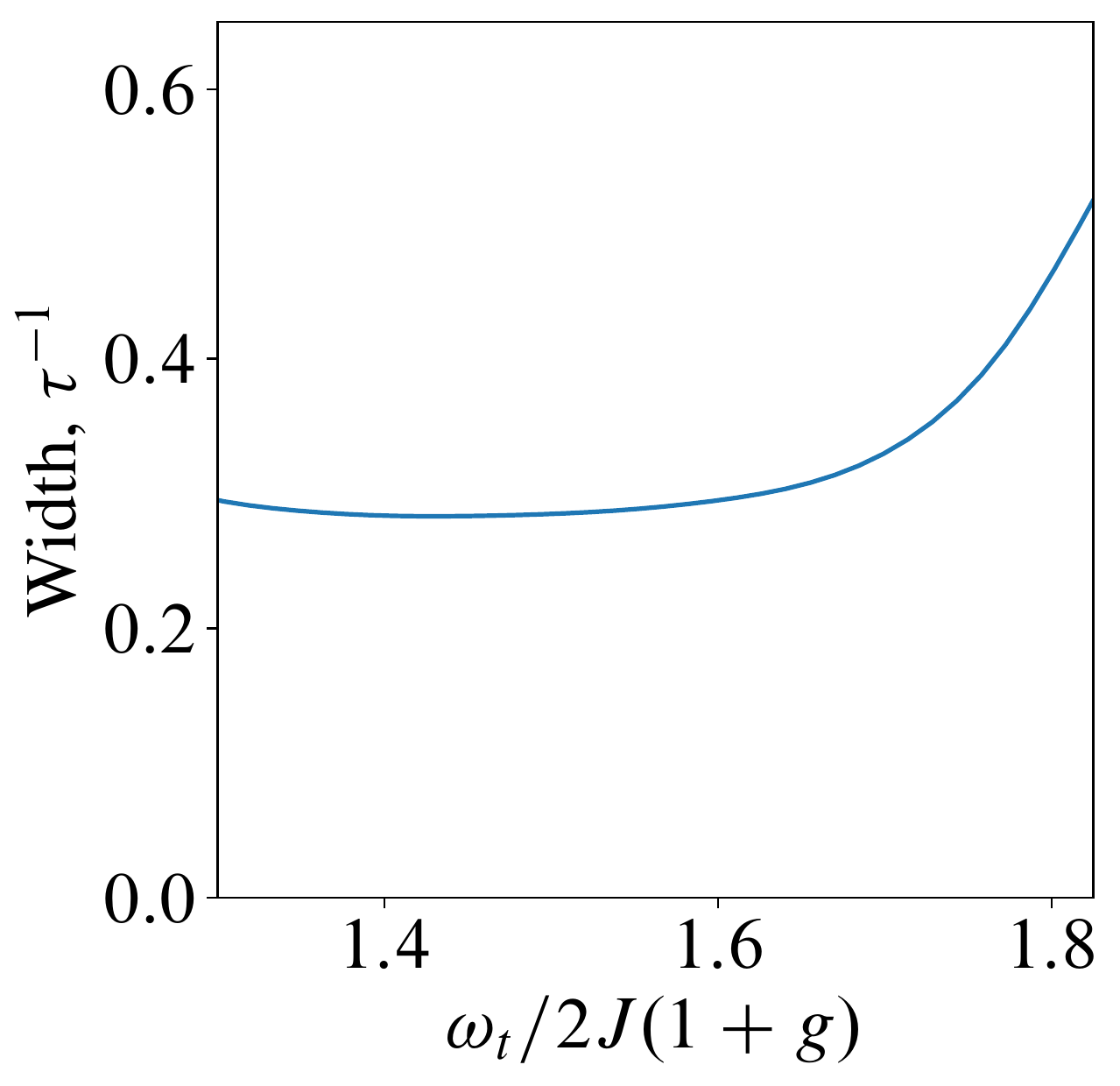}%
    }%
    \caption{Illustration of the ``untwisting'' procedure for the rephasing signal. (a) The raw 2DCS spectrum obtained via the Fourier transform of $\ChiThree(t, t, t+\tau)$ over $t$ and $\tau$, equivalent to the data presented in Fig.~\protect\subref*{fig:interacting-2DCS-spectrum}. (b) The partially untwisted spectrum, in which the time-dependent response $\ChiThree(t, t, t+\tau)$ is symmetrized over $\tau$ prior to taking the Fourier transform. (c) The fully untwisted data, in which each momentum is weighted by the appropriate combination of matrix elements [Eq.~\eqref{eqn:untwisting-weight}], such that the rephasing and non-rephasing signals combine to produce a purely absorptive lineshape. (d) The lifetime extracted from the width of the untwisted rephasing streak in (c), obtained by fitting to a Lorentzian lineshape in the $\omega_t = \omega_\tau$ direction.}
    \label{fig:untwisting}
\end{figure*}

When the lifetime is instead provided by spontaneous decay [Fig.~\subref*{fig:interacting-2DCS-overlap}], the strong \emph{momentum} dependence of the broadening of the single-particle Green's function can be observed. At the momenta where the single-particle Green's function is is most strongly broadened [indicated by the white arrows in Fig.~\subref*{fig:DR-interacting-overlap}], the rephasing signal is also correspondingly broadened at the frequencies $\omega_k = \pm 2E_k$. Nonlinear spectroscopy can therefore be used to extract momentum-dependent lifetimes with ease.
However, when extracting \emph{quantitative} information about the lifetime from the rephasing streak, there are a number of technical considerations that must be discussed [which also contribute to obscuring the anomalous broadening in Fig.~\subref*{fig:interacting-2DCS-spectrum}].
First, the rephasing signal is weighted by the optical matrix element $\sin^4\vartheta_k$~\cite{WanArmitage2019} [see also Eq.~\eqref{eqn:chi-3-sequential}]. This weighting factor vanishes at $k=\pi$, which provides any potential signatures of broadening at the Brillouin zone boundaries with a small relative weight.
Second, the response suffers from ``broadening by convolution''~\cite{Nandkishore2021Spectroscopic}, also known as ``phase twisting'' in the literature~\cite{Khalil2003,Kuehn20112DCS}. Namely, the signal of interest is convolved with $\delta(\omega_t)\delta(\omega_\tau) - \mathcal{P}\frac{1}{\omega_t\omega_\tau} + i[ \delta(\omega_t)\mathcal{P}\frac{1}{\omega_\tau} + \delta(\omega_\tau)\mathcal{P}\frac{1}{\omega_t} ]$ since the response is causal in the sense that $t, \tau > 0$. This leads to a mixing of dispersive and absorptive contributions, i.e., even a pure oscillatory signal $\sim e^{i\nu_t t + t\nu_\tau \tau}$ will exhibit power law smearing of its real part. This latter effect can, however, be ameliorated by ``phase untwisting'' the 2DCS spectrum, a process that we describe in the next section.


\subsection{Phase untwisting}
\label{sec:phase-untwisting}

The mixing of dispersive and absorptive contributions to $\Im \ChiThree(\omega_t, \omega_\tau)$ 
can be reduced by performing a second set of experiments in which one permits $\tau$ to be \emph{negative}.
In this case, the three experiments that comprise the 2DCS protocol are modified as follows. Pulse A remains unaltered, being applied to the system at time $s=0$, $h_\text{A}^{<}(s)= A_0 \delta(s)$. The pulse B is now applied at negative times $s=\tau < 0$, preceding pulse A, $h_\text{B}^{<}(s) = A_\tau \delta(s-\tau) = A_\tau \delta(s+\abs{\tau})$. Finally, the third pulse sequence corresponds to applying A and B in tandem $h_\text{AB}^{<} = A_0\delta(s) + A_\tau\delta(s-\tau)$. In each of the three experiments, the system is measured at a time $t$ after the second pulse A (equivalently, $t+\tau$ after the first pulse, B).
With these definitions, the 2DCS protocol leads to a response
\begin{align}
    M_{\text{2DCS},<}^x(t) = \,
       &A_0 A_\tau \ChiTwo (t, t-\tau) \notag\\
    +  &A_0 A_\tau^2 \ChiThree (t, t-\tau, t-\tau) \notag\\
    +  &A_0^2 A_\tau \ChiThree ( t, t,t-\tau)
    \, ,
    \label{eqn:Mx-2DCS-less}
\end{align}
which mirrors Eq.~\eqref{eqn:Mx-2DCS}, corresponding to the $\tau > 0$ pulse sequence, but with the sign of $\tau$ flipped, $\tau \to -\tau$, and the strengths of the two pulses interchanged, $A_0 \leftrightarrow A_\tau$.
The `untwisting' procedure then works by taking the combination
\begin{equation}
    \theta(t) \theta(\tau)\ChiThree (t, t, t+\tau)
    +
    \theta(t)\theta(-\tau) \ChiThree(t, t, t-\tau) 
    \, ,
    \label{eqn:untwisting-sum}
\end{equation}
which is causal only in the sense that $t>0$, while the argument $\tau$ spans the entire real axis.
Equivalently, since one can view Eq.~\eqref{eqn:untwisting-sum} as symmetrizing the response over $\tau$, the second experiment is not necessary; because the two pulses have the same polarization, the two experiments contain identical information.
The rephasing contribution to the $\tau \to -\tau$ susceptibility now occurs in the upper right and lower left quadrants of the $\omega_t$-$\omega_\tau$ plane and therefore does not interact with the signal of interest, which manifests in the upper left and lower right quadrants.
However, the \emph{third} line of Eq.~\eqref{eqn:chi-3-sequential}, which corresponds to part of the the \emph{nonrephasing} signal, maps under $\tau\to-\tau$ to a response in the appropriate quadrants.
Specifically, combining the third and fourth lines of Eq.~\eqref{eqn:chi-3-sequential}, and assuming a frequency-independent self-energy that gives rise to a momentum-dependent decay rate $\gamma_p$, Eq.~\eqref{eqn:untwisting-sum} contains the signal
\begin{multline}
    -4\theta(t) \sum_p F_p \sin^4\vartheta_p e^{-2\gamma_p t}  \sin[2\epsilon_p (t-\tau)] \times \\
    \left\{ e^{-2\gamma_p \tau}\theta(\tau)+ e^{2\gamma_p \tau}\theta(-\tau) \right\} 
    \, .
    \label{eqn:untwisted}
\end{multline}
The term inside the braces on the second line evaluates to $e^{-2\gamma_p\abs{\tau}}$, whose Fourier transform over $\tau$ has a purely real Lorentzian lineshape.
Hence, after Fourier transformation, the imaginary part of the summand in Eq.~\eqref{eqn:untwisted} maps to a product of two real Lorentzians in the $\omega_t$ and $\omega_\tau$ directions, centered on $(\omega_t, \omega_\tau) = \pm (2\epsilon_p, -2\epsilon_p)$, which no longer mixes real and imaginary parts.
While the third and fourth lines of Eq.~\eqref{eqn:chi-3-sequential} combine to produce a response of purely absorptive character, there is also, however, a contribution from the top line $\propto \sin^2(2\vartheta_p) \sin[2\epsilon_p(t+\tau)]$.
At $\vartheta_p = \pi/2$ (which occurs near the center of the rephasing streak), this contribution vanishes and the spectrum is perfectly untwisted at the corresponding frequencies; away from $\vartheta_p = \pi/2$, the cancellation is imperfect, but the broadening due to phase twisting is still reduced with respect to the Fourier transform of $\ChiThree(t, t, t+\tau)$ alone.
To fully untwist the noninteracting spectrum for all frequencies simultaneously, one must weight the contribution from each momentum $p$ separately to account for the differing matrix elements that accompany the rephasing and nonrephasing signals. As can be seen from Eq.~\eqref{eqn:chi-3-sequential}, the relevant weighting factor, $w_p$, for the non-rephasing signal should be generalized from unity to
\begin{equation}
    w_p = \frac{2\sin^4(\vartheta_p)}{2\sin^4(\vartheta_p) + \sin^2(2\vartheta_p)}
    \, ,
    \label{eqn:untwisting-weight}    
\end{equation}
for momentum $p$. The untwisting procedure is outlined graphically in Fig.~\ref{fig:untwisting}.
Only in the fully untwisted spectrum, Fig.~\subref*{fig:interacting-2DCS-untwisted}, can a Lorentzian lineshape be used to extract a momentum-dependent lifetime\footnote{Integrating the product of two Lorentzians oriented parallel to $\omega_t$ and $\omega_\tau$ along the $\omega_t = -\omega_\tau$ direction produces a simple Lorentzian lineshape in the orthogonal direction, parallel to $\omega_t = \omega_\tau$, as long as the variation of the envelope function, e.g., $\nu(\epsilon) \sin^4\vartheta$, can be neglected over a frequency range set by $\sim\gamma$.}. This procedure is used to extract the effective lifetime from the width of the rephasing streak in Fig.~\subref*{fig:extracted-lifetime}.
The extracted broadening is most pronounced at larger values of $\omega_t$, which correspond via the dispersion relation to momenta near the zone boundaries, consistent with the behavior of the Green's function in Fig.~\subref*{fig:DR-interacting}.


\section{Conclusions}
\label{sec:conclusions}

We have developed a framework based on a real-time path integral approach that is capable of describing the nonlinear response functions of interacting quantum many-body systems, as probed by 2DCS. This framework, in the weak-interaction regime, allows us to apply standard tools of diagrammatic many-body perturbation theory, and to interpret the response in terms of familiar objects such as Green's functions and self-energies. This methodology was applied to the transverse field Ising model in the presence of additional XY interactions between spins, which break the model's integrability and endow the quasiparticles with a nontrivial self-energy.

We confirmed using this methodology that 2DCS is indeed capable of extracting the momentum-dependent line broadening of the single-domain-wall Green's function arising from interactions. In addition, we showed that treating the XY interactions microscopically gives rise to additional spectral features not captured by the phenomenological model of decay studied in Ref.~\cite{WanArmitage2019}.
First, when the single-domain-wall dispersion does not enter the three-domain-wall continuum, but the system contains a nonzero density of thermally excited domain walls,
the self-energy exhibits strong frequency dependence.
As a result, the broadening of the single-particle Green's function becomes ``anomalous'' near the zone boundaries in the sense that the quasiparticle peak splits in two.
We showed that the rephasing signal probed by 2DCS is, in principle, sensitive to the precise lineshape implied by a frequency-dependent self-energy, and could therefore be used to diagnose this anomalous broadening in principle.
Second, when the single-domain-wall dispersion does enter the continuum, but thermally excited domain walls are sparse, the self-energy exhibits strong momentum dependence, since the single-particle mode is broadened only in the regions where spontaneous decay is kinematically permitted. The resulting momentum dependence of the domain-wall lifetime can be observed directly with 2DCS as a frequency-dependent broadening of the rephasing signal.

In practice, however, there are a number of additional complications that can make the interpretation of the 2DCS spectrum challenging. First, since the mapping between the single-particle Green's function and the response probed by 2DCS is highly nontrivial [Eq.~\eqref{eqn:interacting-spinon-echo}], additional spectral features beyond simple lifetime broadening, which derive from a frequency-dependence self-energy, may have rather complex signatures.
Second, the two-dimensional step functions that are necessitated by causality give rise to ``broadening via convolution''~\cite{Nandkishore2021Spectroscopic} (also known as ``phase twisting''~\cite{Khalil2003,Kuehn20112DCS}), which has the potential to smear out such intricate spectral features, and complicates extracting \emph{quantitative} information from the rephasing signal.
We showed, however, that this second issue can be remedied by ``untwisting'' the 2DCS spectrum.
Finally, the optical matrix element, which acts as an envelope function for the rephasing signal, can provide certain spectral features with a small relative weight\footnote{ Experiments on \CoNbO~\cite{Morris2021duality}, however, suggest that the effect of these matrix elements, at least at the level of the linear response absorption spectra, is not as important as the simple calculations presented in this manuscript suggest. Rather than being diminished, the observed absorption is instead \emph{enhanced} at the edges of the spectrum~\cite{ArmitagePrivate}.}.
These findings highlight the fact that, while the nonlinear response of a given system contains a wealth of additional information beyond linear response, the spectrum that is produced can correspondingly be challenging to decipher in more realistic systems. Nevertheless, \emph{in principle}, 2DCS is able not only to reveal the existence of deconfined fractional quasiparticles~\cite{WanArmitage2019, Nandkishore2021Spectroscopic}, but even to reveal details about their self-energy in the presence of interactions, which we illustrated by extracting the momentum-dependent decay rate of domain walls from the 2DCS response. 

A number of challenges present themselves for future work. For one thing, we have confined our present investigation to the simplest possible 2DCS experiment, where only two drive pulses are used, and both have the same polarization. Variations on the 2DCS experiment using three drive pulses and/or crossed polarizations could yield additional information beyond that accessible in the basic experiment. Such generalized 2DCS protocols should be treatable using the same path integral formulation introduced herein. 

Additionally, we have considered the simplest possible example of a phase with deconfined fractionalized excitations -- the one-dimensional transverse field Ising model in its ferromagnetic phase. Generalization to more complicated fractionalized phases, such as spin liquids and models that host fractons, is an important topic for future work. We note that, for systems in greater than one spatial dimension, there will not be a unique mapping between energy and momentum, which may limit the information about the self-energy accessible in the basic 2DCS experiment. Even in one space dimension, however, there are fractionalized phases such as the Luttinger liquid and its spin-incoherent generalization~\cite{FieteSpinIncoherent}. While the application of 2DCS to the former has been discussed~\cite{Li2021Lensing}, the generalization to the latter remains to be performed. Moreover, thus far we have focused on broadening coming from {\it interactions}. Our path integral approach could also be generalized to incorporate disorder, and whether 2DCS can be deployed to reveal disorder-induced self-energies~\cite{Weyl} is also an interesting topic for future work. 

Most broadly, the 2DCS experiment is but a particular example of a general theme -- that of probing the ultrafast quantum {\it dynamics} of solid-state systems to reveal new information inaccessible to conventional experiments. This broader research program is in its infancy, and we anticipate new experimental protocols may arise as it matures. 

\emph{Note added.}
During the preparation of this manuscript, we became aware of an independent study of the effect of interactions on the nonlinear response properties of the transverse field Ising model, although in the paramagnetic phase~\cite{FavaNonperturbative}.

\section*{Acknowledgments}
This work was supported by the U.S.~Department of Energy, Office of Science, Basic Energy Sciences, under Award \#{}DE-SC0021346.
We are indebted to N.~P.~Armitage for useful discussions and particularly helpful feedback on the manuscript. We also acknowledge fruitful discussions with Yuan Wan.


\appendix
\renewcommand{\thesubsection}{\thesection.\arabic{subsection}}
\renewcommand{\thesubsubsection}{\thesubsection.\arabic{subsubsection}}
\renewcommand{\theequation}{\thesection.\arabic{equation}}
%


\section{Majorana Grassmann integration}
\label{sec:integration-rules}

In this Appendix we review for completeness the integration rules associated with $2N$ `Majorana' Grassmann variables $\phi_i$, and outline how the linear and nonlinear response properties of the noninteracting model can be evaluated using derivatives of Pfaffians.
The Grassmann variables anticommute, $\{ \phi_i, \phi_j\} = 0$, square to zero, $\phi^2_i = 0$, and integration is performed using
\begin{equation}
    \int \mathrm{d}\phi_i \, \phi_j = \delta_{ij}
    \, .
\end{equation}
The most important result is the following `Gaussian' integral, which relates the Grassmann integral of an antisymmetric matrix $A_{ij}$ to the Pfaffian:
\begin{equation}
    Z[A] \equiv 
    \int \mathcal{D}[\phi] \exp\left( -\frac12 \sum_{i,j=1}^{2N} \phi_i A_{ij} \phi_j \right)
    =
    \Pf(A)
    \, ,
\end{equation}
where $\mathcal{D}[\phi] \equiv \mathrm{d}\phi_1 \mathrm{d}\phi_2 \cdots \mathrm{d}\phi_{2N}$. An alternative way to derive the noninteracting nonlinear response properties~\eqref{eqn:TFIM-linear-response-xx} and~\eqref{eqn:chi-3-sequential} without explicitly invoking Wick's theorem is therefore to express the generating functional $Z[h^{\text{cl}}, h^{\text{q}}]$ as a Pfaffian, and to subsequently evaluate derivatives thereof.
An expression for the derivative of a Pfaffian of a matrix that depends on the set of variables $\{x_j\}$ can be derived by noting that $\Pf(A)^2 = \det(A)$ and utilizing Jacobi's formula, which gives
\begin{equation}
    \frac{\partial}{\partial x_i} \Pf(A) = \frac12 \Pf(A) \Tr\left( A^{-1} \frac{\partial A}{\partial x_i}  \right)
    \, .
    \label{eqn:Pfaffian-deriv}
\end{equation}
This expression can then be iterated to evaluate higher-order derivatives.
As an example, let's evaluate the second-order derivative, which, in the context of the noninteracting transverse field Ising model, gives rise to the linear susceptibility $\ChiOne(t)$. The exact expression for the generating functional 
is $Z[h^{\text{cl}}, h^{\text{q}}] = \Pf(D^{-1} -2ih^\text{q}M^\text{q} - 2ih^\text{cl}M^{\text{cl}})$, where all matrices possess space-time, Majorana, and Keldysh indices.
The free Green's function is defined in Eqs.~\eqref{eqn:Majorana-DR} and \eqref{eqn:Majorana-DK}, while the magnetization vertices are defined by Eq.~\eqref{eqn:magnetization-vertices}.
Making use of Eq.~\eqref{eqn:Pfaffian-deriv} twice, we find that the second-order derivative evaluates to
\begin{equation}
    \frac{\partial^2}{\partial x_i\partial x_j} \Pf(A) = 
    -\frac12 \Pf(A)  \Tr\left( A^{-1} \frac{\partial A}{\partial x_i}  A^{-1} \frac{\partial A}{\partial x_i}  \right) + \ldots
\end{equation}
The additional terms, denoted by the ellipsis, vanish when evaluated with $A=D^{-1}$ (either they correspond to disconnected diagrams, or they vanish because $A$ depends \emph{linearly} on $\{ x_j \}$ in the present case).
Since the magnetization vertices are time local, we arrive at 
\begin{multline}
    \left. \frac{\delta^2 Z[h^\text{cl}, h^\text{q}]}{\delta h^\text{cl}(t') \delta h^\text{q}(t)}\right\rvert_{h^\text{q}=h^\text{cl}=0}
    = \\
    \Tr\left[
         D(t, t') \circ M^\text{cl} \circ D(t', t) \circ M^\text{q}
    \right]
    \, ,
\end{multline}
where matrix multiplication over momentum, Majorana, and Keldysh indices is denoted by ``$\circ$''; the integration over intermediate times has already been performed.
Once the diagonal nature of the matrices $D$ and $M^{\text{q/cl}}$ with respect to momentum is taken into account ($\propto \delta_{k+k'}$), the expression is equivalent to Eq.~\eqref{eqn:linear-response-unsimplified} presented in the main text.
The same procedure can easily be generalized to higher-order derivatives to obtain the higher-order susceptibilities relevant to nonlinear response.


\begin{figure}[t]
    \centering
    \includegraphics[width=0.75\linewidth]{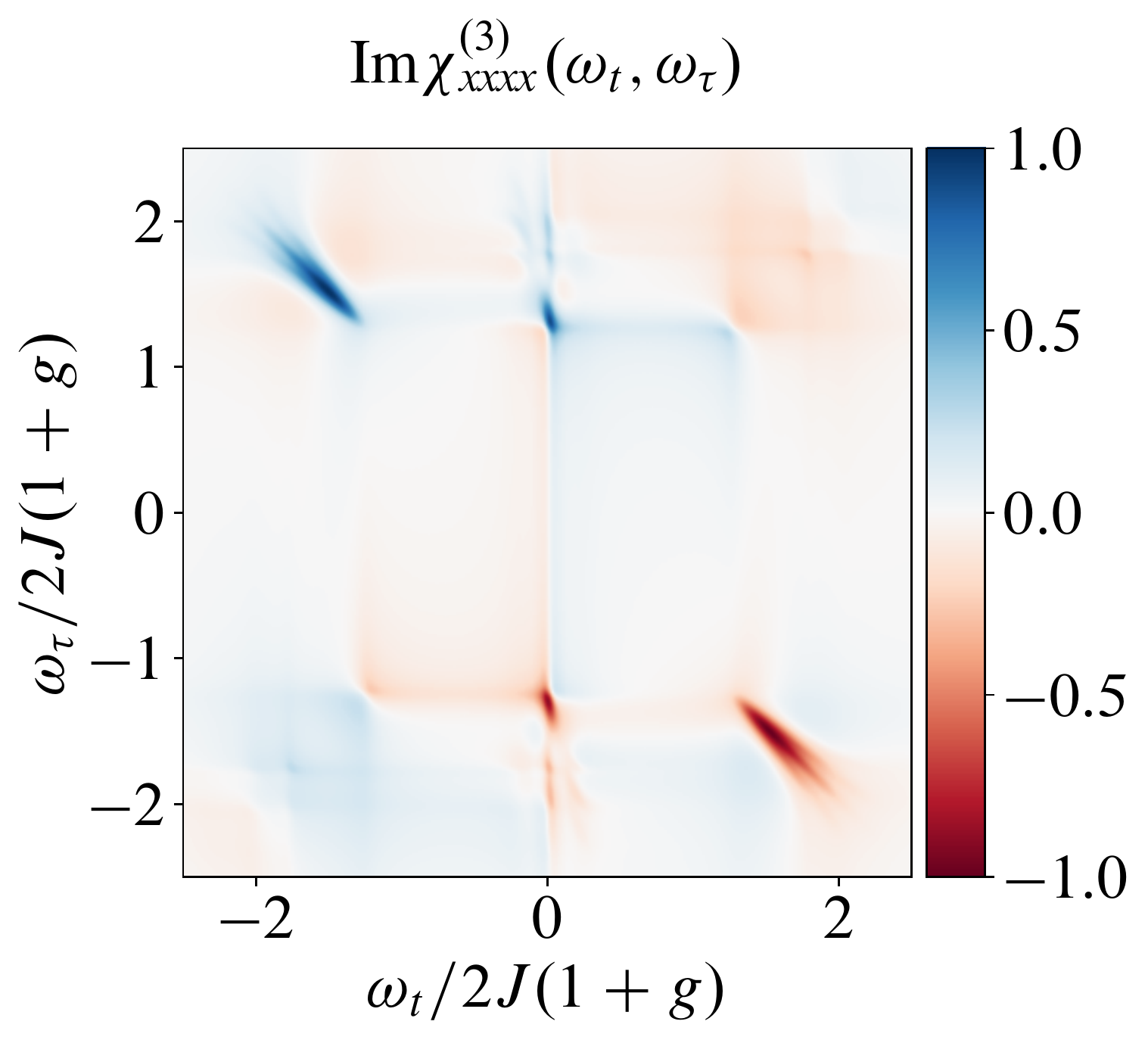}
    \caption{Imaginary part of the Fourier transformed third-order susceptibility, $\ChiThree(t, t, t+\tau)$, from the analytical expression~\eqref{eqn:two-branch} with an exaggerated separation $\delta_k$ between the two branches. The splitting of the anti-diagonal streak into multiple branches can be seen in the vicinity of $(\omega_t, \omega_\tau) = \pm(2, -2)$ in units of $2J(1+g)$. The response is evaluated for a system of size $L=100$.}
    \label{fig:analytical-anomalous}
\end{figure}

\section{Second-order response}
\label{sec:second-order}

In the main text, we focused on the third-order nonlinear response of the magnetization, $\ChiThree$, and contrasted its behavior with that of the linear response, $\ChiOne(t)$.
Here, we derive the second-order response using the formalism developed in the main text.
Our starting point is Eq.~\eqref{eqn:exact-magnetisation}, which can be expanded to second order in the classical component of the generating field, $h^\text{cl}(t)$, to give
\begin{equation}
    \ChiTwo (t_1; t_2, t_3) = \frac{1}{2i} \left.  \frac{\delta^3 Z[h^\text{cl}, h^\text{q}]}{\delta h^\text{cl}_3\delta h^\text{cl}_2\delta h^\text{q}_1} \right\rvert_{h^\text{q} = h^\text{cl} =0}
    \, .
    \label{eqn:chi-2-from-functional}
\end{equation}
Making use of Wick's theorem, or the alternative procedure outlined in Appendix~\ref{sec:integration-rules}, the functional derivative can be evaluated to give
\begin{subequations}
\begin{align}
    &\left. \frac{\delta^3 Z[h^\text{cl}, h^\text{q}]}{\delta h^\text{cl}(t_3)\delta h^\text{cl}(t_2) \delta h^\text{q}(t_1)}  \right\rvert_{h^\text{q} = h^\text{cl} =0}
    =
    i \sum_{p>0} \notag \\
    \Big\{
         &\Tr[\check{\tau}^2 \check{D}_{p}^{\text{K}}(t_{13}) \check{\tau}^2 \check{D}_{p}^{\text{A}}(t_{32}) \check{\tau}^2 \check{D}_{p}^{\text{A}}(t_{21}) ] \label{eqn:chi2-KAA} \\
        +&\Tr[\check{\tau}^2 \check{D}_{p}^{\text{R}}(t_{13}) \check{\tau}^2 \check{D}_{p}^{\text{K}}(t_{32}) \check{\tau}^2 \check{D}_{p}^{\text{A}}(t_{21}) ] \label{eqn:chi2-RKA} \\
        +&\Tr[\check{\tau}^2 \check{D}_{p}^{\text{R}}(t_{13}) \check{\tau}^2 \check{D}_{p}^{\text{R}}(t_{32}) \check{\tau}^2 \check{D}_{p}^{\text{K}}(t_{21}) ] 
    \Big\} \label{eqn:chi2-RRK} \\
    +& \text{permutations of } (t_2, t_3)
    \, . \notag
\end{align}
\label{eqn:nonlinear-functional-deriv-chi2}%
\end{subequations}
Since the times $t_2$ and $t_3$ precede the measurement time $t_1$, the response defined by Eq.~\eqref{eqn:nonlinear-functional-deriv-chi2} is appropriately causal.
Specifically, the $\theta$ functions in Eq.~\eqref{eqn:chi2-KAA} enforce $t_1 > t_2 > t_3$, while Eq.~\eqref{eqn:chi2-RRK} enforces the opposite ordering, $t_1 > t_3 > t_2$. Equation~\eqref{eqn:chi2-RKA}, on the other hand, does not restrict the ordering of $t_2$ and $t_3$; the $\theta$ functions only enforce that $t_1 > t_2$ and $t_1 > t_3$ separately.
The second-order susceptibility in Eq.~\eqref{eqn:nonlinear-functional-deriv-chi2} determines the second-order response of $\hat{M}^x$ via
\begin{equation}
    \langle \hat{M}^x(t_1) \rangle^{(2)}
    =
    \frac{1}{2!}
    \int
    \mathrm{d}t_2 \mathrm{d}t_3 \,
    \ChiTwo (t_1; t_2, t_3)
    \prod_{n=2}^3 h(t_n)
    \, ,
\end{equation}
where both integrations over time span the entire real axis.
As in Eq.~\eqref{eqn:M3-restricted-domain}, we are able to take advantage of the symmetry properties of $\ChiTwo$ to restrict the domain of integration to the region $t_1 > t_2 > t_3$:
\begin{equation}
    \langle \hat{M}^x(t_1) \rangle^{(2)}
    = \!
    \int\limits_{t_n > t_{n+1}} \!\!\!\!
    \mathrm{d}t_2 \mathrm{d}t_3 \,
    \ChiTwo (t_1; t_2, t_3)
    \prod_{n=2}^3 h(t_n)
    \, .
\end{equation}
If we evaluate Eq.~\eqref{eqn:nonlinear-functional-deriv-chi2} in the region $t_1 > t_2 > t_3$, we get contributions from Eqs.~\eqref{eqn:chi2-KAA} and \eqref{eqn:chi2-RKA}.
For the other permutation of $(t_2, t_3)$, Eqs.~\eqref{eqn:chi2-RKA} and \eqref{eqn:chi2-RRK} give rise to a nonzero contribution.
Evaluating the traces over Majorana indices, we find
\begin{multline}
    \ChiTwo(t, t+\tau) =
    -4\theta(t)\theta(\tau) \sum_{p>0}
    F_p \sin^2\vartheta_p \cos\vartheta_p \times \\
    \left[ \cos(2\epsilon_p \tau) - \cos(2\epsilon_p(t+\tau)) \right]
    \, .
    \label{eqn:chi2-final}
\end{multline}
The difference in sign with respect to Ref.~\cite{WanArmitage2019} derives from a different convention for the Jordan-Wigner transformation, and, hence, the Bogoliubov parameter $\vartheta_p$; in Eq.~\eqref{eqn:chi2-final} $\cos\vartheta_p = 2(g-\cos p)/\epsilon_p$.


\section{Pump-probe response}
\label{sec:pump-probe}

While we have focused primarily on the rephasing signal that occurs in the upper left and lower right quadrants in the Fourier-transformed third-order susceptibility $\ChiThree(t, t, t+\tau)$, the other contribution to Eq.~\eqref{eqn:Mx-2DCS}, proportional to $\ChiThree(t, t+\tau, t+\tau)$, also contains useful information, which we briefly explore in this Appendix.
When evaluated with the appropriate time arguments, the general expression~\eqref{eqn:nonlinear-functional-deriv} reduces to
\begin{align}
    &\ChiThree ( t, t+\tau, t+\tau)
    = \frac{i}{2} \theta(t) \theta(\tau) \sum_{p}  \notag \\
    \big\{ & \Tr[\tau^2  D_{-p}^{\text{R}}(t) \tau^2 D_{-p}^{\text{R}}(\tau) \tau^2 D_{-p}^{\text{R}}(0^+) \tau^2  D_{-p}^{\text{K}}(-t-\tau)] \big\} \notag\\
    +&\Tr[\tau^2  D_{-p}^{\text{R}}(t) \tau^2 D_{-p}^{\text{R}}(\tau) \tau^2 D_{-p}^{\text{K}}(0^+) \tau^2  D_{-p}^\text{A}(-t-\tau)] \notag\\
    +&\Tr[\tau^2  D_{-p}^{\text{R}}(t) \tau^2 D_{-p}^{\text{R}}(\tau) \tau^2 D_{-p}^\text{K}(0^-) \tau^2  D_{-p}^\text{A}(-t-\tau)] \notag\\
    +&\Tr[\tau^2  D_{-p}^{\text{R}}(t) \tau^2 D_{-p}^{\text{K}}(\tau) \tau^2 D_{-p}^\text{A}(0^-) \tau^2  D_{-p}^\text{A}(-t-\tau)] \big\}
    \, ,
    \label{eqn:interacting-pump-probe}
\end{align}
where we note that the summation runs over all momenta $p$, and that all permutations of the time arguments have been appropriately accounted for.
The Green's functions evaluated with infinitesimal time arguments can be further simplified by noting that, e.g., $D^\text{R}_p(0^+)=\mathds{I}$.
For the noninteracting model, we obtain the simple expression
\begin{equation}
\begin{aligned}
    \ChiThree (t, t+\tau, t+\tau) = -2 & \theta(t) \theta(\tau) \sum_{p>0} \\
    F_p \big\{&\sin^2 (2\vartheta_p) \sin[2\epsilon_p (t+\tau)] \\
              -&\sin^2 (2\vartheta_p) \sin[2\epsilon_p \tau] \\
              +4&\sin^4 (\,\vartheta_p\,)\, \sin[2\epsilon_p t]\big\}
    \, .
\end{aligned}
\end{equation}
The term on the last line corresponds to the pump-probe response, and produces a streak along the $\omega_\tau = 0$ axis at energies corresponding to twice the quasiparticle spectrum. If the Green's functions exhibit (perhaps momentum-dependent) simple lifetime broadening, leading to exponential decay in time, the pump-probe response will become broadened in the direction orthogonal to the streak, parallel to $\omega_\tau$.

\begin{widetext}


\section{Signatures of anomalous broadening}
\label{sec:anomalous-broadening}

In the main text it was shown that in spite of the anomalous broadening present in the single-particle Green's functions, the 2DCS spectrum depicted in Fig.~\ref{fig:noninteracting-vs-interacting} appeared to show no signatures thereof. Here, we show that the nonlinear response can, at least in principle, indicate the presence of anomalous broadening, but the conditions required for it to be visible differ from the analogous conditions for the single-particle Green's function (depicted in Fig.~\ref{fig:greens-functions}).
The single-particle Green's functions of the interacting system can be approximated by two separate excitation branches.
Suppose that these two branches have mean energy $E_k$ and separation $2\delta_k$. In order to simplify the analysis and to highlight the salient physics, we assume that both branches have identical spectral weight, intrinsic broadening $\gamma_k$, and off-diagonal phase $e^{i\vartheta_k}$. In this case, we obtain a relatively simple analytical approximation for the observed response:
\begin{align}
    &\ChiThree ( t, t, t+\tau)
    = 2  \theta(t) \theta(\tau)  \sum_{p} 
      e^{-2\gamma_p(t+\tau)} \sin^2\vartheta_p \cos(\delta_p t) \cos(\delta_p \tau) \Big[\notag\\
        &\cos(\delta_p t) \cos(\delta_p \tau) \Big(
            \cos^2\vartheta_p \big( 2\sin(2E_p\tau) - \sin[2E_p(t+\tau)] \big) -\sin^2\vartheta_p \sin[2E_p (t-\tau)] 
        \Big)
        - \cos[\delta_p (t+\tau)]  
            \sin[2E_p(t+\tau)]
    \Big]
    \label{eqn:two-branch}
\end{align}
The rephasing signal corresponds to the last term on the second line $\propto \sin[2E_p(t-\tau)]$. The new prefactor $\cos^2(\delta_p t) \cos^2(\delta_p \tau)$, which only differs from unity in the presence of a nonzero splitting $\delta_k > 0$, leads to additional peaks in the 2DCS spectrum.
Explicitly, isolating the rephasing contribution, we write
\begin{multline}
    -2 e^{-2\gamma_p (t+\tau)} 
    \sin^4\vartheta_p 
    \cos^2(\delta_p t) \cos^2(\delta_p \tau) 
    \sin[2E_p (t - \tau)]
    \to \\
    \frac{i}{16} \sin^4\vartheta_p (2+e^{2i\delta_p t} + e^{-2i\delta_p t})(2+e^{2i\delta_p \tau} + e^{-2i\delta_p \tau}) (e^{2iE_p(t-\tau)} - e^{-2iE_p(t-\tau)})
    \, ,
\end{multline}
and expanding out the brackets gives rise to additional peaks located at $(0, \pm 2\delta_p)$, $(\pm 2\delta_p, 0)$, $(\pm 2\delta_p, \mp 2\delta_p)$ with respect to the antidiagonal rephasing streak in the variables $(\omega_t, \omega_\tau)$, Fourier conjugate to $(t, \tau)$.

As shown in Fig.~\ref{fig:analytical-anomalous}, this \emph{can} give rise to signatures of anomalous broadening if the intrinsic broadening $\gamma_k$ is reduced substantially (alternatively, if $\delta_k$ is increased).
While we have the freedom to tune parameters at will in \eqref{eqn:two-branch}, in realistic systems the two contributions $\gamma_k$ and $\delta_k$ cannot be varied independently. In principle of course one could try and `reverse engineer' the interaction term so as to maximize $\delta$ while minimizing $\gamma$, although this may involve introducing long range interactions \cite{SafaviNaini}.


\section{Self-energy: Extra details}
\label{sec:self-energy-details}

In this Appendix we describe how the second-order contribution to the self-energy, whose components are given in Eqs.~\eqref{eqn:Sigma-R-secondorder} and \eqref{eqn:Sigma-K-secondorder}, is evaluated.
Multiplying the interaction vertices with the weights $A_{a_ib_i}^{\sigma_i}(p_i)$, we find that
\begin{multline}
    V_{a a_1 a_2 a_3} (-k, p_1, p_2, p_3) V_{b b_1 b_2 b_3} (k, -p_1, -p_2, -p_3)  A_{a_1 b_1}^{\sigma_1}(p_1) A_{a_2 b_2}^{\sigma_2}(p_2) A_{a_3 b_3}^{\sigma_3}(p_3)= \left(\frac{4\lambda}{3L}\right)^2 \delta_\text{P}\left( k - \sum_{i=1}^3 p_i \right) \\
    B_{ab}(1 \, | \, 2, 3)
    \sin^2\left(\frac{k+p_1}{2}\right) 
    \sin^2\left(\frac{p_2-p_3}{2}\right)
    +
    \sin\left( \frac{k+p_2}{2} \right) \sin\left( \frac{p_3-p_1}{2} \right) \sin\left( \frac{k+p_1}{2} \right) \sin\left( \frac{p_2-p_3}{2} \right)
    C_{ab}(1, 2 \, | \, 3 ) + \\ \text{cyc.~perm.} (1, 2, 3) 
\end{multline}
where we have defined the functions
\begin{align}
    B_{ab}(1 \, | \, 2, 3) &= A_{ab}^{\sigma_1} (p_1) \Tr[\check{A}^{\sigma_2}(p_2)  (\check{A}^{\sigma_3}(p_3))^\mathsf{T}] \, , \label{eqn:B_matrix} \\
    C_{ab}(1,2 \, | \, 3) &= [\check{A}^{\sigma_1}(p_1) (\check{A}^{\sigma_3}(p_3))^\mathsf{T} \check{A}^{\sigma_2}(p_2) + \check{A}^{\sigma_2}(p_2) (\check{A}^{\sigma_3}(p_3))^\mathsf{T} \check{A}^{\sigma_1}(p_1)  ]_{ab} \, .
    \label{eqn:C_matrix}
\end{align}
Using the definition of the weights $A_{a_ib_i}^{\sigma_i}(p_i)$ below Eq.~\eqref{eqn:A-matrix-definition}, the traces and matrix products can be evaluated explicitly. This gives
\begin{equation}
    \Tr[\check{A}^{\sigma_j}(p_j)  (\check{A}^{\sigma_k}(p_k))^\mathsf{T}] =
    2 [\sigma_j \sigma_k -  \cos(\vartheta_{p_j} + \vartheta_{p_k})]
\end{equation}
for the trace factor appearing in Eq.~\eqref{eqn:B_matrix}, and 
\begin{align}
    C_{11}(i, j \, | \, k) &= \phantom{i}2\left\{ \sigma_i\sigma_j \sigma_k + \left[ \sigma_k \cos(\vartheta_i - \vartheta_j) - \sigma_j \cos(\vartheta_i + \vartheta_k) - \sigma_i \cos(\vartheta_j + \vartheta_k) \right]  \right\} \, , \\
    C_{12}(i, j \, | \, k) &=   2i  \left\{  \sigma_j\sigma_k e^{i\vartheta_i} + \sigma_i \sigma_k e^{i\vartheta_j} -  [\sigma_i\sigma_je^{-i\vartheta_k} +  e^{i(\vartheta_i + \vartheta_j+\vartheta_k)} ] \right\} \, .
\end{align}
For the other components, $C_{22} = C_{11}$, $C_{21} = \bar{C}_{12}$, ensuring that the matrix $C_{ab}$ is Hermitian.
Hermiticity of the matrix $B_{ab}$ follows from directly from $\check{A} = \check{A}^\dagger$.

\end{widetext}


\let\oldaddcontentsline\addcontentsline%
\renewcommand{\addcontentsline}[3]{}%
\bibliography{refs}

\begin{thebibliography}{48}%
\makeatletter
\providecommand \@ifxundefined [1]{%
 \@ifx{#1\undefined}
}%
\providecommand \@ifnum [1]{%
 \ifnum #1\expandafter \@firstoftwo
 \else \expandafter \@secondoftwo
 \fi
}%
\providecommand \@ifx [1]{%
 \ifx #1\expandafter \@firstoftwo
 \else \expandafter \@secondoftwo
 \fi
}%
\providecommand \natexlab [1]{#1}%
\providecommand \enquote  [1]{``#1''}%
\providecommand \bibnamefont  [1]{#1}%
\providecommand \bibfnamefont [1]{#1}%
\providecommand \citenamefont [1]{#1}%
\providecommand \href@noop [0]{\@secondoftwo}%
\providecommand \href [0]{\begingroup \@sanitize@url \@href}%
\providecommand \@href[1]{\@@startlink{#1}\@@href}%
\providecommand \@@href[1]{\endgroup#1\@@endlink}%
\providecommand \@sanitize@url [0]{\catcode `\\12\catcode `\$12\catcode
  `\&12\catcode `\#12\catcode `\^12\catcode `\_12\catcode `\%12\relax}%
\providecommand \@@startlink[1]{}%
\providecommand \@@endlink[0]{}%
\providecommand \url  [0]{\begingroup\@sanitize@url \@url }%
\providecommand \@url [1]{\endgroup\@href {#1}{\urlprefix }}%
\providecommand \urlprefix  [0]{URL }%
\providecommand \Eprint [0]{\href }%
\providecommand \doibase [0]{https://doi.org/}%
\providecommand \selectlanguage [0]{\@gobble}%
\providecommand \bibinfo  [0]{\@secondoftwo}%
\providecommand \bibfield  [0]{\@secondoftwo}%
\providecommand \translation [1]{[#1]}%
\providecommand \BibitemOpen [0]{}%
\providecommand \bibitemStop [0]{}%
\providecommand \bibitemNoStop [0]{.\EOS\space}%
\providecommand \EOS [0]{\spacefactor3000\relax}%
\providecommand \BibitemShut  [1]{\csname bibitem#1\endcsname}%
\let\auto@bib@innerbib\@empty
\bibitem [{\citenamefont {Balents}(2010)}]{balents2010frustrated}%
  \BibitemOpen
  \bibfield  {author} {\bibinfo {author} {\bibfnamefont {L.}~\bibnamefont
  {Balents}},\ }\bibfield  {title} {\bibinfo {title} {Spin liquids in
  frustrated magnets},\ }\href {https://doi.org/10.1038/nature08917} {\bibfield
   {journal} {\bibinfo  {journal} {Nature}\ }\textbf {\bibinfo {volume}
  {464}},\ \bibinfo {pages} {199} (\bibinfo {year} {2010})}\BibitemShut
  {NoStop}%
\bibitem [{\citenamefont {Savary}\ and\ \citenamefont
  {Balents}(2016)}]{SavaryBalents2016}%
  \BibitemOpen
  \bibfield  {author} {\bibinfo {author} {\bibfnamefont {L.}~\bibnamefont
  {Savary}}\ and\ \bibinfo {author} {\bibfnamefont {L.}~\bibnamefont
  {Balents}},\ }\bibfield  {title} {\bibinfo {title} {Quantum spin liquids: a
  review},\ }\href {https://doi.org/10.1088/0034-4885/80/1/016502} {\bibfield
  {journal} {\bibinfo  {journal} {Reports on Progress in Physics}\ }\textbf
  {\bibinfo {volume} {80}},\ \bibinfo {pages} {016502} (\bibinfo {year}
  {2016})}\BibitemShut {NoStop}%
\bibitem [{\citenamefont {Zhou}\ \emph {et~al.}(2017)\citenamefont {Zhou},
  \citenamefont {Kanoda},\ and\ \citenamefont {Ng}}]{QSLStates2017}%
  \BibitemOpen
  \bibfield  {author} {\bibinfo {author} {\bibfnamefont {Y.}~\bibnamefont
  {Zhou}}, \bibinfo {author} {\bibfnamefont {K.}~\bibnamefont {Kanoda}},\ and\
  \bibinfo {author} {\bibfnamefont {T.-K.}\ \bibnamefont {Ng}},\ }\bibfield
  {title} {\bibinfo {title} {Quantum spin liquid states},\ }\href
  {https://doi.org/10.1103/RevModPhys.89.025003} {\bibfield  {journal}
  {\bibinfo  {journal} {Rev. Mod. Phys.}\ }\textbf {\bibinfo {volume} {89}},\
  \bibinfo {pages} {025003} (\bibinfo {year} {2017})}\BibitemShut {NoStop}%
\bibitem [{\citenamefont {Knolle}\ and\ \citenamefont
  {Moessner}(2019)}]{KnolleFieldGuide}%
  \BibitemOpen
  \bibfield  {author} {\bibinfo {author} {\bibfnamefont {J.}~\bibnamefont
  {Knolle}}\ and\ \bibinfo {author} {\bibfnamefont {R.}~\bibnamefont
  {Moessner}},\ }\bibfield  {title} {\bibinfo {title} {A field guide to spin
  liquids},\ }\href {https://doi.org/10.1146/annurev-conmatphys-031218-013401}
  {\bibfield  {journal} {\bibinfo  {journal} {Annual Review of Condensed Matter
  Physics}\ }\textbf {\bibinfo {volume} {10}},\ \bibinfo {pages} {451}
  (\bibinfo {year} {2019})}\BibitemShut {NoStop}%
\bibitem [{\citenamefont {Wen}\ and\ \citenamefont
  {Niu}(1990)}]{WenDegeneracy}%
  \BibitemOpen
  \bibfield  {author} {\bibinfo {author} {\bibfnamefont {X.~G.}\ \bibnamefont
  {Wen}}\ and\ \bibinfo {author} {\bibfnamefont {Q.}~\bibnamefont {Niu}},\
  }\bibfield  {title} {\bibinfo {title} {Ground-state degeneracy of the
  fractional quantum hall states in the presence of a random potential and on
  high-genus riemann surfaces},\ }\href
  {https://doi.org/10.1103/PhysRevB.41.9377} {\bibfield  {journal} {\bibinfo
  {journal} {Phys. Rev. B}\ }\textbf {\bibinfo {volume} {41}},\ \bibinfo
  {pages} {9377} (\bibinfo {year} {1990})}\BibitemShut {NoStop}%
\bibitem [{\citenamefont {Hamma}\ \emph {et~al.}(2005)\citenamefont {Hamma},
  \citenamefont {Ionicioiu},\ and\ \citenamefont {Zanardi}}]{HammaBipartite}%
  \BibitemOpen
  \bibfield  {author} {\bibinfo {author} {\bibfnamefont {A.}~\bibnamefont
  {Hamma}}, \bibinfo {author} {\bibfnamefont {R.}~\bibnamefont {Ionicioiu}},\
  and\ \bibinfo {author} {\bibfnamefont {P.}~\bibnamefont {Zanardi}},\
  }\bibfield  {title} {\bibinfo {title} {Bipartite entanglement and entropic
  boundary law in lattice spin systems},\ }\href
  {https://doi.org/10.1103/PhysRevA.71.022315} {\bibfield  {journal} {\bibinfo
  {journal} {Phys. Rev. A}\ }\textbf {\bibinfo {volume} {71}},\ \bibinfo
  {pages} {022315} (\bibinfo {year} {2005})}\BibitemShut {NoStop}%
\bibitem [{\citenamefont {Kitaev}\ and\ \citenamefont
  {Preskill}(2006)}]{KitaevPreskillEE}%
  \BibitemOpen
  \bibfield  {author} {\bibinfo {author} {\bibfnamefont {A.}~\bibnamefont
  {Kitaev}}\ and\ \bibinfo {author} {\bibfnamefont {J.}~\bibnamefont
  {Preskill}},\ }\bibfield  {title} {\bibinfo {title} {Topological entanglement
  entropy},\ }\href {https://doi.org/10.1103/PhysRevLett.96.110404} {\bibfield
  {journal} {\bibinfo  {journal} {Phys. Rev. Lett.}\ }\textbf {\bibinfo
  {volume} {96}},\ \bibinfo {pages} {110404} (\bibinfo {year}
  {2006})}\BibitemShut {NoStop}%
\bibitem [{\citenamefont {Levin}\ and\ \citenamefont {Wen}(2006)}]{LevinWenEE}%
  \BibitemOpen
  \bibfield  {author} {\bibinfo {author} {\bibfnamefont {M.}~\bibnamefont
  {Levin}}\ and\ \bibinfo {author} {\bibfnamefont {X.-G.}\ \bibnamefont
  {Wen}},\ }\bibfield  {title} {\bibinfo {title} {Detecting topological order
  in a ground state wave function},\ }\href
  {https://doi.org/10.1103/PhysRevLett.96.110405} {\bibfield  {journal}
  {\bibinfo  {journal} {Phys. Rev. Lett.}\ }\textbf {\bibinfo {volume} {96}},\
  \bibinfo {pages} {110405} (\bibinfo {year} {2006})}\BibitemShut {NoStop}%
\bibitem [{\citenamefont {Basov}\ \emph {et~al.}(2011)\citenamefont {Basov},
  \citenamefont {Averitt}, \citenamefont {van~der Marel}, \citenamefont
  {Dressel},\ and\ \citenamefont {Haule}}]{BasovRevModPhys}%
  \BibitemOpen
  \bibfield  {author} {\bibinfo {author} {\bibfnamefont {D.~N.}\ \bibnamefont
  {Basov}}, \bibinfo {author} {\bibfnamefont {R.~D.}\ \bibnamefont {Averitt}},
  \bibinfo {author} {\bibfnamefont {D.}~\bibnamefont {van~der Marel}}, \bibinfo
  {author} {\bibfnamefont {M.}~\bibnamefont {Dressel}},\ and\ \bibinfo {author}
  {\bibfnamefont {K.}~\bibnamefont {Haule}},\ }\bibfield  {title} {\bibinfo
  {title} {Electrodynamics of correlated electron materials},\ }\href
  {https://doi.org/10.1103/RevModPhys.83.471} {\bibfield  {journal} {\bibinfo
  {journal} {Rev. Mod. Phys.}\ }\textbf {\bibinfo {volume} {83}},\ \bibinfo
  {pages} {471} (\bibinfo {year} {2011})}\BibitemShut {NoStop}%
\bibitem [{\citenamefont {Giannetti}\ \emph {et~al.}(2016)\citenamefont
  {Giannetti}, \citenamefont {Capone}, \citenamefont {Fausti}, \citenamefont
  {Fabrizio}, \citenamefont {Parmigiani},\ and\ \citenamefont
  {Mihailovic}}]{GiannettiUltrafast}%
  \BibitemOpen
  \bibfield  {author} {\bibinfo {author} {\bibfnamefont {C.}~\bibnamefont
  {Giannetti}}, \bibinfo {author} {\bibfnamefont {M.}~\bibnamefont {Capone}},
  \bibinfo {author} {\bibfnamefont {D.}~\bibnamefont {Fausti}}, \bibinfo
  {author} {\bibfnamefont {M.}~\bibnamefont {Fabrizio}}, \bibinfo {author}
  {\bibfnamefont {F.}~\bibnamefont {Parmigiani}},\ and\ \bibinfo {author}
  {\bibfnamefont {D.}~\bibnamefont {Mihailovic}},\ }\bibfield  {title}
  {\bibinfo {title} {Ultrafast optical spectroscopy of strongly correlated
  materials and high-temperature superconductors: a non-equilibrium approach},\
  }\href {https://doi.org/10.1080/00018732.2016.1194044} {\bibfield  {journal}
  {\bibinfo  {journal} {Advances in Physics}\ }\textbf {\bibinfo {volume}
  {65}},\ \bibinfo {pages} {58} (\bibinfo {year} {2016})}\BibitemShut {NoStop}%
\bibitem [{\citenamefont {Kuehn}\ \emph {et~al.}(2011)\citenamefont {Kuehn},
  \citenamefont {Reimann}, \citenamefont {Woerner}, \citenamefont {Elsaesser},\
  and\ \citenamefont {Hey}}]{Kuehn20112DCS}%
  \BibitemOpen
  \bibfield  {author} {\bibinfo {author} {\bibfnamefont {W.}~\bibnamefont
  {Kuehn}}, \bibinfo {author} {\bibfnamefont {K.}~\bibnamefont {Reimann}},
  \bibinfo {author} {\bibfnamefont {M.}~\bibnamefont {Woerner}}, \bibinfo
  {author} {\bibfnamefont {T.}~\bibnamefont {Elsaesser}},\ and\ \bibinfo
  {author} {\bibfnamefont {R.}~\bibnamefont {Hey}},\ }\bibfield  {title}
  {\bibinfo {title} {Two-dimensional terahertz correlation spectra of
  electronic excitations in semiconductor quantum wells},\ }\href
  {https://doi.org/10.1021/jp1099046} {\bibfield  {journal} {\bibinfo
  {journal} {The Journal of Physical Chemistry B}\ }\textbf {\bibinfo {volume}
  {115}},\ \bibinfo {pages} {5448} (\bibinfo {year} {2011})}\BibitemShut
  {NoStop}%
\bibitem [{\citenamefont {Woerner}\ \emph {et~al.}(2013)\citenamefont
  {Woerner}, \citenamefont {Kuehn}, \citenamefont {Bowlan}, \citenamefont
  {Reimann},\ and\ \citenamefont {Elsaesser}}]{Woerner20132DCS}%
  \BibitemOpen
  \bibfield  {author} {\bibinfo {author} {\bibfnamefont {M.}~\bibnamefont
  {Woerner}}, \bibinfo {author} {\bibfnamefont {W.}~\bibnamefont {Kuehn}},
  \bibinfo {author} {\bibfnamefont {P.}~\bibnamefont {Bowlan}}, \bibinfo
  {author} {\bibfnamefont {K.}~\bibnamefont {Reimann}},\ and\ \bibinfo {author}
  {\bibfnamefont {T.}~\bibnamefont {Elsaesser}},\ }\bibfield  {title} {\bibinfo
  {title} {Ultrafast two-dimensional terahertz spectroscopy of elementary
  excitations in solids},\ }\href
  {https://doi.org/10.1088/1367-2630/15/2/025039} {\bibfield  {journal}
  {\bibinfo  {journal} {New Journal of Physics}\ }\textbf {\bibinfo {volume}
  {15}},\ \bibinfo {pages} {025039} (\bibinfo {year} {2013})}\BibitemShut
  {NoStop}%
\bibitem [{\citenamefont {Aue}\ \emph {et~al.}(1976)\citenamefont {Aue},
  \citenamefont {Bartholdi},\ and\ \citenamefont {Ernst}}]{Aue1976Two}%
  \BibitemOpen
  \bibfield  {author} {\bibinfo {author} {\bibfnamefont {W.}~\bibnamefont
  {Aue}}, \bibinfo {author} {\bibfnamefont {E.}~\bibnamefont {Bartholdi}},\
  and\ \bibinfo {author} {\bibfnamefont {R.~R.}\ \bibnamefont {Ernst}},\
  }\bibfield  {title} {\bibinfo {title} {Two-dimensional spectroscopy.
  {A}pplication to nuclear magnetic resonance},\ }\href
  {https://doi.org/10.1063/1.432450} {\bibfield  {journal} {\bibinfo  {journal}
  {The Journal of Chemical Physics}\ }\textbf {\bibinfo {volume} {64}},\
  \bibinfo {pages} {2229} (\bibinfo {year} {1976})}\BibitemShut {NoStop}%
\bibitem [{\citenamefont {Mukamel}(1999)}]{Mukamel1999Principles}%
  \BibitemOpen
  \bibfield  {author} {\bibinfo {author} {\bibfnamefont {S.}~\bibnamefont
  {Mukamel}},\ }\href@noop {} {\emph {\bibinfo {title} {Principles of Nonlinear
  Optical Spectroscopy}}},\ Oxford series in optical and imaging sciences\
  (\bibinfo  {publisher} {Oxford University Press},\ \bibinfo {year}
  {1999})\BibitemShut {NoStop}%
\bibitem [{\citenamefont {Hamm}\ and\ \citenamefont
  {Zanni}(2011)}]{HammZanni2011}%
  \BibitemOpen
  \bibfield  {author} {\bibinfo {author} {\bibfnamefont {P.}~\bibnamefont
  {Hamm}}\ and\ \bibinfo {author} {\bibfnamefont {M.}~\bibnamefont {Zanni}},\
  }\href {https://doi.org/10.1017/CBO9780511675935} {\emph {\bibinfo {title}
  {Concepts and Methods of 2D Infrared Spectroscopy}}}\ (\bibinfo  {publisher}
  {Cambridge University Press},\ \bibinfo {year} {2011})\BibitemShut {NoStop}%
\bibitem [{\citenamefont {Cundiff}\ and\ \citenamefont
  {Mukamel}(2013)}]{CundiffMukamel}%
  \BibitemOpen
  \bibfield  {author} {\bibinfo {author} {\bibfnamefont {S.~T.}\ \bibnamefont
  {Cundiff}}\ and\ \bibinfo {author} {\bibfnamefont {S.}~\bibnamefont
  {Mukamel}},\ }\bibfield  {title} {\bibinfo {title} {Optical multidimensional
  coherent spectroscopy},\ }\href {https://doi.org/10.1063/PT.3.2047}
  {\bibfield  {journal} {\bibinfo  {journal} {Physics Today}\ }\textbf
  {\bibinfo {volume} {66}},\ \bibinfo {pages} {44} (\bibinfo {year}
  {2013})}\BibitemShut {NoStop}%
\bibitem [{\citenamefont {Lu}\ \emph {et~al.}(2016)\citenamefont {Lu},
  \citenamefont {Zhang}, \citenamefont {Hwang}, \citenamefont {Ofori-Okai},
  \citenamefont {Fleischer},\ and\ \citenamefont {Nelson}}]{Lu2016_2DCS}%
  \BibitemOpen
  \bibfield  {author} {\bibinfo {author} {\bibfnamefont {J.}~\bibnamefont
  {Lu}}, \bibinfo {author} {\bibfnamefont {Y.}~\bibnamefont {Zhang}}, \bibinfo
  {author} {\bibfnamefont {H.~Y.}\ \bibnamefont {Hwang}}, \bibinfo {author}
  {\bibfnamefont {B.~K.}\ \bibnamefont {Ofori-Okai}}, \bibinfo {author}
  {\bibfnamefont {S.}~\bibnamefont {Fleischer}},\ and\ \bibinfo {author}
  {\bibfnamefont {K.~A.}\ \bibnamefont {Nelson}},\ }\bibfield  {title}
  {\bibinfo {title} {Nonlinear two-dimensional terahertz photon echo and
  rotational spectroscopy in the gas phase},\ }\href
  {https://doi.org/10.1073/pnas.1609558113} {\bibfield  {journal} {\bibinfo
  {journal} {Proceedings of the National Academy of Sciences}\ }\textbf
  {\bibinfo {volume} {113}},\ \bibinfo {pages} {11800} (\bibinfo {year}
  {2016})}\BibitemShut {NoStop}%
\bibitem [{\citenamefont {Lu}\ \emph {et~al.}(2017)\citenamefont {Lu},
  \citenamefont {Li}, \citenamefont {Hwang}, \citenamefont {Ofori-Okai},
  \citenamefont {Kurihara}, \citenamefont {Suemoto},\ and\ \citenamefont
  {Nelson}}]{Lu2017SpinWaves}%
  \BibitemOpen
  \bibfield  {author} {\bibinfo {author} {\bibfnamefont {J.}~\bibnamefont
  {Lu}}, \bibinfo {author} {\bibfnamefont {X.}~\bibnamefont {Li}}, \bibinfo
  {author} {\bibfnamefont {H.~Y.}\ \bibnamefont {Hwang}}, \bibinfo {author}
  {\bibfnamefont {B.~K.}\ \bibnamefont {Ofori-Okai}}, \bibinfo {author}
  {\bibfnamefont {T.}~\bibnamefont {Kurihara}}, \bibinfo {author}
  {\bibfnamefont {T.}~\bibnamefont {Suemoto}},\ and\ \bibinfo {author}
  {\bibfnamefont {K.~A.}\ \bibnamefont {Nelson}},\ }\bibfield  {title}
  {\bibinfo {title} {Coherent two-dimensional terahertz magnetic resonance
  spectroscopy of collective spin waves},\ }\href
  {https://doi.org/10.1103/PhysRevLett.118.207204} {\bibfield  {journal}
  {\bibinfo  {journal} {Phys. Rev. Lett.}\ }\textbf {\bibinfo {volume} {118}},\
  \bibinfo {pages} {207204} (\bibinfo {year} {2017})}\BibitemShut {NoStop}%
\bibitem [{\citenamefont {Mahmood}\ \emph {et~al.}(2021)\citenamefont
  {Mahmood}, \citenamefont {Chaudhuri}, \citenamefont {Gopalakrishnan},
  \citenamefont {Nandkishore},\ and\ \citenamefont
  {Armitage}}]{Mahmood2021Observation}%
  \BibitemOpen
  \bibfield  {author} {\bibinfo {author} {\bibfnamefont {F.}~\bibnamefont
  {Mahmood}}, \bibinfo {author} {\bibfnamefont {D.}~\bibnamefont {Chaudhuri}},
  \bibinfo {author} {\bibfnamefont {S.}~\bibnamefont {Gopalakrishnan}},
  \bibinfo {author} {\bibfnamefont {R.}~\bibnamefont {Nandkishore}},\ and\
  \bibinfo {author} {\bibfnamefont {N.}~\bibnamefont {Armitage}},\ }\bibfield
  {title} {\bibinfo {title} {Observation of a marginal fermi glass},\ }\href
  {https://doi.org/10.1038/s41567-020-01149-0} {\bibfield  {journal} {\bibinfo
  {journal} {Nature Physics}\ }\textbf {\bibinfo {volume} {17}},\ \bibinfo
  {pages} {627} (\bibinfo {year} {2021})}\BibitemShut {NoStop}%
\bibitem [{\citenamefont {Choi}\ \emph {et~al.}(2020)\citenamefont {Choi},
  \citenamefont {Lee},\ and\ \citenamefont {Kim}}]{Choi2DCSKitaev}%
  \BibitemOpen
  \bibfield  {author} {\bibinfo {author} {\bibfnamefont {W.}~\bibnamefont
  {Choi}}, \bibinfo {author} {\bibfnamefont {K.~H.}\ \bibnamefont {Lee}},\ and\
  \bibinfo {author} {\bibfnamefont {Y.~B.}\ \bibnamefont {Kim}},\ }\bibfield
  {title} {\bibinfo {title} {Theory of two-dimensional nonlinear spectroscopy
  for the kitaev spin liquid},\ }\href
  {https://doi.org/10.1103/PhysRevLett.124.117205} {\bibfield  {journal}
  {\bibinfo  {journal} {Phys. Rev. Lett.}\ }\textbf {\bibinfo {volume} {124}},\
  \bibinfo {pages} {117205} (\bibinfo {year} {2020})}\BibitemShut {NoStop}%
\bibitem [{\citenamefont {Parameswaran}\ and\ \citenamefont
  {Gopalakrishnan}(2020)}]{Parameswaran2020RandomTFIM}%
  \BibitemOpen
  \bibfield  {author} {\bibinfo {author} {\bibfnamefont {S.~A.}\ \bibnamefont
  {Parameswaran}}\ and\ \bibinfo {author} {\bibfnamefont {S.}~\bibnamefont
  {Gopalakrishnan}},\ }\bibfield  {title} {\bibinfo {title} {Asymptotically
  exact theory for nonlinear spectroscopy of random quantum magnets},\ }\href
  {https://doi.org/10.1103/PhysRevLett.125.237601} {\bibfield  {journal}
  {\bibinfo  {journal} {Phys. Rev. Lett.}\ }\textbf {\bibinfo {volume} {125}},\
  \bibinfo {pages} {237601} (\bibinfo {year} {2020})}\BibitemShut {NoStop}%
\bibitem [{\citenamefont {Nandkishore}\ \emph {et~al.}(2021)\citenamefont
  {Nandkishore}, \citenamefont {Choi},\ and\ \citenamefont
  {Kim}}]{Nandkishore2021Spectroscopic}%
  \BibitemOpen
  \bibfield  {author} {\bibinfo {author} {\bibfnamefont {R.~M.}\ \bibnamefont
  {Nandkishore}}, \bibinfo {author} {\bibfnamefont {W.}~\bibnamefont {Choi}},\
  and\ \bibinfo {author} {\bibfnamefont {Y.~B.}\ \bibnamefont {Kim}},\
  }\bibfield  {title} {\bibinfo {title} {Spectroscopic fingerprints of gapped
  quantum spin liquids, both conventional and fractonic},\ }\href
  {https://doi.org/10.1103/PhysRevResearch.3.013254} {\bibfield  {journal}
  {\bibinfo  {journal} {Phys. Rev. Research}\ }\textbf {\bibinfo {volume}
  {3}},\ \bibinfo {pages} {013254} (\bibinfo {year} {2021})}\BibitemShut
  {NoStop}%
\bibitem [{\citenamefont {Li}\ \emph {et~al.}(2021)\citenamefont {Li},
  \citenamefont {Oshikawa},\ and\ \citenamefont {Wan}}]{Li2021Lensing}%
  \BibitemOpen
  \bibfield  {author} {\bibinfo {author} {\bibfnamefont {Z.-L.}\ \bibnamefont
  {Li}}, \bibinfo {author} {\bibfnamefont {M.}~\bibnamefont {Oshikawa}},\ and\
  \bibinfo {author} {\bibfnamefont {Y.}~\bibnamefont {Wan}},\ }\bibfield
  {title} {\bibinfo {title} {Photon echo from lensing of fractional excitations
  in {T}omonaga-{L}uttinger spin liquid},\ }\href
  {https://doi.org/10.1103/PhysRevX.11.031035} {\bibfield  {journal} {\bibinfo
  {journal} {Phys. Rev. X}\ }\textbf {\bibinfo {volume} {11}},\ \bibinfo
  {pages} {031035} (\bibinfo {year} {2021})}\BibitemShut {NoStop}%
\bibitem [{\citenamefont {Fava}\ \emph {et~al.}(2021)\citenamefont {Fava},
  \citenamefont {Biswas}, \citenamefont {Gopalakrishnan}, \citenamefont
  {Vasseur},\ and\ \citenamefont {Parameswaran}}]{Fava2021Hydrodynamic}%
  \BibitemOpen
  \bibfield  {author} {\bibinfo {author} {\bibfnamefont {M.}~\bibnamefont
  {Fava}}, \bibinfo {author} {\bibfnamefont {S.}~\bibnamefont {Biswas}},
  \bibinfo {author} {\bibfnamefont {S.}~\bibnamefont {Gopalakrishnan}},
  \bibinfo {author} {\bibfnamefont {R.}~\bibnamefont {Vasseur}},\ and\ \bibinfo
  {author} {\bibfnamefont {S.~A.}\ \bibnamefont {Parameswaran}},\ }\bibfield
  {title} {\bibinfo {title} {Hydrodynamic nonlinear response of interacting
  integrable systems},\ }\href {https://doi.org/10.1073/pnas.2106945118}
  {\bibfield  {journal} {\bibinfo  {journal} {Proceedings of the National
  Academy of Sciences}\ }\textbf {\bibinfo {volume} {118}},\ \bibinfo {pages}
  {e2106945118} (\bibinfo {year} {2021})}\BibitemShut {NoStop}%
\bibitem [{\citenamefont {Gerken}\ \emph {et~al.}(2022)\citenamefont {Gerken},
  \citenamefont {Posske}, \citenamefont {Mukamel},\ and\ \citenamefont
  {Thorwart}}]{GerkenTopoPhases}%
  \BibitemOpen
  \bibfield  {author} {\bibinfo {author} {\bibfnamefont {F.}~\bibnamefont
  {Gerken}}, \bibinfo {author} {\bibfnamefont {T.}~\bibnamefont {Posske}},
  \bibinfo {author} {\bibfnamefont {S.}~\bibnamefont {Mukamel}},\ and\ \bibinfo
  {author} {\bibfnamefont {M.}~\bibnamefont {Thorwart}},\ }\href
  {https://doi.org/10.48550/ARXIV.2202.00030} {\bibinfo {title} {Unique
  signatures of topological phases in two-dimensional {THz} spectroscopy}}
  (\bibinfo {year} {2022})\BibitemShut {NoStop}%
\bibitem [{\citenamefont {Phuc}\ and\ \citenamefont
  {Trung}(2021)}]{PhucManyBody}%
  \BibitemOpen
  \bibfield  {author} {\bibinfo {author} {\bibfnamefont {N.~T.}\ \bibnamefont
  {Phuc}}\ and\ \bibinfo {author} {\bibfnamefont {P.~Q.}\ \bibnamefont
  {Trung}},\ }\bibfield  {title} {\bibinfo {title} {Direct and ultrafast
  probing of quantum many-body interactions through coherent two-dimensional
  spectroscopy: {F}rom weak- to strong-interaction regimes},\ }\href
  {https://doi.org/10.1103/PhysRevB.104.115105} {\bibfield  {journal} {\bibinfo
   {journal} {Phys. Rev. B}\ }\textbf {\bibinfo {volume} {104}},\ \bibinfo
  {pages} {115105} (\bibinfo {year} {2021})}\BibitemShut {NoStop}%
\bibitem [{\citenamefont {Wan}\ and\ \citenamefont
  {Armitage}(2019)}]{WanArmitage2019}%
  \BibitemOpen
  \bibfield  {author} {\bibinfo {author} {\bibfnamefont {Y.}~\bibnamefont
  {Wan}}\ and\ \bibinfo {author} {\bibfnamefont {N.~P.}\ \bibnamefont
  {Armitage}},\ }\bibfield  {title} {\bibinfo {title} {Resolving continua of
  fractional excitations by spinon echo in {THz} {2D} coherent spectroscopy},\
  }\href {https://doi.org/10.1103/PhysRevLett.122.257401} {\bibfield  {journal}
  {\bibinfo  {journal} {Phys. Rev. Lett.}\ }\textbf {\bibinfo {volume} {122}},\
  \bibinfo {pages} {257401} (\bibinfo {year} {2019})}\BibitemShut {NoStop}%
\bibitem [{\citenamefont {Coldea}\ \emph {et~al.}(2010)\citenamefont {Coldea},
  \citenamefont {Tennant}, \citenamefont {Wheeler}, \citenamefont {Wawrzynska},
  \citenamefont {Prabhakaran}, \citenamefont {Telling}, \citenamefont
  {Habicht}, \citenamefont {Smeibidl},\ and\ \citenamefont
  {Kiefer}}]{coldea2010quantum}%
  \BibitemOpen
  \bibfield  {author} {\bibinfo {author} {\bibfnamefont {R.}~\bibnamefont
  {Coldea}}, \bibinfo {author} {\bibfnamefont {D.}~\bibnamefont {Tennant}},
  \bibinfo {author} {\bibfnamefont {E.}~\bibnamefont {Wheeler}}, \bibinfo
  {author} {\bibfnamefont {E.}~\bibnamefont {Wawrzynska}}, \bibinfo {author}
  {\bibfnamefont {D.}~\bibnamefont {Prabhakaran}}, \bibinfo {author}
  {\bibfnamefont {M.}~\bibnamefont {Telling}}, \bibinfo {author} {\bibfnamefont
  {K.}~\bibnamefont {Habicht}}, \bibinfo {author} {\bibfnamefont
  {P.}~\bibnamefont {Smeibidl}},\ and\ \bibinfo {author} {\bibfnamefont
  {K.}~\bibnamefont {Kiefer}},\ }\bibfield  {title} {\bibinfo {title} {Quantum
  criticality in an {I}sing chain: experimental evidence for emergent {E8}
  symmetry},\ }\href {https://doi.org/10.1126/science.1180085} {\bibfield
  {journal} {\bibinfo  {journal} {Science}\ }\textbf {\bibinfo {volume}
  {327}},\ \bibinfo {pages} {177} (\bibinfo {year} {2010})}\BibitemShut
  {NoStop}%
\bibitem [{\citenamefont {Morris}\ \emph {et~al.}(2014)\citenamefont {Morris},
  \citenamefont {Vald\'es~Aguilar}, \citenamefont {Ghosh}, \citenamefont
  {Koohpayeh}, \citenamefont {Krizan}, \citenamefont {Cava}, \citenamefont
  {Tchernyshyov}, \citenamefont {McQueen},\ and\ \citenamefont
  {Armitage}}]{Morris2014}%
  \BibitemOpen
  \bibfield  {author} {\bibinfo {author} {\bibfnamefont {C.~M.}\ \bibnamefont
  {Morris}}, \bibinfo {author} {\bibfnamefont {R.}~\bibnamefont
  {Vald\'es~Aguilar}}, \bibinfo {author} {\bibfnamefont {A.}~\bibnamefont
  {Ghosh}}, \bibinfo {author} {\bibfnamefont {S.~M.}\ \bibnamefont
  {Koohpayeh}}, \bibinfo {author} {\bibfnamefont {J.}~\bibnamefont {Krizan}},
  \bibinfo {author} {\bibfnamefont {R.~J.}\ \bibnamefont {Cava}}, \bibinfo
  {author} {\bibfnamefont {O.}~\bibnamefont {Tchernyshyov}}, \bibinfo {author}
  {\bibfnamefont {T.~M.}\ \bibnamefont {McQueen}},\ and\ \bibinfo {author}
  {\bibfnamefont {N.~P.}\ \bibnamefont {Armitage}},\ }\bibfield  {title}
  {\bibinfo {title} {Hierarchy of bound states in the one-dimensional
  ferromagnetic {I}sing chain {${\mathrm{CoNb}}_{2}{\mathrm{O}}_{6}$}
  investigated by high-resolution time-domain terahertz spectroscopy},\ }\href
  {https://doi.org/10.1103/PhysRevLett.112.137403} {\bibfield  {journal}
  {\bibinfo  {journal} {Phys. Rev. Lett.}\ }\textbf {\bibinfo {volume} {112}},\
  \bibinfo {pages} {137403} (\bibinfo {year} {2014})}\BibitemShut {NoStop}%
\bibitem [{\citenamefont {Cabrera}\ \emph {et~al.}(2014)\citenamefont
  {Cabrera}, \citenamefont {Thompson}, \citenamefont {Coldea}, \citenamefont
  {Prabhakaran}, \citenamefont {Bewley}, \citenamefont {Guidi}, \citenamefont
  {Rodriguez-Rivera},\ and\ \citenamefont {Stock}}]{Cabrera2014Excitations}%
  \BibitemOpen
  \bibfield  {author} {\bibinfo {author} {\bibfnamefont {I.}~\bibnamefont
  {Cabrera}}, \bibinfo {author} {\bibfnamefont {J.~D.}\ \bibnamefont
  {Thompson}}, \bibinfo {author} {\bibfnamefont {R.}~\bibnamefont {Coldea}},
  \bibinfo {author} {\bibfnamefont {D.}~\bibnamefont {Prabhakaran}}, \bibinfo
  {author} {\bibfnamefont {R.~I.}\ \bibnamefont {Bewley}}, \bibinfo {author}
  {\bibfnamefont {T.}~\bibnamefont {Guidi}}, \bibinfo {author} {\bibfnamefont
  {J.~A.}\ \bibnamefont {Rodriguez-Rivera}},\ and\ \bibinfo {author}
  {\bibfnamefont {C.}~\bibnamefont {Stock}},\ }\bibfield  {title} {\bibinfo
  {title} {{Excitations in the quantum paramagnetic phase of the
  quasi-one-dimensional Ising magnet ${\mathrm{CoNb}}_{2}$${\mathrm{O}}_{6}$ in
  a transverse field: Geometric frustration and quantum renormalization
  effects}},\ }\href {https://doi.org/10.1103/PhysRevB.90.014418} {\bibfield
  {journal} {\bibinfo  {journal} {Phys. Rev. B}\ }\textbf {\bibinfo {volume}
  {90}},\ \bibinfo {pages} {014418} (\bibinfo {year} {2014})}\BibitemShut
  {NoStop}%
\bibitem [{\citenamefont {Kinross}\ \emph {et~al.}(2014)\citenamefont
  {Kinross}, \citenamefont {Fu}, \citenamefont {Munsie}, \citenamefont
  {Dabkowska}, \citenamefont {Luke}, \citenamefont {Sachdev},\ and\
  \citenamefont {Imai}}]{Kinross2014}%
  \BibitemOpen
  \bibfield  {author} {\bibinfo {author} {\bibfnamefont {A.~W.}\ \bibnamefont
  {Kinross}}, \bibinfo {author} {\bibfnamefont {M.}~\bibnamefont {Fu}},
  \bibinfo {author} {\bibfnamefont {T.~J.}\ \bibnamefont {Munsie}}, \bibinfo
  {author} {\bibfnamefont {H.~A.}\ \bibnamefont {Dabkowska}}, \bibinfo {author}
  {\bibfnamefont {G.~M.}\ \bibnamefont {Luke}}, \bibinfo {author}
  {\bibfnamefont {S.}~\bibnamefont {Sachdev}},\ and\ \bibinfo {author}
  {\bibfnamefont {T.}~\bibnamefont {Imai}},\ }\bibfield  {title} {\bibinfo
  {title} {Evolution of quantum fluctuations near the quantum critical point of
  the transverse field {I}sing chain system
  {${\mathrm{CoNb}}_{2}{\mathrm{O}}_{6}$}},\ }\href
  {https://doi.org/10.1103/PhysRevX.4.031008} {\bibfield  {journal} {\bibinfo
  {journal} {Phys. Rev. X}\ }\textbf {\bibinfo {volume} {4}},\ \bibinfo {pages}
  {031008} (\bibinfo {year} {2014})}\BibitemShut {NoStop}%
\bibitem [{\citenamefont {Robinson}\ \emph {et~al.}(2014)\citenamefont
  {Robinson}, \citenamefont {Essler}, \citenamefont {Cabrera},\ and\
  \citenamefont {Coldea}}]{Robinson2014breakdown}%
  \BibitemOpen
  \bibfield  {author} {\bibinfo {author} {\bibfnamefont {N.~J.}\ \bibnamefont
  {Robinson}}, \bibinfo {author} {\bibfnamefont {F.~H.~L.}\ \bibnamefont
  {Essler}}, \bibinfo {author} {\bibfnamefont {I.}~\bibnamefont {Cabrera}},\
  and\ \bibinfo {author} {\bibfnamefont {R.}~\bibnamefont {Coldea}},\
  }\bibfield  {title} {\bibinfo {title} {{Quasiparticle breakdown in the
  quasi-one-dimensional Ising ferromagnet
  ${\mathrm{CoNb}}_{2}{\mathrm{O}}_{6}$}},\ }\href
  {https://doi.org/10.1103/PhysRevB.90.174406} {\bibfield  {journal} {\bibinfo
  {journal} {Phys. Rev. B}\ }\textbf {\bibinfo {volume} {90}},\ \bibinfo
  {pages} {174406} (\bibinfo {year} {2014})}\BibitemShut {NoStop}%
\bibitem [{\citenamefont {Liang}\ \emph {et~al.}(2015)\citenamefont {Liang},
  \citenamefont {Koohpayeh}, \citenamefont {Krizan}, \citenamefont {McQueen},
  \citenamefont {Cava},\ and\ \citenamefont {Ong}}]{liang2015heat}%
  \BibitemOpen
  \bibfield  {author} {\bibinfo {author} {\bibfnamefont {T.}~\bibnamefont
  {Liang}}, \bibinfo {author} {\bibfnamefont {S.}~\bibnamefont {Koohpayeh}},
  \bibinfo {author} {\bibfnamefont {J.}~\bibnamefont {Krizan}}, \bibinfo
  {author} {\bibfnamefont {T.}~\bibnamefont {McQueen}}, \bibinfo {author}
  {\bibfnamefont {R.}~\bibnamefont {Cava}},\ and\ \bibinfo {author}
  {\bibfnamefont {N.~P.}\ \bibnamefont {Ong}},\ }\bibfield  {title} {\bibinfo
  {title} {Heat capacity peak at the quantum critical point of the transverse
  {I}sing magnet {${\mathrm{CoNb}}_{2}{\mathrm{O}}_{6}$}},\ }\href
  {https://doi.org/10.1038/ncomms8611} {\bibfield  {journal} {\bibinfo
  {journal} {Nature communications}\ }\textbf {\bibinfo {volume} {6}},\
  \bibinfo {pages} {7611} (\bibinfo {year} {2015})}\BibitemShut {NoStop}%
\bibitem [{\citenamefont {Morris}\ \emph {et~al.}(2021)\citenamefont {Morris},
  \citenamefont {Desai}, \citenamefont {Viirok}, \citenamefont {H{\"u}vonen},
  \citenamefont {Nagel}, \citenamefont {Room}, \citenamefont {Krizan},
  \citenamefont {Cava}, \citenamefont {McQueen}, \citenamefont {Koohpayeh}
  \emph {et~al.}}]{Morris2021duality}%
  \BibitemOpen
  \bibfield  {author} {\bibinfo {author} {\bibfnamefont {C.}~\bibnamefont
  {Morris}}, \bibinfo {author} {\bibfnamefont {N.}~\bibnamefont {Desai}},
  \bibinfo {author} {\bibfnamefont {J.}~\bibnamefont {Viirok}}, \bibinfo
  {author} {\bibfnamefont {D.}~\bibnamefont {H{\"u}vonen}}, \bibinfo {author}
  {\bibfnamefont {U.}~\bibnamefont {Nagel}}, \bibinfo {author} {\bibfnamefont
  {T.}~\bibnamefont {Room}}, \bibinfo {author} {\bibfnamefont {J.}~\bibnamefont
  {Krizan}}, \bibinfo {author} {\bibfnamefont {R.}~\bibnamefont {Cava}},
  \bibinfo {author} {\bibfnamefont {T.}~\bibnamefont {McQueen}}, \bibinfo
  {author} {\bibfnamefont {S.}~\bibnamefont {Koohpayeh}}, \emph {et~al.},\
  }\bibfield  {title} {\bibinfo {title} {Duality and domain wall dynamics in a
  twisted {K}itaev chain},\ }\href {https://doi.org/0.1038/s41567-021-01208-0}
  {\bibfield  {journal} {\bibinfo  {journal} {Nature Physics}\ }\textbf
  {\bibinfo {volume} {17}},\ \bibinfo {pages} {832} (\bibinfo {year}
  {2021})}\BibitemShut {NoStop}%
\bibitem [{\citenamefont {Fava}\ \emph {et~al.}(2020)\citenamefont {Fava},
  \citenamefont {Coldea},\ and\ \citenamefont {Parameswaran}}]{Fava}%
  \BibitemOpen
  \bibfield  {author} {\bibinfo {author} {\bibfnamefont {M.}~\bibnamefont
  {Fava}}, \bibinfo {author} {\bibfnamefont {R.}~\bibnamefont {Coldea}},\ and\
  \bibinfo {author} {\bibfnamefont {S.}~\bibnamefont {Parameswaran}},\
  }\bibfield  {title} {\bibinfo {title} {Glide symmetry breaking and ising
  criticality in the quasi-1d magnet conb2o6},\ }\href@noop {} {\bibfield
  {journal} {\bibinfo  {journal} {Proceedings of the National Academy of
  Sciences}\ }\textbf {\bibinfo {volume} {117}},\ \bibinfo {pages} {25219}
  (\bibinfo {year} {2020})}\BibitemShut {NoStop}%
\bibitem [{\citenamefont {Lieb}\ \emph {et~al.}(1961)\citenamefont {Lieb},
  \citenamefont {Schultz},\ and\ \citenamefont {Mattis}}]{Lieb1961Soluble}%
  \BibitemOpen
  \bibfield  {author} {\bibinfo {author} {\bibfnamefont {E.}~\bibnamefont
  {Lieb}}, \bibinfo {author} {\bibfnamefont {T.}~\bibnamefont {Schultz}},\ and\
  \bibinfo {author} {\bibfnamefont {D.}~\bibnamefont {Mattis}},\ }\bibfield
  {title} {\bibinfo {title} {Two soluble models of an antiferromagnetic
  chain},\ }\href
  {https://doi.org/https://doi.org/10.1016/0003-4916(61)90115-4} {\bibfield
  {journal} {\bibinfo  {journal} {Annals of Physics}\ }\textbf {\bibinfo
  {volume} {16}},\ \bibinfo {pages} {407} (\bibinfo {year} {1961})}\BibitemShut
  {NoStop}%
\bibitem [{\citenamefont {Pfeuty}(1970)}]{Pfeuty1970}%
  \BibitemOpen
  \bibfield  {author} {\bibinfo {author} {\bibfnamefont {P.}~\bibnamefont
  {Pfeuty}},\ }\bibfield  {title} {\bibinfo {title} {The one-dimensional ising
  model with a transverse field},\ }\href
  {https://doi.org/https://doi.org/10.1016/0003-4916(70)90270-8} {\bibfield
  {journal} {\bibinfo  {journal} {Annals of Physics}\ }\textbf {\bibinfo
  {volume} {57}},\ \bibinfo {pages} {79} (\bibinfo {year} {1970})}\BibitemShut
  {NoStop}%
\bibitem [{\citenamefont {Hilker}\ \emph {et~al.}(2017)\citenamefont {Hilker},
  \citenamefont {Salomon}, \citenamefont {Grusdt}, \citenamefont {Omran},
  \citenamefont {Boll}, \citenamefont {Demler}, \citenamefont {Bloch},\ and\
  \citenamefont {Gross}}]{BlochStringOrder}%
  \BibitemOpen
  \bibfield  {author} {\bibinfo {author} {\bibfnamefont {T.~A.}\ \bibnamefont
  {Hilker}}, \bibinfo {author} {\bibfnamefont {G.}~\bibnamefont {Salomon}},
  \bibinfo {author} {\bibfnamefont {F.}~\bibnamefont {Grusdt}}, \bibinfo
  {author} {\bibfnamefont {A.}~\bibnamefont {Omran}}, \bibinfo {author}
  {\bibfnamefont {M.}~\bibnamefont {Boll}}, \bibinfo {author} {\bibfnamefont
  {E.}~\bibnamefont {Demler}}, \bibinfo {author} {\bibfnamefont
  {I.}~\bibnamefont {Bloch}},\ and\ \bibinfo {author} {\bibfnamefont
  {C.}~\bibnamefont {Gross}},\ }\bibfield  {title} {\bibinfo {title} {Revealing
  hidden antiferromagnetic correlations in doped hubbard chains via string
  correlators},\ }\href {https://doi.org/10.1126/science.aam8990} {\bibfield
  {journal} {\bibinfo  {journal} {Science}\ }\textbf {\bibinfo {volume}
  {357}},\ \bibinfo {pages} {484} (\bibinfo {year} {2017})}\BibitemShut
  {NoStop}%
\bibitem [{\citenamefont {Sachdev}(2011)}]{SachdevQPT}%
  \BibitemOpen
  \bibfield  {author} {\bibinfo {author} {\bibfnamefont {S.}~\bibnamefont
  {Sachdev}},\ }\href {https://doi.org/10.1017/CBO9780511973765} {\emph
  {\bibinfo {title} {Quantum Phase Transitions}}},\ \bibinfo {edition} {2nd}\
  ed.\ (\bibinfo  {publisher} {Cambridge University Press},\ \bibinfo {year}
  {2011})\BibitemShut {NoStop}%
\bibitem [{\citenamefont {Kamenev}(2011)}]{Kamenev2011}%
  \BibitemOpen
  \bibfield  {author} {\bibinfo {author} {\bibfnamefont {A.}~\bibnamefont
  {Kamenev}},\ }\href {https://doi.org/10.1017/CBO9781139003667} {\emph
  {\bibinfo {title} {Field Theory of Non-Equilibrium Systems}}}\ (\bibinfo
  {publisher} {Cambridge University Press},\ \bibinfo {year}
  {2011})\BibitemShut {NoStop}%
\bibitem [{\citenamefont {Shankar}(2017)}]{ShankarQFT}%
  \BibitemOpen
  \bibfield  {author} {\bibinfo {author} {\bibfnamefont {R.}~\bibnamefont
  {Shankar}},\ }\href {https://doi.org/10.1017/9781139044349} {\emph {\bibinfo
  {title} {Quantum Field Theory and Condensed Matter: An Introduction}}}\
  (\bibinfo  {publisher} {Cambridge University Press},\ \bibinfo {year}
  {2017})\BibitemShut {NoStop}%
\bibitem [{\citenamefont {Chou}\ \emph {et~al.}(1985)\citenamefont {Chou},
  \citenamefont {Su}, \citenamefont {Hao},\ and\ \citenamefont
  {Yu}}]{ChouUnified}%
  \BibitemOpen
  \bibfield  {author} {\bibinfo {author} {\bibfnamefont {K.-C.}\ \bibnamefont
  {Chou}}, \bibinfo {author} {\bibfnamefont {Z.-B.}\ \bibnamefont {Su}},
  \bibinfo {author} {\bibfnamefont {B.-L.}\ \bibnamefont {Hao}},\ and\ \bibinfo
  {author} {\bibfnamefont {L.}~\bibnamefont {Yu}},\ }\bibfield  {title}
  {\bibinfo {title} {Equilibrium and nonequilibrium formalisms made unified},\
  }\href {https://doi.org/https://doi.org/10.1016/0370-1573(85)90136-X}
  {\bibfield  {journal} {\bibinfo  {journal} {Physics Reports}\ }\textbf
  {\bibinfo {volume} {118}},\ \bibinfo {pages} {1} (\bibinfo {year}
  {1985})}\BibitemShut {NoStop}%
\bibitem [{\citenamefont {Khalil}\ \emph {et~al.}(2003)\citenamefont {Khalil},
  \citenamefont {Demird\"{o}ven},\ and\ \citenamefont
  {Tokmakoff}}]{Khalil2003}%
  \BibitemOpen
  \bibfield  {author} {\bibinfo {author} {\bibfnamefont {M.}~\bibnamefont
  {Khalil}}, \bibinfo {author} {\bibfnamefont {N.}~\bibnamefont
  {Demird\"{o}ven}},\ and\ \bibinfo {author} {\bibfnamefont {A.}~\bibnamefont
  {Tokmakoff}},\ }\bibfield  {title} {\bibinfo {title} {Coherent {2D} {IR}
  spectroscopy: {M}olecular structure and dynamics in solution},\ }\href
  {https://doi.org/10.1021/jp0219247} {\bibfield  {journal} {\bibinfo
  {journal} {The Journal of Physical Chemistry A}\ }\textbf {\bibinfo {volume}
  {107}},\ \bibinfo {pages} {5258} (\bibinfo {year} {2003})}\BibitemShut
  {NoStop}%
\bibitem [{Arm()}]{ArmitagePrivate}%
  \BibitemOpen
  \href@noop {} {\bibinfo {title} {{N.~P.~Armitage, private
  communication}}}\BibitemShut {NoStop}%
\bibitem [{\citenamefont {Fiete}(2007)}]{FieteSpinIncoherent}%
  \BibitemOpen
  \bibfield  {author} {\bibinfo {author} {\bibfnamefont {G.~A.}\ \bibnamefont
  {Fiete}},\ }\bibfield  {title} {\bibinfo {title} {Colloquium: The
  spin-incoherent luttinger liquid},\ }\href
  {https://doi.org/10.1103/RevModPhys.79.801} {\bibfield  {journal} {\bibinfo
  {journal} {Rev. Mod. Phys.}\ }\textbf {\bibinfo {volume} {79}},\ \bibinfo
  {pages} {801} (\bibinfo {year} {2007})}\BibitemShut {NoStop}%
\bibitem [{\citenamefont {Pixley}\ \emph {et~al.}(2017)\citenamefont {Pixley},
  \citenamefont {Chou}, \citenamefont {Goswami}, \citenamefont {Huse},
  \citenamefont {Nandkishore}, \citenamefont {Radzihovsky},\ and\ \citenamefont
  {Das~Sarma}}]{Weyl}%
  \BibitemOpen
  \bibfield  {author} {\bibinfo {author} {\bibfnamefont {J.~H.}\ \bibnamefont
  {Pixley}}, \bibinfo {author} {\bibfnamefont {Y.-Z.}\ \bibnamefont {Chou}},
  \bibinfo {author} {\bibfnamefont {P.}~\bibnamefont {Goswami}}, \bibinfo
  {author} {\bibfnamefont {D.~A.}\ \bibnamefont {Huse}}, \bibinfo {author}
  {\bibfnamefont {R.}~\bibnamefont {Nandkishore}}, \bibinfo {author}
  {\bibfnamefont {L.}~\bibnamefont {Radzihovsky}},\ and\ \bibinfo {author}
  {\bibfnamefont {S.}~\bibnamefont {Das~Sarma}},\ }\bibfield  {title} {\bibinfo
  {title} {Single-particle excitations in disordered weyl fluids},\ }\href
  {https://doi.org/10.1103/PhysRevB.95.235101} {\bibfield  {journal} {\bibinfo
  {journal} {Phys. Rev. B}\ }\textbf {\bibinfo {volume} {95}},\ \bibinfo
  {pages} {235101} (\bibinfo {year} {2017})}\BibitemShut {NoStop}%
\bibitem [{\citenamefont {Fava}\ \emph {et~al.}(2022)\citenamefont {Fava},
  \citenamefont {Gopalakrishnan}, \citenamefont {Vasseur}, \citenamefont
  {Essler},\ and\ \citenamefont {Parameswaran}}]{FavaNonperturbative}%
  \BibitemOpen
  \bibfield  {author} {\bibinfo {author} {\bibfnamefont {M.}~\bibnamefont
  {Fava}}, \bibinfo {author} {\bibfnamefont {S.}~\bibnamefont
  {Gopalakrishnan}}, \bibinfo {author} {\bibfnamefont {R.}~\bibnamefont
  {Vasseur}}, \bibinfo {author} {\bibfnamefont {F.~H.~L.}\ \bibnamefont
  {Essler}},\ and\ \bibinfo {author} {\bibfnamefont {S.~A.}\ \bibnamefont
  {Parameswaran}},\ }\href@noop {} {\bibinfo {title} {Divergent nonlinear
  response from quasiparticle interactions}} (\bibinfo {year} {2022}),\
  \bibinfo {note} {to appear}\BibitemShut {NoStop}%
\bibitem [{\citenamefont {Safavi-Naini}\ \emph {et~al.}(2019)\citenamefont
  {Safavi-Naini}, \citenamefont {Wall}, \citenamefont {Acevedo}, \citenamefont
  {Rey},\ and\ \citenamefont {Nandkishore}}]{SafaviNaini}%
  \BibitemOpen
  \bibfield  {author} {\bibinfo {author} {\bibfnamefont {A.}~\bibnamefont
  {Safavi-Naini}}, \bibinfo {author} {\bibfnamefont {M.~L.}\ \bibnamefont
  {Wall}}, \bibinfo {author} {\bibfnamefont {O.~L.}\ \bibnamefont {Acevedo}},
  \bibinfo {author} {\bibfnamefont {A.~M.}\ \bibnamefont {Rey}},\ and\ \bibinfo
  {author} {\bibfnamefont {R.~M.}\ \bibnamefont {Nandkishore}},\ }\bibfield
  {title} {\bibinfo {title} {Quantum dynamics of disordered spin chains with
  power-law interactions},\ }\href {https://doi.org/10.1103/PhysRevA.99.033610}
  {\bibfield  {journal} {\bibinfo  {journal} {Phys. Rev. A}\ }\textbf {\bibinfo
  {volume} {99}},\ \bibinfo {pages} {033610} (\bibinfo {year}
  {2019})}\BibitemShut {NoStop}%
\end{thebibliography}%
\let\addcontentsline\oldaddcontentsline%

\end{document}